\documentclass[11pt]{article}

\newcommand{\cmp}{Comm. Math. Phys.~}

\newcommand{\jmp}{J. Math. Phys.~}

\newcommand{\prl}{Phys. Rev. Lett.~}
\newcommand{\pra}{Phys. Rev. A~}

\newcommand{\pla}{Phys. Lett. A~}

\usepackage{young}
\usepackage[latin1]{inputenc}
\usepackage{appendix}
\usepackage{epstopdf}
\usepackage{xcolor}
\usepackage{subfigure}
\usepackage[T1]{fontenc}
\usepackage[sc]{mathpazo}
\usepackage{amsmath}
\usepackage{amssymb}
\usepackage{graphicx,color}
\usepackage{framed}
\usepackage{multirow}
\usepackage{enumerate}
\usepackage{amsthm}
\usepackage{amsfonts,mathrsfs}
\usepackage{geometry} 
\usepackage{eepic}
\usepackage{ifthen}
\usepackage[vcentermath]{youngtab}
\usepackage[unicode=true,pdfusetitle, bookmarks=true,bookmarksnumbered=false,bookmarksopen=false, breaklinks=false,pdfborder={0 0 0},backref=false,colorlinks=false] {hyperref}
\hypersetup{
colorlinks,linkcolor=myurlcolor,citecolor=myurlcolor,urlcolor=myurlcolor}
\definecolor{myurlcolor}{rgb}{0,0,0.7}

\usepackage{color}

\newcommand{\blue}{\textcolor{blue}}

\newcommand{\proj}[1]{| #1\rangle\!\langle #1 |}

\usepackage{hyperref}
\hypersetup{pdfpagemode=UseNone}

\newcommand{\tinyspace}{\mspace{1mu}}

\newcommand{\op}[1]{\operatorname{#1}}

\newcommand{\abs}[1]{\left\lvert\tinyspace #1 \tinyspace\right\rvert}

\newcommand{\norm}[1]{\left\lVert\tinyspace #1 \tinyspace\right\rVert}

\renewcommand{\det}{\operatorname{det}}
\renewcommand{\t}{{\scriptscriptstyle\mathsf{T}}}

\newcommand{\setft}[1]{\mathrm{#1}}
\newcommand{\lin}[1]{\setft{L}\left(#1\right)}

\newcommand{\pd}[1]{\setft{Pd}\left(#1\right)}

\newcommand{\rank}{\op{rank}}

\def\haar{\mathrm{Haar}}

\def\ee{\mathrm{e}}

\def \dif {\mathrm{d}}
\def \diag {\mathrm{diag}}

\def\I{\mathbb{1}}

\def\zero{\mathbf{0}}

\def\bdlambda{\boldsymbol{\lambda}}
\def\bdalpha{\boldsymbol{\alpha}}
\def\bdbeta{\boldsymbol{\beta}}

\def\bdsigma{\boldsymbol{\sigma}}
\def\bdmu{\boldsymbol{\mu}}

\def\bdxi{\boldsymbol{\xi}}

\newenvironment{mylist}[1]{\begin{list}{}{
    \setlength{\leftmargin}{#1}
    \setlength{\rightmargin}{0mm}
    \setlength{\labelsep}{2mm}
    \setlength{\labelwidth}{8mm}
    \setlength{\itemsep}{0mm}}}
    {\end{list}}

\def\ot{\otimes}

\newcommand{\inner}[2]{\langle #1 , #2\rangle}

\newcommand{\out}[2]{| #1\rangle\langle #2 |}
\newcommand{\Inner}[2]{\left\langle #1 , #2\right\rangle}
\newcommand{\Innerm}[3]{\left\langle #1 \left| #2 \right| #3 \right\rangle}

\newcommand{\Herm}{\mathrm{Herm}}

\newcommand{\pa}[1]{(#1)}
\newcommand{\Pa}[1]{\left(#1\right)}

\newcommand{\Br}[1]{\left[#1\right]}
\newcommand{\set}[1]{\{#1\}}
\newcommand{\Set}[1]{\left\{#1\right\}}

\newcommand{\bra}[1]{\langle#1|}

\newcommand{\ket}[1]{|#1\rangle}

\DeclareMathOperator{\trace}{Tr}
\newcommand{\ptr}[2]{\trace_{#1}\pa{#2}}
\newcommand{\Ptr}[2]{\trace_{#1}\Pa{#2}}

\newcommand{\Tr}[1]{\Ptr{}{#1}}

\newcommand{\Abs}[1]{\left|\tinyspace#1\tinyspace\right|}

\def\cC{\mathcal{C}}\def\cE{\mathcal{E}}
\def\cH{\mathcal{H}}\def\cI{\mathcal{I}}\def\cJ{\mathcal{J}}
\def\cN{\mathcal{N}}
\def\cP{\mathcal{P}}\def\cR{\mathcal{R}}\def\cT{\mathcal{T}}

\def\cZ{\mathcal{Z}}

\def\bsA{\boldsymbol{A}}\def\bsB{\boldsymbol{B}}\def\bsC{\boldsymbol{C}}\def\bsD{\boldsymbol{D}}\def\bsE{\boldsymbol{E}}
\def\bsG{\boldsymbol{G}}\def\bsH{\boldsymbol{H}}\def\bsI{\boldsymbol{I}}
\def\bsK{\boldsymbol{K}}\def\bsM{\boldsymbol{M}}
\def\bsP{\boldsymbol{P}}\def\bsQ{\boldsymbol{Q}}\def\bsR{\boldsymbol{R}}\def\bsT{\boldsymbol{T}}
\def\bsU{\boldsymbol{U}}\def\bsV{\boldsymbol{V}}\def\bsW{\boldsymbol{W}}\def\bsX{\boldsymbol{X}}\def\bsY{\boldsymbol{Y}}
\def\bsZ{\boldsymbol{Z}}

\def\bsa{\boldsymbol{a}}\def\bsb{\boldsymbol{b}}\def\bsc{\boldsymbol{c}}\def\bse{\boldsymbol{e}}
\def\bsg{\boldsymbol{g}}

\def\bsp{\boldsymbol{p}}\def\bsq{\boldsymbol{q}}\def\bsr{\boldsymbol{r}}\def\bst{\boldsymbol{t}}
\def\bsu{\boldsymbol{u}}\def\bsv{\boldsymbol{v}}\def\bsx{\boldsymbol{x}}\def\bsy{\boldsymbol{y}}
\def\bsz{\boldsymbol{z}}

\def\bu{\mathbf{u}}

\def\rD{\mathrm{D}}
\def\rG{\mathrm{G}}
\def\rL{\mathrm{L}}

\def\rU{\mathrm{U}}

\def\sE{\mathscr{E}}
\def\sF{\mathscr{F}}
\def\sL{\mathscr{L}}

\def\B{\textsf{B}}\def\C{\textsf{C}}\def\D{\textsf{D}}
\def\G{\textsf{G}}\def\H{\textsf{H}}                  
\def\K{\textsf{K}}\def\L{\textsf{L}}
\def\Q{\textsf{Q}}\def\S{\textsf{S}}
\def\V{\textsf{V}}\def\W{\textsf{W}}

\def\bbC{\mathbb{C}}\def\bbE{\mathbb{E}}

\def\bbR{\mathbb{R}}\def\bbT{\mathbb{T}}

\def\bbZ{\mathbb{Z}}

\def\sfB{\mathsf{B}}\def\sfC{\mathsf{C}}
\def\sfG{\mathsf{G}}\def\sfI{\mathsf{I}}

\def\sfQ{\mathsf{Q}}\def\sfS{\mathsf{S}}
\def\sfV{\mathsf{V}}

\newtheorem{thrm}{Theorem}[section]
\newtheorem{lem}[thrm]{Lemma}
\newtheorem{prop}[thrm]{Proposition}
\newtheorem{cor}[thrm]{Corollary}
\theoremstyle{definition}
\newtheorem{definition}[thrm]{Definition}
\newtheorem{remark}[thrm]{Remark}

\numberwithin{equation}{section}

\newcounter{questionnumber}

\usepackage{esint}

\begin{document}

\title{Subadditivity and the failure of strong subadditivity \\ in the Nichols--Wootters family of entropies}

\author{\blue{Lin Zhang}$^{1}$\footnote{E-mail: godyalin@163.com}, \quad \blue{Wenlin Zhang}$^{2}$\footnote{E-mail: e1327962@u.nus.edu}, \quad\blue{Haobo Ma}$^{3}$\footnote{E-mail: auric@aelf.io}\\
  {\it\small $^{1}$School of Mathematical Sciences, Hangzhou Dianzi University, Hangzhou 310018, China}\\
{\it\small $^{2}$The Omega Institute, National University of Singapore, Singapore}\\
{\it\small $^{3}$The Omega Institute, ChronoAI Pte Ltd, Singapore}
}

\date{}
\maketitle

\begin{abstract}
We prove subadditivity for the Nichols--Wootters family of entropies interpolating between the von Neumann
entropy and the subentropy, with all spectra evaluated in one common ambient dimension. Beyond the von Neumann
entropy, equality holds exactly when at least one marginal is pure. This settles the subadditivity conjecture concerning subentropy. Every member except the von Neumann entropy fails strong subadditivity, already on a maximally
entangled two-qubit state tensored with a mixed qubit. Its strong-subadditivity defect takes both signs on quantum
Markov states. For a pure bipartite state tensored with an ancillary state, we identify the defect with a separately
concave product gap and determine its exact maximum under rank constraints. For arbitrary bipartite states, we
obtain an explicit lower bound in the marginal purities. For two-qubit states, we determine the optimal coefficient
multiplying the product of the smaller marginal eigenvalues in every common ambient dimension, including the
subentropy coefficient $1/3$. A positive representation by Haar averages of compressed matrix entropies gives the
pinching inequality underlying these results and also yields entropy-gain bounds for unital quantum channels.
As an application, each entropy equals the classical communication capacity retained after a specified isotropic partial
projective measurement, at a prescribed average input state, even when entangled block encodings are allowed.
\end{abstract}
\newpage
\tableofcontents
\newpage
%
\section{Introduction}

Does subadditivity extend from the von Neumann entropy to the subentropy and to the entropies between them? We answer this question for the Nichols-Wootters family of entropies: subadditivity holds throughout the family when the spectra are evaluated in one common ambient dimension. We also characterize equality and prove that strong subadditivity fails for every member except the von Neumann entropy. The subentropy endpoint therefore has a definite position: it is subadditive on arbitrary bipartite states, with equality exactly when a marginal is pure, and it is not strongly subadditive.

The two endpoints arise from different information-theoretic questions. The von Neumann entropy $\S(\rho)=-\Tr{\rho\log\rho}$ supplies the familiar
entropy inequalities of quantum information theory. Subentropy $\Q(\rho)$ was introduced by Jozsa, Robb and Wootters~\cite{Jozsa1994}
as a lower bound on the accessible information of a pure-state ensemble that depends only on its average state. Nichols and
Wootters~\cite{Nichols2003} placed these quantities in a family $\S_1\geqslant \S_2\geqslant\cdots\geqslant \S_d$, with $\S_1=\S$ and $\S_d=\Q$.
Their construction uses the elementary symmetric polynomials of the resolvent; Jozsa and Mitchison~\cite{Jozsa2015} developed the related
symmetric-polynomial description of the endpoints.

Datta, Dorlas, Jozsa and Benatti proved concavity and product-state subadditivity of subentropy and conjectured subadditivity
for arbitrary bipartite states~\cite[Conjecture 1 and Proposition 3]{Datta2014}. Zhang, Singh and Pati~\cite[Appendix 7.1]{Zhang2017a} subsequently asked both the bipartite
and the strong tripartite subadditivity questions. The familiar von Neumann inequalities~\cite{Araki1970,Lieb1973} do not settle
these questions: the intermediate entropies are not additive on product states, and a fixed intermediate index changes under the addition of
zero eigenvalues. The ambient dimension must therefore be specified before formulating an entropy inequality.

Our first result is subadditivity for every finite-dimensional bipartite state and every member of the family. At $k=1$, equality has
the usual characterization by product states. For $k\geqslant2$, equality requires a pure marginal. The same argument gives multipartite
subadditivity and a triangle inequality, with a complete equality characterization for the latter. At the subentropy endpoint, zero-padding invariance makes these statements inequalities for the original physical states.

The second result identifies the obstruction to strong subadditivity. For $\rho_{ABC}=\proj{\psi}_{AB}\otimes\tau_C$, the defect is exactly the product gap of the Schmidt spectrum of $\ket{\psi}$ and the spectrum of $\tau$. As we show in Section~\ref{sect:3}, it is positive for every $k\geqslant2$ precisely when $\ket{\psi}$ is
entangled and $\tau$ is mixed. In particular, $\proj{\Omega_2}_{AB}\otimes(\frac12\I_2)_C$ has subentropy defect $1/12$, although its von Neumann conditional mutual information is zero. A three-factor product state with the same two mixed spectra has the opposite defect. Thus the defect has both signs even on quantum Markov states. Separate strict concavity determines the largest defect on the pure-bipartite-factor family under Schmidt-rank and ancilla-rank bounds. The same positive product gap thus measures strict bipartite subadditivity and produces a violation of the tripartite inequality.

We also quantify the strictness of bipartite subadditivity. For two-qubit states with $a=\lambda_{\min}(\rho_A)$ and $b=\lambda_{\min}(\rho_B)$,
the subentropy gap is at least $ab/3$, with optimal coefficient $1/3$. In common dimension four, the optimal coefficients for $\S_2$ and $\S_3$
are both $2/3$. The augmentation formula then determines the optimal coefficient in every common dimension $D\geqslant4$. In arbitrary local
dimensions, a Gaussian representation of the subentropy gap of the locally pinched state as an averaged classical mutual information gives an explicit positive lower
bound in the two marginal purity deficits. A trace-distance consequence shows that near equality forces the state to be close to a product state with a pure marginal.

The common mechanism is a positive representation of $\S_k$ by Haar averages of compressed matrix entropies. Joint convexity of relative
entropy makes the pinching gain convex, which gives an upper bound in terms of the block-mass distribution. A Gram-matrix construction then
proves subadditivity. The rank-one compression component transfers strict positivity from subentropy to every $k\geqslant2$. The representation also gives an adapted divergence and a sharp comparison of entropy gains for unital quantum channels.

The same representation provides an operational application of the entropy inequalities. We construct a hierarchy of randomly oriented partial projective measurements, with uniformly chosen cuts of an ordered basis. When the receiver retains the setting, the measurement outcome and the surviving quantum state, its classical communication capacity at prescribed average input $\rho$ is exactly $\S_k^{(d)}(\rho)$. The proof includes arbitrary entangled block encodings. Adding a uniformly chosen cut is a receiver-side degradation, so the interpolation describes the loss of reliable communication under progressively finer measurement. In a common ambient dimension, our subadditivity and stability results therefore compare the capacities of this same detector at different input means.

Section~\ref{sect:2} reviews the Nichols--Wootters (NW) family of entropies and introduces the common-dimension convention. The ordering, spectral concavity, uniform-spectrum maximum and general augmentation formula belong to the established structure of the family~\cite{Nichols2003}; Appendix~\ref{app:A} supplies derivations and boundary extensions in the notation used here. The positive compression representation and pinching argument in Section~\ref{subsubsect:4.1.1} provide the mechanism for our main inequalities. Section~\ref{sect:3} collects the main results, and Section~\ref{sect:application} presents their operational application. Section~\ref{sect:proofs} explains the proof mechanisms of these results and their dependencies, and provides complete proofs motivated by the method given in the appendices. Section~\ref{sect:5} discusses the resulting distinction between the von Neumann entropy and the remaining family and presents open questions.

\section{Preliminaries}\label{sect:2}

\subsection{Notation and the entropy family}

All Hilbert spaces carrying quantum inputs or residual systems are finite-dimensional, and all logarithms are natural. We write $\Herm(\bbC^d)$ for the Hermitian operators and $\rD(\bbC^d)$ for the density operators on $\bbC^d$, respectively, $\bdlambda(\rho)=(\lambda_1,\ldots,\lambda_d)$ for the spectrum in \emph{nonincreasing order}, and $H_k=\sum_{j=1}^k1/j$, with $H_0=0$ throughout this paper. Set $\eta(x)=-x\log x$ for $x>0$ and $\eta(0)=0$. For a complex $d\times d$ matrix $\bsM$, the elementary symmetric
polynomials $\ee_k(\bsM)$ are defined by 
$$
 \det(t\I_d-\bsM)=\sum_{k=0}^d(-1)^k\ee_k(\bsM)t^{d-k}.
$$
In particular, $\ee_0(\bsM)=1, \ee_1(\bsM)=\Tr{\bsM}$ and $\ee_d(\bsM)=\det(\bsM)$.

For $z\neq0$ that is not an eigenvalue of $\rho$, put $\bsR_\rho(z)=(\I_d-z^{-1}\rho)^{-1}$.
\begin{definition}[$k$-entropy]\label{def:kentropy}
For $\rho\in\rD(\bbC^d)$ and $1\leqslant k\leqslant d$, define
\begin{eqnarray}\label{eq:EntkSk}
\S_k^{(d)}(\rho):=\binom{d-1}{k-1}^{-1}\op{Ent}_k^{(d)}(\rho),\qquad 
 \op{Ent}_k^{(d)}(\rho):=-\frac1{2\pi\mathrm i}
 \oint \ee_k[\bsR_\rho(z)](\log z)\dif z.
\end{eqnarray}
The positively oriented contour encloses all nonzero eigenvalues and avoids zero and the principal logarithm's cut along the negative real
axis. Repeated and zero eigenvalues are included by continuous extension.
\end{definition}
This is the normalized Nichols--Wootters family~\cite{Nichols2003}. Its endpoints are $\S_1^{(d)}=\S$ and $\S_d^{(d)}=\Q$.
The binomial factor is part of the definition throughout this paper.

\subsection{Structural identities and properties}

The following representations will be used in the results and proofs; the properties listed below them are consequences of these representations and are proved in Appendix~\ref{app:A}. For $\bsx=(x_1,\ldots,x_k)\in\bbR^k_{>0}$, put
$$
 \Phi_k(\bsx):=-[x_1,\ldots,x_k](x^k\ln x),\qquad  \S_k^{(d)}(\rho)=\binom{d-1}{k-1}^{-1}  \sum_{|\cI|=k}\Phi_k(\bdlambda_{\cI}).
$$
Here the divided difference is interpreted confluently at coincident
arguments. If $\bsT$ is uniform on the $(k-1)$-dimensional probability simplex $\Delta_{k-1}$ and $\cI$ is an
independent uniform $k$-subset of $[d]:=\set{1,\ldots,d}$, then
$$
\S_k^{(d)}(\rho)=d\cdot \bbE_{\cI,\bsT}\Br{ \eta(\Inner{\bsT}{\bdlambda_{\cI}})} - (H_k-1).
$$
From these representations, the family is continuous, unitarily invariant, strictly concave and Schur concave. Its range and hierarchy are
$$
 0\leqslant \S_k^{(d)}(\rho)\leqslant\ln d-H_k+1,\qquad  \S(\rho)=\S_1^{(d)}(\rho)\geqslant\cdots\geqslant \S_d^{(d)}(\rho)=\Q(\rho).
$$
The minimum occurs exactly at pure states and the maximum exactly at $\I_d/d$. Every adjacent inequality is strict for a mixed state.

\subsection{The common ambient dimension}

For a state $\sigma\in\rD(\bbC^\ell)$ and $D\geqslant \ell$, write
$\sigma^{[D]}=\sigma\oplus0_{D-\ell}$.
We occasionally write $\S_k^{(D)}(\sigma)$ with this padding understood. That is,
$$
\S_k^{(D)}(\sigma)\equiv \S_k^{(D)}(\sigma^{[D]}).
$$
For a probability vector, the same notation refers to the diagonal
state with that spectrum.
Appending a zero eigenvalue gives, for $1\leqslant k\leqslant d$,
$$
\S_k^{(d+1)}(\rho\oplus0)  =\frac{d-k+1}{d}\S_k^{(d)}(\rho)  +  \frac{k-1}{d}\S_{k-1}^{(d)}(\rho),
$$
with the second term omitted at $k=1$. The general augmentation formula is \eqref{eq:padding-normalized}. Thus $\S$ and $\Q$ are invariant under zero padding, with their respective endpoint indices, whereas a fixed intermediate $\S^{(d)}_k$ generally is not.

For a bipartite state on $\bbC^m\ot\bbC^n$, fix $D\geqslant mn$ and define
\begin{eqnarray}\label{eq:gap-definition}
 \widetilde\Delta_k^{(D)}(\rho_{AB}):= \S_k^{(D)}(\rho_A)+\S_k^{(D)}(\rho_B) - \S_k^{(D)}(\rho_{AB}),\qquad 1\leqslant k\leqslant D.
\end{eqnarray}
At $k=1$, this is the mutual information; at $k=D$, it is
$\Delta_{\Q}:=\Q(\rho_A) + \Q(\rho_B) - \Q(\rho_{AB})$.
The convention is essential. For $2\leqslant k\leqslant\min(m,n)$,
\begin{eqnarray}\label{eq:native-counterexample}
\S_k^{(m)}\Pa{\frac{\I_m}m}+ \S_k^{(n)}\Pa{\frac{\I_n}n}- \S_k^{(mn)}\Pa{\frac{\I_{mn}}{mn}} =1-H_k<0.
\end{eqnarray}
Thus a common ambient dimension is required for the deficit to be nonnegative, and all multipartite deficits below therefore use one common dimension.

\subsection{Compressions and the adapted divergence}

Sections~\ref{sect:3} and \ref{sect:proofs} use two tools that are not needed for the previous statements but are used repeatedly in the proofs: Haar averages of compressed matrix entropies over the complex Stiefel manifold \cite{Coulembier2015}, and a Bregman divergence \cite{Bregman1967,Petz2007,Pitrik2015} adapted to $\S_k^{(d)}$. We collect their definitions and the properties used later here, so that Section~\ref{sect:proofs} can cite them rather than introduce them. The positive compression representation is proved in Section~\ref{subsubsect:4.1.1}; the underlying Stiefel-integral facts and the Bregman-divergence identities are proved in Appendices~\ref{app:B} and \ref{app:C}, respectively.

\paragraph{Compressions and the positive compression representation.}

For $1\leqslant \ell\leqslant d$, the \emph{complex Stiefel manifold} is
$$
\op{St}(d,\ell):=\Set{\bsV\in M_{d,\ell}(\bbC): \bsV^\dagger \bsV=\I_\ell},
$$
equipped with its normalized Haar probability measure $\mu_\ell$. For $\bsX\geqslant\zero$ acting on $\bbC^d$, define the \emph{compressed entropy functional}
\begin{eqnarray}\label{eq:compressedentropy}
\G_\ell(\bsX):=\int_{\op{St}(d,\ell)}\Tr{\eta(\bsV^\dagger \bsX \bsV)}\dif \mu_\ell(\bsV),\qquad 1\leqslant \ell\leqslant d, 
\end{eqnarray}
where $\eta(x)=-x\log x$ is applied by functional calculus and the compressed matrix $\bsV^\dagger \bsX \bsV$ is not divided by its trace. The dimension $d$ is fixed in the notation $\G_\ell$. Indeed,
$\zero\leqslant\bsV^\dagger\bsX\bsV\leqslant\norm{\bsX}\I_\ell$
for every $\bsV\in\op{St}(d,\ell)$. Continuity of $\eta$ on
$[0,\norm{\bsX}]$, including at zero, gives
\begin{eqnarray*}
\Abs{\Tr{\eta(\bsV^\dagger\bsX\bsV)}}
\leqslant\ell\max_{0\leqslant x\leqslant\norm{\bsX}}|\eta(x)|.
\end{eqnarray*}
The integrand is continuous on the compact Stiefel manifold and is thus
integrable also when $\bsX$ or a compression is singular. If $\bsX=\zero$,
it vanishes identically. Extend $\S_k^{(d)}$ to the positive semidefinite cone by the same normalized spectral formula as in Proposition~\ref{prop:eigfor}; throughout this positive-cone extension, no renormalization by $\Tr{\bsX}$ is performed. On trace-one matrices it agrees with Definition~\ref{def:kentropy}.

\begin{prop}[Positive compression representation] \label{prop:posrep}
For $2\leqslant k\leqslant d$ and every $\bsX\geqslant\zero$ on $\bbC^d$,
\begin{eqnarray}\label{eq:kentropy-intrep}
\sfS_k^{(d)}(\bsX)=\sum_{\ell=1}^{d-k+1}\mathfrak{C}_{d,k,\ell}\sfG_\ell(\bsX)-(H_k-1)\Tr{\bsX},
\end{eqnarray}
with positive coefficients
\begin{eqnarray}\label{eq:constantC}
\mathfrak{C}_{d,k,\ell}=\frac{k}{\binom{d-1}{k-1}}\binom{d-\ell-1}{k-2},\qquad 1\leqslant \ell\leqslant d-k+1. 
\end{eqnarray}
For $k=1$, the corresponding identity is $\sfS_1^{(d)}(\bsX)=\sfG_d(\bsX)=\Tr{\eta(\bsX)}$.
\end{prop}
The representation \eqref{eq:kentropy-intrep} is the common mechanism behind subadditivity, the pinching inequality, and the quantitative bounds. Its proof rests on the continuous-function identity
\begin{eqnarray}\label{eq:continuous-function-identity}
\sum_{|\cI|=k}\bbE_{\bsT} \Br{f\Pa{\Inner{\bsT}{\bdlambda_{\cI}}}} =
\sum_{\ell=1}^{d-k+1}\binom{d-\ell-1}{k-2}\int_{\op{St}(d,\ell)}\Tr{f(\bsV^\dagger \bsX \bsV)}\dif\mu_\ell(\bsV), 
\end{eqnarray}
valid for $2\leqslant k\leqslant d$ and every continuous $f$ on $[0,\norm{\bsX}]$, where $\bsT$ is uniform on $\Delta_{k-1}$ and $\cI$ is a uniform $k$-subset of $[d]$. The Haar integral in \eqref{eq:continuous-function-identity} is evaluated by the Schur-Weyl formula
\begin{eqnarray}\label{eq:Schur-Weylformula}
\int_{\op{St}(d,\ell)}s_\nu(\bsV^\dagger \bsX \bsV)\dif\mu_\ell(\bsV)=\frac{s_\nu(1^\ell)}{s_\nu(1^d)} s_\nu(\bsX), 
\end{eqnarray}
for every partition $\nu$ with at most $d$ rows; both sides vanish when $\nu$ has more than $\ell$ rows. Formula \eqref{eq:Schur-Weylformula} extends to singular $\bsX$ by polynomial representation theory. The derivation of Eqs.~\eqref{eq:kentropy-intrep}--\eqref{eq:Schur-Weylformula}, including the coefficient identity that produces \eqref{eq:constantC}, is given in Section~\ref{subsubsect:4.1.1}; the Stiefel-integral facts underlying \eqref{eq:Schur-Weylformula} are collected in Appendix~\ref{app:B}.

Two consequences are used repeatedly:
\begin{itemize}
\item $\G_\ell$ is \emph{unitarily invariant} and \emph{concave}, since compression is linear and $\bsX\mapsto \Tr{\eta(\bsX)}$ is concave;
\item for a pure state $\proj{\psi}$ on $\bbC^d$,
\begin{eqnarray}\label{eq:value@purestate}
g_\ell:=\G_\ell(\proj{\psi})=\frac{\ell}{d}(H_d-H_\ell), 
\end{eqnarray}
so the functional $\widetilde\G_\ell(\rho):=\G_\ell(\rho)-g_\ell$ vanishes on every pure state and is nonnegative and concave on $\rD(\bbC^d)$. On density matrices,
\begin{eqnarray}\label{eq:Sk-Jell}
\S_k^{(d)}=\sum_{\ell=1}^{d-k+1}\mathfrak{C}_{d,k,\ell}\widetilde\G_\ell, 
\end{eqnarray}
which is the form used in the equality analysis and in the quantitative bounds.
\end{itemize}

\paragraph{The generalized relative entropy $\B_+$.}

For positive semidefinite matrices of the same size, set
\begin{eqnarray}\label{eq:GRE}
\B_+(\bsX||\bsY):=\Tr{\bsX(\log \bsX-\log \bsY)}-\Tr{\bsX-\bsY}, 
\end{eqnarray}
with $0\log 0=0$ and $\B_+(\bsX||\bsY)=+\infty$ if $\op{supp}[\bsX]\not\subseteq \op{supp}[\bsY]$. The trace correction $-\Tr{\bsX-\bsY}$ is essential here: compressed matrices $\bsV^\dagger \bsX\bsV, \bsV^\dagger \bsY\bsV$ generally have different traces, and without the correction the divergence would not be nonnegative. For trace-one arguments \eqref{eq:GRE} reduces to the ordinary relative entropy.

\begin{prop}\label{prop:GREproperty} 
The generalized relative entropy $\sfB_+$ is nonnegative, jointly convex, and lower semicontinuous on pairs of positive semidefinite matrices, with the support convention above; joint convexity holds for matrices of arbitrary trace. It is finite whenever $\op{supp}[\bsX]\subseteq\op{supp}[\bsY]$, and
$$
\sfB_+(\bsX||\bsY)=\lim_{\epsilon\downarrow 0}\sfB_+(\bsX+\epsilon\I||\bsY+\epsilon\I).
$$
\end{prop}
This result is proved in Appendix~\ref{app:C}.

\paragraph{The adapted divergence $\B_k^{(d)}$.}

Before introducing the adapted divergence, we recall the general notion. For a convex, differentiable function $F$ on a real inner-product space, the Bregman divergence of $F$ is 
\begin{eqnarray}\label{eq:Def-for-Breg-Div}
\B_F(\bsX||\bsY):=F(\bsX) - F(\bsY) - \Inner{\nabla F(\bsY)}{\bsX-\bsY}.
\end{eqnarray}
We apply this construction with $F=F_k:=-\S_k^{(d)}$ and the real Hilbert--Schmidt inner product on Hermitian matrices; the resulting divergence is denoted $\B_k^{(d)}$. The definition \eqref{eq:Def-for-Breg-Div} is the standard one; its operator-valued extension \cite{Petz2007} and the derivation of the compressed representation below are given in Appendix~\ref{app:C}.

Rather than take \eqref{eq:Def-for-Breg-Div} as the operative definition, we use the following structural identity, which exhibits $\B_k^{(d)}$ as a positive weighted Haar average of the standard divergence $\B_+$ over compressed matrices. This identity is what makes the divergence match the entropy gain throughout the family, and it is the form cited in Section~\ref{proof:entropygains}.

\begin{prop}[Bregman divergence of the negative $k$-entropy]\label{prop:BD-for-kentropy}
For $2\leqslant k\leqslant d$ and positive definite $\bsX,\bsY$ on $\bbC^d$,
\begin{eqnarray}\label{eq:BregDivforkentropy}
\sfB_k^{(d)}(\bsX||\bsY) =\sum_{\ell=1}^{d-k+1}\mathfrak{C}_{d,k,\ell}\int_{\op{St}(d,\ell)}
\sfB_+(\bsV^\dagger \bsX \bsV||\bsV^\dagger \bsY \bsV)\dif\mu_\ell(\bsV), 
\end{eqnarray}
with the coefficients \eqref{eq:constantC}. For $k=1,\sfB_1^{(d)}=\sfB_+$. Equivalently, $\sfB_k^{(d)}$ is the Bregman divergence
\begin{eqnarray}\label{eq:adaptedBD}
\sfB_k^{(d)}(\bsX||\bsY)=F_k(\bsX)-F_k(\bsY)-\Inner{\sL_k(\bsY)}{\bsX-\bsY},\qquad \sL_k:=\nabla F_k, 
\end{eqnarray}
and the two expressions agree because the affine trace term in \eqref{eq:kentropy-intrep} cancels in the Bregman difference. 
\end{prop}
The following properties follow from \eqref{eq:BregDivforkentropy}, Proposition~\ref{prop:GREproperty} and the representation in Proposition~\ref{prop:posrep}, and are recorded for use in Section~\ref{proof:entropygains}:
\begin{enumerate}
\item \textbf{Nonnegativity and joint convexity:}  $\B_k^{(d)}$ is nonnegative, jointly convex, and lower semicontinuous on pairs of positive semidefinite matrices, with the support convention of $\B_+$ applied separately to each compressed pair in \eqref{eq:BregDivforkentropy}; joint convexity holds for arbitrary trace.
\item \textbf{Unitary invariance:} $\B_k^{(d)}(\bsU\bsX\bsU^\dagger||\bsU\bsY\bsU^\dagger)=\B_k^{(d)}(\bsX||\bsY)$ for every unitary $\bsU$, by the change of variables $\bsV\mapsto \bsU^\dagger \bsV$ in the Haar integral.
\item \textbf{Bregman identity:} For $\bsX,\bsY>\zero$, \eqref{eq:adaptedBD} holds with $\sL_k(\bsY)=\nabla F_k(\bsY)$. At the endpoints, $\sL_1(\bsY)=\log \bsY+\I$ and $\B_1^{(d)}=\B_+$.
\item \textbf{Endpoint:} At $k=d$, \eqref{eq:BregDivforkentropy} is the scalar-compression formula
$$
\B_d^{(d)}(\bsX||\bsY)=d\int_{\op{St}(d,1)}\B_+(\Innerm{\bsv}{\bsX}{\bsv}||\Innerm{\bsv}{\bsY}{\bsv})\dif\mu_1(\bsv).
$$
\end{enumerate}
The proof of Proposition~\ref{prop:GREproperty}, together with the Bregman identity \eqref{eq:adaptedBD} and the endpoint formulas, is given in Appendix~\ref{app:C}.

\paragraph{Why the adapted divergence, and not ordinary relative entropy.}

The divergence $\B_k^{(d)}$ cannot be replaced by the ordinary relative entropy $\B_+$ without losing the lower bounds of Section~\ref{proof:entropygains}. For computational dephasing of a qubit, take $\rho=\proj{+}$ with $\ket{+}=\frac{\ket{0}+\ket{1}}{\sqrt2}$; then $\cE(\rho)=\frac12\I_2=\cE^\dagger(\cE(\rho))$ and
$$
\B_+(\rho||\cE^\dagger\circ\cE(\rho))=\log 2, \qquad \Q(\cE(\rho))-\Q(\rho)=\log 2-\tfrac12 .
$$
Thus the ordinary relative entropy exceeds the subentropy gain, and the lower bound in Theorem~\ref{th:channel} fails at $k=d$ if $\B_k^{(d)}$ is replaced by $\B_+$. The trace correction in \eqref{eq:GRE} and the Haar averaging in \eqref{eq:BregDivforkentropy} are exactly what make the adapted divergence match the entropy gain throughout the family. This is discussed further in Section~\ref{subsubsect:4.6.7}.
\paragraph{What is used where.} 

For the reader's orientation, the tools of this subsection enter the later results as follows.
\begin{itemize}
\item \emph{Pinching lemma (Lemma~\ref{lem:pinching})} The lower bound is Schur concavity; the upper bound uses \eqref{eq:BregDivforkentropy} to write the pinching gain as a positive sum of compressed divergences, hence as a convex functional of the input, and then evaluates it on pure states via \eqref{eq:value@purestate}. See Section~\ref{proof:compression-pinching}.
\item \emph{Subadditivity (Theorem~\ref{th:subadditivity}) and equality (Theorem~\ref{th:equality})} The two-pinch argument of Section~\ref{proof:subadditivityequality} uses Lemma~\ref{lem:pinching} together with the decomposition \eqref{eq:Sk-Jell} and the rank-one term $\widetilde \G_1=\Q/d$; the latter transfers strict positivity from subentropy to every $k\geqslant 2$.
\item \emph{Quantitative bounds (Section~\ref{subsect:QB})} The comparison \eqref{eq:transfer-main} between $\widetilde\Delta_k^{(D)}$ and the subentropy deficit is proved by summing the exact remainders $\delta_\ell\geqslant 0$ obtained from \eqref{eq:Sk-Jell} and the pinching inequality.
\item \emph{Unital channels (Theorem~\ref{th:channel})} Section~\ref{subsubsect:4.6.2} proves the general pinched recovery bound, Section~\ref{subsubsect:4.6.3} establishes the optimal gain comparison, Section~\ref{subsubsect:4.6.4} gives the unpinched bounds at the endpoint indices, and Section~\ref{subsubsect:4.6.6} proves the entropy-exchange bounds.
\end{itemize}
Accordingly, the statements of this subsection are cited in Section~\ref{sect:proofs} only by number, and their proofs are deferred to Appendices \ref{app:B} and \ref{app:C}.

\section{Main results}\label{sect:3}

Subadditivity holds throughout the family, while strong subadditivity
fails beyond the von Neumann endpoint. The product gap connects these
two results: it is positive on products of mixed spectra and becomes
the SSA defect on a pure bipartite factor tensored with an ancilla.
We then maximize this defect under rank constraints, quantify the
bipartite gap, and apply the common pinching mechanism to channels.
An operational application then identifies the family with the classical
communication rates retained after partial measurements.
The corresponding subsections of Section~\ref{sect:proofs} provide full proofs of these statements.

\subsection{Subadditivity and equality}\label{subsect:SAE}

\begin{thrm}[Subadditivity]\label{th:subadditivity}
For every $\rho_{AB}\in\rD(\bbC^m\otimes\bbC^n)$, every $D\geqslant mn$, and $1\leqslant k\leqslant D$,
\begin{eqnarray}\label{eq:saturationsub}
\sfS_k^{(D)}(\rho_{AB}) \leqslant \sfS_k^{(D)}(\rho_A) + \sfS_k^{(D)}(\rho_B).
\end{eqnarray}
\end{thrm}

\begin{thrm}[Equality]\label{th:equality}
Under the hypotheses of Theorem~\ref{th:subadditivity}, it holds that
\begin{enumerate}[(i)]
\item For $k=1$, the inequality \eqref{eq:saturationsub} is saturated if and only if $\rho_{AB}=\rho_A\otimes\rho_B$;
\item For $2\leqslant k\leqslant D$, the inequality \eqref{eq:saturationsub} is saturated if and only if at least one marginal
is pure. In that case, $\rho_{AB}$ is a product state.
\end{enumerate}
\end{thrm}

\begin{cor}[Multipartite subadditivity]\label{cor:multipartite}
Let $\rho_{1\ldots r}\in\rD(\ot_{i=1}^r\bbC^{d_i})$, let
$\rho_i$ be its $i$-th marginal, and let $D\geqslant\prod^r_{i=1} d_i$.
For $1\leqslant k\leqslant D$,
$$
\sfS_k^{(D)}(\rho_{1\ldots r})  \leqslant\sum^r_{i=1} \sfS_k^{(D)}(\rho_i).
$$
\end{cor}

\begin{cor}[Triangle inequality]\label{cor:triangle}
Let $\rho_{AB}\in \rD(\bbC^m\ot\bbC^n)$ and let
$r=\rank(\rho_{AB})$. For $D\geqslant mnr$ and $1\leqslant k\leqslant D$,
\begin{eqnarray}\label{eq:Araki-Lieb}
 \Abs{\sfS_k^{(D)}(\rho_A) - \sfS_k^{(D)}(\rho_B)}  \leqslant \sfS_k^{(D)}(\rho_{AB}).
\end{eqnarray}
For $k\geqslant2$, the equality holds exactly when $\rho_{AB}$ is pure or a
marginal is pure. At the subentropy endpoint this gives, in the original
physical dimensions,
$$
\Abs{\sfQ(\rho_A) - \sfQ(\rho_B)}\leqslant \sfQ(\rho_{AB})
$$ 
with the same equality cases.
\end{cor}

\begin{cor}[Pure tripartite states]\label{cor:SSApure}
For a pure state $\rho_{ABC}\in\rD(\bbC^m\ot\bbC^n\ot\bbC^r)$, $D\geqslant mnr$ and $1\leqslant k\leqslant D$,
\begin{eqnarray}\label{eq:SSApure}
\sfS_k^{(D)}(\rho_{ABC}) + \sfS_k^{(D)}(\rho_B)  \leqslant \sfS_k^{(D)}(\rho_{AB}) + \sfS_k^{(D)}(\rho_{BC}).
\end{eqnarray}
For $k\geqslant2$, equality holds exactly when $\rho_A$ or $\rho_C$ is pure.
\end{cor}

\subsection{The $k$-product gap}\label{subsect:KPG}

In this section, the number $D$ is fixed, for spectra $\bdlambda,\bdmu$ of lengths $m,n$ with $mn\leqslant D$, and the product gap is
\begin{eqnarray}\label{eq:Dklambdamu}
\widetilde\Delta_k^{(D),\op{prod}}(\bdlambda,\bdmu)= \S_k^{(D)}(\bdlambda) + \S_k^{(D)}(\bdmu) - \S_k^{(D)}(\bdlambda\ot\bdmu).
\end{eqnarray}
The product gap is nonnegative:
$$
\widetilde \Delta_k^{(D),\mathrm{prod}}(\bdlambda,\bdmu)\geqslant0.
$$
At $k=1$ the gap is identically zero.

\begin{thrm}[Separate concavity]\label{th:sepconcavity}
The product gap is separately concave. For $2\leqslant k\leqslant D$, it is strictly
concave in either argument whenever the other argument is mixed.
\end{thrm}
For $k\geqslant2$, Theorem~\ref{th:equality} applied to the diagonal product
state gives
$$
\widetilde \Delta_k^{(D),\mathrm{prod}}(\bdlambda,\bdmu)=0 \quad\iff\quad \bdlambda\text{ or }\bdmu\text{ is pure}.
$$

\begin{thrm}[Fixed-rank maximum]\label{th:rank-max}
Let $\alpha,\beta\in\mathbb N$ with $\alpha,\beta\geqslant2$ and
$\alpha\beta\leqslant D$, and let $2\leqslant k\leqslant D$. Among probability vectors
with support sizes at most $\alpha,\beta$, respectively, represented as vectors
of lengths $\alpha,\beta$ by removing or adjoining zero entries,
$$
\widetilde \Delta_k^{(D),\mathrm{prod}}(\bdlambda,\bdmu) \leqslant d_{\alpha\beta}^{(D,k)}:=T_{D,k}(\alpha)+T_{D,k}(\beta)-T_{D,k}(\alpha\beta),
$$
with equality exactly at the uniform spectra of ranks $\alpha,\beta$, up to
permutations. The maximum is positive. If one rank bound is one, the gap
is identically zero.
\end{thrm}
Here
\begin{eqnarray}\label{eq:padded-uniform}
T_{D,k}(\ell)=\frac{1}{\binom{D-1}{k-1}}  \sum_{j=1}^{\min(k,\ell)}\binom{D-\ell}{k-j}\binom{\ell-1}{j-1}  (\ln \ell -H_j+1).
\end{eqnarray}
At $k=D$, formula \eqref{eq:padded-uniform} reduces to
$d_{\alpha\beta}^{(D,D)}=H_{\alpha\beta}-H_\alpha-H_\beta+1$.
At $k=1$ every pair attains the zero maximum, so uniqueness is not asserted.

\subsection{Failure of strong subadditivity}\label{subsect:FSSA}

Fix $D\geqslant mnr$ and evaluate every reduced spectrum in dimension $D$.
The same convention that gives bipartite subadditivity yields a simple
mixed-state obstruction to strong subadditivity.

\begin{definition}[$k$-SSA defect]
For a state $\rho=\rho_{ABC}\in\rD(\bbC^m\ot\bbC^n\ot\bbC^r)$, define
\begin{eqnarray}\label{eq:kSSAdefect}
\V_k^{(D)}(\rho)= \S_k^{(D)}(\rho_{ABC}) + \S_k^{(D)}(\rho_B) - \S_k^{(D)}(\rho_{AB}) - \S_k^{(D)}(\rho_{BC}).
\end{eqnarray}
Thus $\V_k^{(D)}>0$ means violation of strong subadditivity.
\end{definition}
At $k=1$, $\V_1^{(D)}=- \sfI(A:C|B)\leqslant0$ by von Neumann strong
subadditivity~\cite{Lieb1973}. The defect is symmetric in $A,C$.

\paragraph{Reduction to the $k$-product gap.} Let
\begin{eqnarray}\label{eq:counterexample}
 \rho_{ABC}=\proj{\psi}_{AB}\ot\tau_C, \qquad \ket{\psi}=\sum_i\sqrt{\lambda_i}\ket{ii},
 \quad\bdmu=\bdlambda(\tau).
\end{eqnarray}
\begin{prop}[Defect equals $k$-product gap]\label{prop:cexam}
For \eqref{eq:counterexample} and every $1\leqslant k\leqslant D$,
$$
\sfV_k^{(D)}(\rho)=\widetilde\Delta_k^{(D),\op{prod}}(\bdlambda,\bdmu).
$$
\end{prop}

\begin{thrm}[Failure of SSA]\label{th:FSSA}
Suppose $m,n,r\geqslant2$ and $D\geqslant mnr$. For every $2\leqslant k\leqslant D$,
strong subadditivity fails on the state
\begin{eqnarray}\label{eq:ABotC}
 \proj{\Omega_2}_{AB}\ot(\tfrac12 \I_2)_C,  \qquad \ket{\Omega_2}=\frac1{\sqrt2}(\ket{00}+\ket{11}),
\end{eqnarray}
embedded in the stated physical spaces. More generally,
\eqref{eq:counterexample} has $\sfV_k^{(D)}>0$ exactly when
$\ket{\psi}$ is entangled and $\tau$ is mixed.
\end{thrm}

\paragraph{Exact examples.} The padded uniform value $T_{D,k}(\ell)$ in \eqref{eq:padded-uniform} satisfies
$T_{D,k}(\ell)=\S_k^{(D)}\Pa{\tfrac{\I_\ell}\ell}$. Consequently the Bell-family defect is $2T_{D,k}(2)-T_{D,k}(4)>0$
for $k\geqslant2$. For $D=8$, exact simplification gives
$$
\begin{array}{c|rrrrrrrr}
k&1&2&3&4&5&6&7&8\\\hline
\V_k^{(8)}&0&\tfrac1{14}&\tfrac5{42}&\tfrac{61}{420}&\tfrac{16}{105}&\tfrac17&\tfrac5{42}&\tfrac1{12}
\end{array}
$$
These rational values follow by substituting $D=8$ in the finite
binomial formula \eqref{eq:padded-uniform}.
At the endpoint $k=D$ one obtains independently
$$
\V_D^{(D)}=2\Q\Pa{\tfrac{\I_2}2} -  \Q\Pa{\tfrac{\I_4}4}=\frac1{12}.
$$
For uniform Schmidt rank $\alpha$ and uniform ancilla rank $\beta$, the corresponding
exact expression is $T_{D,k}(\alpha)+T_{D,k}(\beta)-T_{D,k}(\alpha\beta)$.

\paragraph{Weak monotonicity and conditional mutual information.} Define the weak-monotonicity left-hand side in the same ambient dimension by
$$
\W_k^{(D)}:=\S_k^{(D)}(\rho_{AB}) + \S_k^{(D)}(\rho_{BC})  - \S_k^{(D)}(\rho_A) - \S_k^{(D)}(\rho_C).
$$
On \eqref{eq:counterexample}, $\W_k^{(D)}=-\widetilde\Delta_k^{(D),\mathrm{prod}}=-\V_k^{(D)}$. Also $\sfI(A:C|B)=0$. The violations above therefore occur already at zero von Neumann conditional mutual information.

\begin{cor}[Both signs on quantum Markov states]\label{cor:markov-signs}
Let $m,n,r\geqslant2$, $D\geqslant mnr$, and $2\leqslant k\leqslant D$.
The defect $\sfV_k^{(D)}$ takes both strictly positive and strictly negative
values on states with $\sfI(A:C|B)=0$. More precisely, for
$\rho^+_{ABC}=\proj{\psi}_{AB}\otimes\tau_C$ as in
\eqref{eq:counterexample}, let
$\rho^-_{ABC}=\rho_A^+\otimes\proj{\bsb}_B\otimes\tau_C$, where $\bsb$ is
any unit vector, and $\rho^+_A=\ptr{B}{\proj{\psi}_{AB}}$. Both states have zero conditional mutual information, and
\begin{eqnarray}\label{eq:markov-opposite}
\sfV_k^{(D)}(\rho^+_{ABC})=- \sfV_k^{(D)}(\rho^-_{ABC})  =\widetilde\Delta_k^{(D),\mathrm{prod}}(\bdlambda,\bdmu).
\end{eqnarray}
Writing $q=\rank(\rho_A^+)\rank(\tau_C)$, their common magnitude satisfies
\begin{eqnarray}\label{eq:markov-quantitative}
\sfV_k^{(D)}(\rho^+_{ABC})\geqslant  \frac{k(k-1)(1-\Tr{(\rho_A^+)^2})(1-\Tr{\tau^2_C})}
 {8D(D-1)q(q+1)(q+2)(q+3)}.
\end{eqnarray}
It is strictly positive exactly when $\proj{\psi}_{AB}$ is entangled and $\tau_C$ is mixed.
\end{cor}
For the Bell and uniform-qubit spectra, the negative witness gives exactly
the negative of every entry in the table above. The defect has no fixed
sign even within the quantum Markov class.

\paragraph{Maximum strong-subadditivity defect on the pure-factor family.}

For \eqref{eq:counterexample}, impose Schmidt rank at most $\alpha$ and
ancilla rank at most $\beta$, with $\alpha\leqslant\min(m,n)$ and $\beta\leqslant r$.
Theorem~\ref{th:rank-max} and Proposition~\ref{prop:cexam} give the exact
maximum $d_{\alpha\beta}^{(D,k)}$. For $k\geqslant2$ and $\alpha,\beta\geqslant2$, it is attained precisely when
$\ket{\psi}$ is maximally entangled on a rank-$\alpha$ subspace and
$\tau$ is uniform on a rank-$\beta$ subspace, up to local unitary changes
and the choice of supports. If either rank bound is one, the defect is
identically zero and every permitted state is a maximizer. This maximization concerns the stated
pure-factor family; no reduction of arbitrary tripartite states to this
family is assumed.

\subsection{Quantitative bounds}\label{subsect:QB}

The endpoint
$\widetilde\Delta^{(d)}_d=\Delta_{\Q}$ is \emph{independent of all zero padding}.
The equality characterization identifies when this gap vanishes.
The following bounds quantify its size for arbitrary bipartite states
in terms of the marginal spectra.

\begin{thrm}[The sharp two-qubit subentropy bound]\label{th:lowerboundfor2-qubit}
For every two-qubit state, let $a,b\in[0,\frac12]$ be given by two minimal eigenvalues $a=\lambda_{\min}(\rho_A)$ and
$b=\lambda_{\min}(\rho_B)$, respectively. Then
\begin{eqnarray}\label{eq:sharp-note}
 \widetilde\Delta_4^{(4)}=\sfQ(\rho_A) + \sfQ(\rho_B) - \sfQ(\rho_{AB})\geqslant\frac13ab.
\end{eqnarray}
The coefficient $\frac13$ is optimal. Equality holds exactly when $ab=0$ or
$\rho_{AB}=\frac14\I_4$.
\end{thrm}

\begin{cor}[Sharp intermediate two-qubit bounds]\label{cor:twoqubit-allk}
For a two-qubit state with smaller marginal eigenvalues $a,b$,
\begin{eqnarray}\label{eq:sharp-intermediate}
 \widetilde\Delta_2^{(4)}\geqslant\frac23ab,\qquad
 \widetilde\Delta_3^{(4)}\geqslant\frac23ab.
\end{eqnarray}
Both coefficients are optimal. In either inequality, equality holds
exactly when $ab=0$ or $\rho_{AB}=\frac14\I_4$.
\end{cor}

\begin{cor}[Optimal two-qubit coefficient in every common dimension]
\label{cor:twoqubit-padded-sharp}
For a two-qubit state, $D\geqslant4$ and $2\leqslant k\leqslant D$,
\begin{equation}\label{eq:twoqubit-all-D}
 \widetilde\Delta_k^{(D)}\geqslant c_{D,k}ab,\qquad
 c_{D,k}=\frac{2\binom{D-3}{k-2}+
 \frac13\binom{D-4}{k-4}}{\binom{D-1}{k-1}}.
\end{equation}
Binomial coefficients outside their range are zero. The coefficient is
optimal, and equality holds exactly when $ab=0$ or
$\rho_{AB}=\frac14\I_4$.
Equivalently,
$$
c_{D,k}=\frac{2(k-1)(D-k)}{(D-1)(D-2)} + \frac{(k-1)(k-2)(k-3)}{3(D-1)(D-2)(D-3)}.
$$
For fixed $k\geqslant2$, $c_{D,k}\sim\frac2D(k-1)$ as $D\to\infty$.
\end{cor}

\begin{thrm}[A bound in arbitrary local dimensions]\label{th:purity}
For $d=mn\geqslant2$ and $2\leqslant k\leqslant d$,
\begin{eqnarray}\label{eq:purity-main}
 \widetilde\Delta^{(d)}_k(\rho_{AB})\geqslant\frac{k(k-1)}{d(d-1)}\Delta_{\sfQ}(\rho_{AB}),\qquad \Delta_{\sfQ}(\rho_{AB})\geqslant\frac{ (1-\Tr{\rho_A^2})(1-\Tr{\rho_B^2})}{8d(d+1)(d+2)(d+3)}.
\end{eqnarray}
Moreover $\widetilde\Delta^{(d)}_k=0$ if and only if at least one marginal is pure.
\end{thrm}

For $k=d$, the displayed lower-bound coefficient becomes
$\frac1{8d(d+1)(d+2)(d+3)}$.
For $k=1$, the best nonnegative coefficient multiplying the purity product
is zero when $m,n\geqslant2$, since mixed product states have zero mutual
information. If a local dimension is one, both quantities vanish.
Determining the optimal purity-product constants in general dimensions
is an open problem. Formula \eqref{eq:native-counterexample} rules out a positive
bound with the native-local-dimension expression at intermediate $k$.

\begin{cor}[Stability of equality]\label{cor:stability}
Let $D\geqslant mn=:d\geqslant2$, and $2\leqslant k\leqslant D$.
Let $\mathcal Z_{m,n}$ be the set of bipartite states with at least one
pure marginal; each such state is a product state. Put
\begin{eqnarray}\label{eq:stability-constant}
 \beta_{D,k,d}:=\frac{k(k-1)}
 {8D(D-1)d(d+1)(d+2)(d+3)}.
\end{eqnarray}
Every state $\rho_{AB}\in\rD(\bbC^m\otimes\bbC^n)$ satisfies
\begin{eqnarray}\label{eq:stability-distance}
 \inf_{\omega\in\cZ_{m,n}}\frac12\norm{\rho_{AB}-\omega}_1
 \leqslant\min\Set{1,\Pa{\frac{\widetilde\Delta_k^{(D)}(\rho_{AB})}
 {\beta_{D,k,d}}}^{\frac14}}.
\end{eqnarray}
\end{cor}
Thus, in fixed dimensions, a vanishing deficit forces convergence in
trace distance to the equality set.

\subsection{Pinching and entropy gains of unital channels}\label{subsect:EG}

The preceding inequalities follow from a common bound on the entropy
increase under pinching. We state that bound here together with its
extensions to unital channels; Section~\ref{sect:proofs} explains how
it yields subadditivity and the quantitative estimates.

Let $\cH=\bigoplus^r_{i=1}\cH_i$ be an orthogonal decomposition, denote $\lin{\cH}$ the set of all linear operators acting on $\cH$, $\Pi_i$ the projection onto $\cH_i$, and $\cP$ the \emph{pinching}
$$
\cP(\bsX) = \sum^r_{i=1}\Pi_i\bsX\Pi_i,\qquad\bsX\in\lin{\cH},
$$
where $\sum^r_{j=1}\Pi_j=\I_{\cH}$. Discard zero projectors, so that
$1\leqslant r\leqslant d$ when $\cH\cong\bbC^d$; nonzero projectors with zero
state mass are retained. For $\rho\in\rD(\cH)$, denote $p_i=\Tr{\Pi_i\rho}$, so that $\bsp=(p_1,\ldots,p_r)$ is an $r$-dimensional probability vector and $\cP(\rho)$ has block masses $p_i$. Recall that, for von Neumann entropy $\S=\S_1$, we have \cite{Zhang2015,Buscemi2016}
\begin{eqnarray*}
\K\L(\rho||\cE^\dagger\circ\cE(\rho))\leqslant \S(\cE(\rho)) - \S(\rho)\leqslant \S(\widehat\cE(\rho))
\end{eqnarray*}
for any state $\rho\in\rD(\cH)$ and any \emph{unital} quantum channel
$\cE(*)=\sum_\mu \bsK_\mu(*)\bsK^\dagger_\mu$; the first inequality uses unitality so that
$\cE^\dagger\circ\cE(\rho)$ has trace one. Here
$\widehat \cE(*)=\sum_{\mu,\nu}\Tr{\bsK_\mu(*)\bsK^\dagger_\nu}\out{\mu}{\nu}$ is the complementary channel,
and $\K\L$ is the ordinary relative entropy. The entropy-gain upper bound by
$\S(\widehat\cE(\rho))$ is valid for an arbitrary quantum channel $\cE$. For a pinching channel, the following bound holds throughout the family:

\begin{lem}[Pinching lemma]\label{lem:pinching}
For every $\rho\in\rD(\bbC^d)$ and $k\in[d]$, we have
\begin{equation}
0\leqslant \sfS^{(d)}_k(\cP(\rho)) - \sfS^{(d)}_k(\rho)\leqslant \sfS^{(d)}_k(\diag(\bsp)),\quad \diag(\bsp)=\diag(p_1,\ldots,p_r).
\end{equation}
The right-hand side is the $k$-entropy of the diagonal matrix with entries $p_j$ (zero-padded to dimension $d$ if $r<d$).
\end{lem}
Write $\C^{(d)}_k(\cP,\rho):=\S_k^{(d)}(\cP(\rho)) - \S_k^{(d)}(\rho)$ for the pinching $\cP$ gain. The lower bound is the statement that pinching cannot decrease $\S^{(d)}_k$. The upper bound says the increase is controlled by the $k$-entropy of the block-mass distribution alone.

\begin{remark}
For $k=1$, the above Lemma~\ref{lem:pinching} reduces to the elementary fact that pinching increases von Neumann entropy by at most the entropy of the block masses. For $k=d$, it reduces to the subentropy analysis.
The intermediate cases interpolate between these two statements.
\end{remark}

\paragraph{General unital quantum channels.} The general-channel extension uses a divergence adapted to $\S^{(d)}_k$.
For positive semidefinite matrices of the same size, set
$$
\B_+(\bsX||\bsY)=\Tr{\bsX(\log \bsX-\log \bsY)}-\Tr{\bsX-\bsY}.
$$
We use $0\log0=0$ and set $\B_+(\bsX||\bsY)=+\infty$ if
$\op{supp}\bsX\nsubseteq\op{supp}\bsY$.
The Stiefel manifold is denoted by
$$
\op{St}(d,\ell):=\Set{\bsV\in M_{d,\ell}(\bbC): \bsV^\dagger \bsV=\I_\ell}
$$
which is equipped with its normalized Haar probability
measure $\mu_\ell$. For $k\geqslant2$, put
\begin{eqnarray*}
 \frak{C}_{d,k,\ell}:=k\binom{d-1}{k-1}^{-1}\binom{d-\ell-1}{k-2}, \qquad 1\leqslant\ell\leqslant d-k+1.
\end{eqnarray*}
Set $\B_1^{(d)}=\B_+$ and we will see that
\begin{eqnarray}\label{eq:Dk}
 \B_k^{(d)}(\bsX||\bsY)=\sum_{\ell=1}^{d-k+1} \frak{C}_{d,k,\ell} \int_{\op{St}(d,\ell)}
\B_+(\bsV^\dagger \bsX\bsV||\bsV^\dagger\bsY\bsV)\dif \mu_\ell(\bsV).
\end{eqnarray}
Let $\cE:M_d(\bbC)\to M_d(\bbC)$ be trace-preserving, completely positive
and unital. Write 
$$
\C^{(d)}_k(\cE,\rho):=\S_k^{(d)}(\cE(\rho)) - \S_k^{(d)}(\rho),
$$
where $\cE^\dagger$ is the dual map with respect to Hilbert-Schmidt inner product and
$\cP_\rho$ pinches onto the distinct eigenspaces of $\rho$.
\begin{thrm}\label{th:channel}
For $d\geqslant2$ and every $1\leqslant k\leqslant d$,
\begin{eqnarray}\label{eq:channel-chain}
 0\leqslant \sfB_k^{(d)}\Pa{\rho||\cP_\rho\Pa{\cE^\dagger\circ\cE(\rho)}}
 \leqslant \sfC^{(d)}_k(\cE,\rho)\leqslant \alpha_{d,k}\sfC^{(d)}_1(\cE,\rho)\leqslant \alpha_{d,k}\sfS(\widehat\cE(\rho)).
\end{eqnarray}
where $\alpha_{d,k}:=1-\frac{2(k-1)}{3(d-1)}$.
The coefficient in 
$$
\sfC^{(d)}_k(\cE,\rho)\leqslant \alpha_{d,k}\sfC^{(d)}_1(\cE,\rho)
$$ 
is optimal uniformly over states and unital channels. The lower-bound pinching can be omitted if
$\cE$ is mixed-unitary, if $[\rho,\cE^\dagger\circ\cE(\rho)]=0$, or if $k\in\set{1,d-1,d}$.
If $\cE$ is mixed-unitary and the complement has dimension $r$, then,
with $N=\max(d,r)$,
\begin{eqnarray}\label{eq:exchange-Sk}
\sfC^{(d)}_k(\cE,\rho)\leqslant \sfS^{(N)}_k\Pa{\widehat\cE(\rho)}.
\end{eqnarray}
At the subentropy endpoint every unital channel satisfies
\begin{eqnarray}\label{eq:exchange-Q}
0\leqslant \sfB_d^{(d)}(\rho||\cE^\dagger\circ\cE(\rho))\leqslant \sfQ(\cE(\rho)) - \sfQ(\rho)\leqslant \sfQ\Pa{\widehat\cE(\rho)}.
\end{eqnarray}
\end{thrm}

The ordinary relative entropy in the von Neumann lower bound cannot be
used unchanged for every $k$: dephasing $\ket{+}$ on a qubit gives
$\B_+(\rho||\cE^\dagger\circ\cE(\rho))=\ln2$, whereas the subentropy gain
is $\ln2-\frac12$. The arbitrary-unital unpinched lower bound at
$d\geqslant4$, $2\leqslant k\leqslant d-2$, and the intermediate-$k$ exchange bound for
arbitrary unital channels remain separate open questions.

\section{Application}\label{sect:application}\label{subsect:operational}

The positive compression representation also gives a communication meaning to
the whole entropy family. Consider a detector that distinguishes $k$
orthogonal subspaces of $\bbC^d$ and preserves the state within the observed
subspace. We show that, for the isotropic measurement law below,
$\S_k^{(d)}(\rho)$ is exactly the classical communication capacity retained
by the measurement record and the surviving quantum system, when the
average input state is prescribed to be $\rho$. Information extracted by a
measurement and information remaining in its posterior states have been
studied in~\cite{Schumacher1996,Jacobs2003}; here the coefficients in
Proposition~\ref{prop:posrep} identify a particular measurement hierarchy
with the Nichols--Wootters family of entropies. At the subentropy endpoint,
Haar-averaged measurement-information formulas and a classical-channel
interpretation were established in~\cite[Proposition 1 and Section 2.6 of the arXiv version]{Datta2014}.
Our application concerns the whole family and gives an exact capacity
identity at a prescribed average input for the detector specified below.

\paragraph{The measurement channel.}
Choose a Haar-distributed unitary $\bsU\in\op{U}(d)$ and, independently,
a uniformly distributed $(k-1)$-subset $C$ of the $d-1$ gaps between
successive coordinate vectors. The cuts partition $[d]$ into $k$
nonempty consecutive intervals $I_j(C)$, of lengths $m_j(C)$.
Write $L_{d,k}=\binom{d-1}{k-1}$ and let
$\bsV_{C,j}:\bbC^{m_j(C)}\to\bbC^d$ be the coordinate inclusion.
For an input state $\omega$, the unnormalized branch is
\begin{eqnarray}\label{eq:operational-branch}
 \bsA_{\bsU,C,j}(\omega)
 :=\bsV_{C,j}^\dagger\bsU^\dagger\omega\bsU\bsV_{C,j}.
\end{eqnarray}
Its trace is the probability of outcome $j$, conditional on $(\bsU,C)$.
The channel $\cN_{d,k}$ gives the receiver $(\bsU,C,j)$ and the normalized
residual state on each nonzero branch. Every outcome is retained.
The identity
$\sum_j\bsU\bsV_{C,j}\bsV_{C,j}^\dagger\bsU^\dagger=\I_d$
ensures trace preservation. The settings are sampled independently on
every use, after encoding, and their realizations are unknown to the
sender when preparing the inputs. The uniform law on cut sets, and hence
on ordered positive compositions of $d$, is part of the channel definition.

For $n$ uses, allow $M_n$ equiprobable messages, arbitrary input states
$\omega_w^{(n)}\in\rD((\bbC^d)^{\ot n})$, and a measurable collective decoder on
all records and residual systems. In particular, codewords may be
entangled across uses. There is no preshared entanglement, feedback, or
additional communication. Define the actual aggregate input mean by
\begin{eqnarray}\label{eq:operational-input-mean}
 \tau_n:=\frac1{nM_n}\sum_{w=1}^{M_n}\sum_{t=1}^n
                  (\omega_w^{(n)})_t,
\end{eqnarray}
where the subscript $t$ denotes the one-site marginal. Let
$C_\rho(\cN_{d,k})$ be the supremum of rates
$\liminf_{n\to\infty}n^{-1}\log M_n$ attainable with average decoding
error tending to zero and $\norm{\tau_n-\rho}_1\to0$.

\begin{thrm}[Retained classical communication capacity]\label{th:operational}
For every $\rho\in\rD(\bbC^d)$ and $1\leqslant k\leqslant d$,
\begin{eqnarray}\label{eq:operational-capacity}
 C_\rho(\cN_{d,k})=\sfS_k^{(d)}(\rho).
\end{eqnarray}
For every finite pure-state ensemble
$\{p_a,\proj{\psi_a}\}$ with mean $\rho$, the Holevo information
of its complete output is also
\begin{eqnarray}\label{eq:operational-holevo}
 \chi(\{p_a,\cN_{d,k}(\proj{\psi_a})\})=\sfS_k^{(d)}(\rho).
\end{eqnarray}
For $k<d$, a receiver-side quantum channel $\cT_k$ satisfies
\begin{eqnarray}\label{eq:operational-degradation}
 \cN_{d,k+1}=\cT_k\circ\cN_{d,k}.
\end{eqnarray}
\end{thrm}
The Holevo quantity in \eqref{eq:operational-holevo} averages the
finite-dimensional conditional Holevo information over the independent
Haar setting; its precise definition and the proof of the theorem are
given in Section~\ref{proof:operational}. Collective decoding supplies
the attainable communication rate in \eqref{eq:operational-capacity}.

At $k=1$, the receiver can undo the reported rotation and recover the
input, giving capacity $\S(\rho)$. At $k=d$, every residual is
one-dimensional, so the complete random-basis measurement record has
capacity $\Q(\rho)$. Thus $\S(\rho)-\S_k^{(d)}(\rho)$ is the loss of
reliable classical communication rate caused by the partial measurement.
For a maximally mixed mean, the value in Section~\ref{sect:2} gives
$C_{\I_d/d}(\cN_{d,k})=\log d-H_k+1$; each additional cut costs
$1/(k+1)$ nats per use at that mean. In the first intermediate example,
$d=3,k=2$, one measures a Haar-random ray and its orthogonal plane and
retains the chosen ray, the outcome and the state in the observed subspace. Its capacity
at $\rho=\I_3/3$ is $\log3-\frac12$ nats per use.

\paragraph{Operational meaning of the common-dimension inequalities.}
For any state $\sigma$ on a space of dimension at most $D$, write
$\sigma^{[D]}$ for its zero-padded embedding into $\bbC^D$.
Fix $D,k$ and abbreviate
$C_\sigma:=C_{\sigma^{[D]}}(\cN_{D,k})$ in the following corollary.
These capacities refer to separate preparations for the same
$D$-dimensional detector. In particular, the detector applied to a
padded joint state is specified by $\cN_{D,k}$, independently of the
tensor-factor decomposition of that state.

\begin{cor}[Capacity differences in a common dimension]\label{cor:operational-gap}
For $\rho_{AB}\in\rD(\bbC^m\ot\bbC^n)$, $D\geqslant mn$ and
$1\leqslant k\leqslant D$,
\begin{eqnarray}\label{eq:operational-gap}
 C_{\rho_A}+C_{\rho_B}-C_{\rho_{AB}}
 =\widetilde\Delta_k^{(D)}(\rho_{AB})\geqslant0.
\end{eqnarray}
For $k\geqslant2$, equality holds exactly when a marginal is pure.
The quantitative bounds and stability statement of
Section~\ref{subsect:QB} apply to this same capacity difference.
For a tripartite state and $D$ at least its total physical dimension,
\begin{eqnarray}\label{eq:operational-ssa}
 C_{\rho_{ABC}}+C_{\rho_B}-C_{\rho_{AB}}-C_{\rho_{BC}}
 =\sfV_k^{(D)}(\rho_{ABC}).
\end{eqnarray}
For $k\geqslant2$ and local dimensions at least two, this four-capacity
combination takes both signs even on quantum Markov states.
\end{cor}
This gives the entropy inequalities a direct application without
changing their ambient-dimension convention: they compare the
communication rates retained by one specified detector at different
prescribed input means.

\section{Proofs of main results}\label{sect:proofs}

Before proving the results, we recall that the spectral formula also defines $\S_k^{(d)}(\bsX)$ for a positive semidefinite matrix $\bsX$
of arbitrary trace. In that setting the scaling rule of Appendix~\ref{app:scaling} is retained; no compression is implicitly
normalized. All ambient dimensions below are fixed before applying an entropy inequality.

We begin with the positive compression representation and the pinching lemma, which provide the common mechanism. We then prove the results of Section~\ref{sect:3} in their displayed order: subadditivity and its equality cases, the $k$-product gap, failure of strong subadditivity,
the quantitative bounds, and the channel inequalities. The proof of the operational application in Section~\ref{sect:application} follows these arguments. The structural identities used in these proofs are established in Appendix~\ref{app:A}.
The detailed derivations of the compression coefficients, the polynomial comparisons, the purity estimate and the singular-state limits are
included below.

\subsection{Positive compressions and the pinching inequality}\label{proof:compression-pinching}

We first prove Lemma~\ref{lem:pinching}. Fix the ambient dimension $d$
and the orthogonal block pinching
$\cP(\bsX)=\sum_{j=1}^r\Pi_j\bsX\Pi_j$ on $\bbC^d$.
Zero projectors are discarded, whereas nonzero projectors with zero state
mass are retained. Thus $1\leqslant r\leqslant d$.
The main step is a positive representation of $\S_k^{(d)}$ by Haar
averages of compressed matrix entropies. Each compression is linear in
the input matrix, which will allow us to prove convexity of the pinching
gain from joint convexity of $\B_+$.

\subsubsection{A positive compression representation}\label{subsubsect:4.1.1}

Fix $2\leqslant k\leqslant d$. Let
$\bdlambda=\bdlambda(\bsX)$, and let $\bsT$ be uniformly distributed
on the probability simplex $\Delta_{k-1}$. For every continuous function $f$ on
$[0,\norm{\bsX}]$, we will prove \eqref{eq:continuous-function-identity}, i.e.,
\begin{eqnarray*}
 \sum_{|\cI|=k}\bbE_{\bsT}\Br{ f(\Inner{\bsT}{\bdlambda_{\cI}})}  =\sum_{\ell=1}^{d-k+1}\binom{d-\ell-1}{k-2}
 \int_{\op{St}(d,\ell)}  \Tr{f(\bsV^\dagger\bsX\bsV)}\dif\mu_\ell(\bsV).
\end{eqnarray*}
Throughout the calculation, binomial coefficients outside their natural
range are zero. We first take $f(x)=x^n$ with $n\geqslant1$. Let $h_n$ be the complete homogeneous symmetric polynomial of degree $n$.
The Dirichlet moments give
\begin{eqnarray}\label{eq:B1-Dirichlet}
 \bbE_{\bsT}\Br{(\Inner{\bsT}{\bdlambda_{\cI}})^n} =\binom{n+k-1}{n}^{-1}\cdot h_n(\bdlambda_{\cI}).
\end{eqnarray}
Indeed, for nonnegative integers $\alpha_i$ with
$\alpha_1+\cdots+\alpha_k=n$, let $\bsu=\bdlambda_{\cI}$ for fixed $\cI$, we see that
\begin{eqnarray*}
&&\bbE_{\bsT}\Br{(\Inner{\bsT}{\bsu})^n}  =  \bbE_{\bsT}\Br{ \sum_{\substack{\alpha_1,\ldots,\alpha_k\geqslant0\\ \alpha_1+\cdots+\alpha_k=n} } \binom{n}{\alpha_1,\ldots,\alpha_k}\prod^k_{i=1}T_i^{\alpha_i}\prod^k_{i=1}u_i^{\alpha_i}} \\
&&=\sum_{\substack{\alpha_1,\ldots,\alpha_k\geqslant0\\ \alpha_1+\cdots+\alpha_k=n} } \binom{n}{\alpha_1,\ldots,\alpha_k}\bbE_{\bsT}\Br{ \prod^k_{i=1}T_i^{\alpha_i}}\prod^k_{i=1}u_i^{\alpha_i},
\end{eqnarray*}
where 
\begin{eqnarray*}
 \bbE_{\bsT}\Br{\prod_{i=1}^kT_i^{\alpha_i}} =\Gamma(k)\int_{\bbR^k_{\geqslant0}}\Br{\prod_{i=1}^kT_i^{\alpha_i}}\delta\Pa{1-\sum^k_{i=1}T_i}[\dif\bsT]=\frac{(k-1)!\prod_{i=1}^k\alpha_i!}{(n+k-1)!}.
\end{eqnarray*}
Multiplying by the multinomial coefficient
$\binom{n}{\alpha_1,\ldots,\alpha_k}=\frac{n!}{\prod_i\alpha_i!}$ leaves the same coefficient
$\binom{n+k-1}{n}^{-1}$ for every monomial in 
$$
h_n(\bsu)=\sum_{\substack{\alpha_1,\ldots,\alpha_k\geqslant0\\ \alpha_1+\cdots+\alpha_k=n} }\prod^k_{i=1}u_i^{\alpha_i}.
$$
For $0\leqslant j\leqslant\min(n-1,d-1)$, write
$s_{n,j}=s_{(n-j,1^j)}$ for the Schur polynomial of hook shape
$(n-j,1^j)$. A monomial of degree $n$ with exactly $q$ distinct variables
has coefficient $\binom{q-1}{j}$ in $s_{n,j}$. In a semistandard tableau
with this prescribed content, the upper-left cell contains the smallest
label in the support. The remaining $j$ labels of the first column can
be chosen in $\binom{q-1}{j}$ ways from the other labels. Once those
labels have been chosen, their order in the column and the rest of the
first row are determined. Consequently,
\begin{eqnarray}\label{eq:B1-hook-expansion}
 \sum_{|\cI|=k}h_n(\bdlambda_{\cI})
 =\sum_{j=0}^{\min(n-1,d-1)}(-1)^j
 \binom{d-j-1}{k-1}s_{n,j}(\bdlambda).
\end{eqnarray}
To verify the coefficients, fix a monomial with support size $q$.
Its coefficient on the left is $\binom{d-q}{k-q}$.
On the right the required equality follows by taking the coefficient
of $z^{k-1}$ in
$$
 \sum_{j=0}^{q-1}(-1)^j\binom{q-1}{j}(1+z)^{d-j-1} =z^{q-1}(1+z)^{d-q}.
$$
The same tableau count, followed by Vandermonde's identity, yields
\begin{eqnarray}\label{eq:B1-hook-dimension}
 s_{n,j}(1^\ell)=\sum_{q=j+1}^{\min(n,\ell)} \binom{\ell}{q}\binom{n-1}{q-1}\binom{q-1}{j} =\binom{n-1}{j}\binom{\ell+n-j-1}{n}.
\end{eqnarray}
Here $1^\ell$ means $\ell$ unit entries. In the first sum,
$\binom{\ell}{q}$ chooses the support, $\binom{n-1}{q-1}$ chooses the
positive exponents, and $\binom{q-1}{j}$ counts the tableaux.
For the last equality, use
$\binom{n-1}{q-1}\binom{q-1}{j}
=\binom{n-1}{j}\binom{n-j-1}{q-j-1}$ and then Vandermonde's identity.
The same coefficient argument also proves the power-sum expansion
\begin{eqnarray*}
\Tr{\bsX^n}= \sum_i\lambda_i^n=\sum_{j=0}^{\min(n-1,d-1)}(-1)^js_{n,j}(\bdlambda),
\end{eqnarray*}
since a monomial of support size $q$ has coefficient
$\sum_{j=0}^{q-1}(-1)^j\binom{q-1}{j}$, which is one when $q=1$ and
zero otherwise.

We next calculate the Haar moments in \eqref{eq:continuous-function-identity}.
Recall that, for a partition $\nu$ with at most $d$ rows, \eqref{eq:Schur-Weylformula} is
\begin{equation}\label{eq:B1-Schur-average}
 \int_{\op{St}(d,\ell)}
 s_\nu(\bsV^\dagger\bsX\bsV)\dif\mu_\ell(\bsV)
 =\frac{s_\nu(1^\ell)}{s_\nu(1^d)}s_\nu(\bsX).
\end{equation}
Here is a representation-theoretic proof that also covers singular
$\bsX$. Put $\Pi_\ell^0=\diag(\I_\ell,0)$ and
$\Pi=\bsU\Pi_\ell^0\bsU^\dagger$, where $\bsU$ is Haar distributed
in $\rU(d)$. Its first $\ell$ columns form $\bsV$, so
$\Pi=\bsV\bsV^\dagger$. Let $\pi_\nu$ be the irreducible polynomial
representation of $\op{GL}_d(\bbC)$ whose character is $s_\nu$,
as supplied by the Schur-functor construction
\cite[Sections~6.1--6.2]{Fulton1991}.
Its restriction to $\rU(d)$ is irreducible. Thus the Haar average of
$\pi_\nu(\Pi)$ is a scalar operator by Schur's lemma, and its trace
determines the scalar:
\begin{eqnarray*}
 \int\pi_\nu(\Pi)\dif\mu_{\haar}(\bsU) =\frac{s_\nu(1^\ell)}{s_\nu(1^d)}\I.
\end{eqnarray*}
The nonzero eigenvalues of $\bsV^\dagger\bsX\bsV$ and
$\bsX^{\frac12}\Pi\bsX^{\frac12}$ coincide. By multiplicativity and cyclicity
of the trace,
\begin{eqnarray*}
 s_\nu(\bsV^\dagger\bsX\bsV)=\Tr{\pi_\nu(\bsX^{\frac12})\pi_\nu(\Pi) \pi_\nu(\bsX^{\frac12})}=\Tr{\pi_\nu(\bsX)\pi_\nu(\Pi)}.
\end{eqnarray*}
Averaging proves \eqref{eq:B1-Schur-average}: 
\begin{eqnarray*}
 \int_{\op{St}(d,\ell)}s_\nu(\bsV^\dagger\bsX\bsV)\dif\mu_\ell(\bsV)&=&\Tr{\pi_\nu(\bsX) \int_{\op{St}(d,\ell)}\pi_\nu(\Pi)\dif\mu_\ell(\bsV)} \\
&=& \frac{s_\nu(1^\ell)}{s_\nu(1^d)}\Tr{\pi_\nu(\bsX)} = \frac{s_\nu(1^\ell)}{s_\nu(1^d)}s_\nu(\bsX).
\end{eqnarray*}
Polynomial representations
extend to singular matrices; moreover, if $\nu$ has more than $\ell$
rows, both the Schur polynomial on $\ell$ variables and $s_\nu(1^\ell)$
vanish. These observations justify the identity at every rank.
Using the power-sum expansion and \eqref{eq:B1-hook-dimension}, we obtain
\begin{equation}\label{eq:B1-compression-moment}
 \int_{\op{St}(d,\ell)}
 \Tr{(\bsV^\dagger\bsX\bsV)^n}\dif\mu_\ell(\bsV)
 =\sum_{j=0}^{\min(n-1,d-1)}(-1)^j
 \frac{\binom{\ell+n-j-1}{n}}{\binom{d+n-j-1}{n}}
 s_{n,j}(\bdlambda).
\end{equation}

The remaining coefficient identity is
\begin{equation}\label{eq:B1-convolution}
 \sum_{\ell=1}^{d-k+1}\binom{d-\ell-1}{k-2} \frac{\binom{\ell+n-j-1}{n}}{\binom{d+n-j-1}{n}} =\frac{\binom{d-j-1}{k-1}}{\binom{n+k-1}{n}}.
\end{equation}
Remove the denominator independent of $\ell$. Terms with
$\ell\leqslant j$ vanish. On writing $\ell=j+1+u$, the numerator sum
is the coefficient of $z^{d-k-j}$ in
$$
 (1-z)^{-(k-1)}(1-z)^{-(n+1)}=(1-z)^{-(n+k)},
$$
and therefore equals $\binom{d+n-j-1}{n+k-1}$.
Dividing by $\binom{d+n-j-1}{n}$ gives
\eqref{eq:B1-convolution}. If $j>d-k$, both sides are zero.
Equations~\eqref{eq:B1-Dirichlet}, \eqref{eq:B1-hook-expansion},
\eqref{eq:B1-compression-moment} and \eqref{eq:B1-convolution}
prove \eqref{eq:continuous-function-identity} for all monomials of positive degree.
For the constant function $f=1$, that identity becomes
$$
 \binom{d}{k} =\sum_{\ell=1}^{d-k+1}\ell\binom{d-\ell-1}{k-2}.
$$
This counts $k$-subsets by their second-smallest element $\ell+1$:
there are $\ell$ choices for the smallest element and
$\binom{d-\ell-1}{k-2}$ choices for the other elements.
Both sides of \eqref{eq:continuous-function-identity} are positive linear functionals
of $f$ with total mass $\binom{d}{k}$. Consequently a uniform
approximation error $\varepsilon$ changes either side by at most
$\binom{d}{k}\varepsilon$. Uniform polynomial approximation proves
\eqref{eq:continuous-function-identity} for continuous $f$.

Finally, sum the simplex identity \eqref{eq:Phik2} over the $k$-subsets.
Every eigenvalue belongs to $\binom{d-1}{k-1}$ such subsets, so
$$
 \sum_{|\cI|=k}\sum_{i\in\cI}\lambda_i =\binom{d-1}{k-1}\Tr{\bsX}.
$$
Taking $f=\eta$ in \eqref{eq:continuous-function-identity} and dividing by
$\binom{d-1}{k-1}$ gives \eqref{eq:kentropy-intrep}.
Continuity of $\eta$ at zero includes repeated and zero eigenvalues
throughout this argument; no distinctness or full-rank assumption is used.

\subsubsection{Relative entropy and convexity of the gain}\label{subsubsect:4.1.2}

Recall the generalized relative entropy from Section~\ref{subsect:EG}:
\begin{eqnarray}\label{eq:B1-Dplus}
 \B_+(\bsX||\bsY) &=&\Tr{\bsX(\log\bsX-\log\bsY)} - \Tr{\bsX-\bsY} \notag\\
&=& -\Tr{\eta(\bsX)} -\Tr{\bsX\log\bsY} - \Tr{\bsX-\bsY}.
\end{eqnarray}
It is nonnegative and jointly convex on positive definite arguments
\cite[Definition~2 and Facts~3--4]{Tropp2012}, see also Appendix~\ref{app:C}. The extension to positive
semidefinite arguments, with the stated support convention, satisfies
$\B_+(\bsX||\bsY)=\lim_{\varepsilon\downarrow0} \B_+(\bsX+\varepsilon\I||\bsY+\varepsilon\I)$.

For each $1\leqslant\ell\leqslant d$, we \emph{claim} the exact gain identity
\begin{eqnarray}\label{eq:B1-compression-gain}
 \G_\ell(\cP(\rho)) - \G_\ell(\rho) =\int_{\op{St}(d,\ell)} \B_+(\bsV^\dagger\rho\bsV||  \bsV^\dagger\cP(\rho)\bsV)\dif\mu_\ell(\bsV).
\end{eqnarray}
Relabel the nonzero projectors by $j=0,\ldots,r-1$, and set
\begin{eqnarray}\label{eq:B1-twirl}
 \omega=\ee^{2\pi\mathrm{i}/r},\qquad
 \bsU_i=\sum_{j=0}^{r-1}\omega^{ij}\Pi_j,\qquad
 \cP(\bsX)=\frac1r\sum_{i=0}^{r-1}\bsU_i\bsX\bsU_i^\dagger.
\end{eqnarray}
Each $\bsU_i\in\rU(d)$ fixes $\cP(\rho)$ under conjugation. For fixed
$\bsV\in\op{St}(d,\ell)$, put
\begin{eqnarray*}
 \bsA_{i,\bsV}=\bsV^\dagger\bsU_i\rho\bsU_i^\dagger\bsV, \qquad \bsB_{\bsV}=\bsV^\dagger\cP(\rho)\bsV   =\frac1r\sum_{i=0}^{r-1}\bsA_{i,\bsV}.
\end{eqnarray*}
Expansion of \eqref{eq:B1-Dplus} gives
\begin{eqnarray}\label{eq:B1-Jensen}
 \Tr{\eta(\bsB_{\bsV})} -\frac1r\sum_{i=0}^{r-1}\Tr{\eta(\bsA_{i,\bsV})} =\frac1r\sum_{i=0}^{r-1}  \B_+(\bsA_{i,\bsV}||\bsB_{\bsV}).
\end{eqnarray}
Indeed, apply the expansion to the pair $(\bsA_{i,\bsV},\bsB_{\bsV})$:
\begin{eqnarray*}
\B_+(\bsA_{i,\bsV}||\bsB_{\bsV}) = -\Tr{\eta(\bsA_{i,\bsV})} - \Tr{\bsA_{i,\bsV}\log \bsB_{\bsV}} - \Tr{\bsA_{i,\bsV} - \bsB_{\bsV}}.
\end{eqnarray*}
Average over $i=0,\ldots,r-1$ with weights $\frac1r$:
\begin{eqnarray*}
\frac1r\sum^{r-1}_{i=0}\B_+(\bsA_{i,\bsV}||\bsB_{\bsV}) &=& -\frac1r\sum^{r-1}_{i=0}\Tr{\eta(\bsA_{i,\bsV})} - \frac1r\sum^{r-1}_{i=0}\Tr{\bsA_{i,\bsV}\log \bsB_{\bsV}} - \frac1r\sum^{r-1}_{i=0}\Tr{\bsA_{i,\bsV} - \bsB_{\bsV}}\\
&=& -\frac1r\sum^{r-1}_{i=0}\Tr{\eta(\bsA_{i,\bsV})} + \Tr{\eta(\bsB_{\bsV})} - 0,
\end{eqnarray*}
where the cross terms average to $-\Tr{\bsB_{\bsV}\log\bsB_{\bsV}} = \Tr{\eta(\bsB_{\bsV})}$, and the linear trace terms cancel.
The change of variable $\bsV\mapsto\bsU_i^\dagger\bsV$ preserves
$\mu_\ell$. Since
$\bsU_i\cP(\rho)\bsU^\dagger_i=\cP(\rho)$, it follows that
\begin{eqnarray*}
 \int_{\op{St}(d,\ell)}\Tr{\eta(\bsA_{i,\bsV})}\dif\mu_\ell(\bsV)
 &=&\G_\ell(\rho),\\
 \int_{\op{St}(d,\ell)} \B_+(\bsA_{i,\bsV}||\bsB_{\bsV})\dif\mu_\ell(\bsV)
 &=&\int\B_+(\bsV^\dagger\rho\bsV||  \bsV^\dagger\cP(\rho)\bsV)\dif\mu_\ell(\bsV).
\end{eqnarray*}
Integration of \eqref{eq:B1-Jensen} proves
\eqref{eq:B1-compression-gain}:
\begin{eqnarray*}
\G_\ell(\cP(\rho))  - \G_\ell(\rho) &=&  \int_{\op{St}(d,\ell)} \Tr{\eta(\bsB_{\bsV})}\dif\mu_\ell(\bsV) - \frac1r\sum^{r-1}_{i=0}\int_{\op{St}(d,\ell)}\Tr{\eta(\bsA_{i,\bsV})}\dif\mu_\ell(\bsV)\\
&=& \int_{\op{St}(d,\ell)}\Pa{\Tr{\eta(\bsB_{\bsV})} - \frac1r\sum^{r-1}_{i=0}\Tr{\eta(\bsA_{i,\bsV})}}\dif\mu_\ell(\bsV)\\
&=& \int_{\op{St}(d,\ell)}\Pa{\frac1r\sum_{i=0}^{r-1}  \B_+(\bsA_{i,\bsV}||\bsB_{\bsV})}\dif\mu_\ell(\bsV)=\frac1r\sum_{i=0}^{r-1}  \int_{\op{St}(d,\ell)}\B_+(\bsA_{i,\bsV}||\bsB_{\bsV})\dif\mu_\ell(\bsV)\\
&=& \frac1r\sum_{i=0}^{r-1}  \int\B_+(\bsV^\dagger\rho\bsV||  \bsV^\dagger\cP(\rho)\bsV)\dif\mu_\ell(\bsV) = \int\B_+(\bsV^\dagger\rho\bsV||  \bsV^\dagger\cP(\rho)\bsV)\dif\mu_\ell(\bsV).
\end{eqnarray*}
These calculations also justify the identity for singular states.
The relation $\bsA_{i,\bsV}\leqslant r\bsB_{\bsV}$ implies
$\op{supp}[\bsA_{i,\bsV}]\subseteq\op{supp}[\bsB_{\bsV}]$, so each relative
entropy in \eqref{eq:B1-Jensen} is finite. Evaluate
$\log\bsB_{\bsV}$ on its support and use continuous functional calculus
for $\bsA_{i,\bsV}\log\bsA_{i,\bsV}$, with $0\log0=0$.
All compressed spectra lie in $[0,1]$, making the trace entropies on
the left of \eqref{eq:B1-Jensen} bounded. Since every term on the right
is nonnegative, their integrals are finite. Thus the argument does not
require an interchange of an unbounded logarithm and an integral.

For $2\leqslant k\leqslant d$, using \eqref{eq:kentropy-intrep} in Proposition~\ref{prop:posrep}, the trace correction in \eqref{eq:kentropy-intrep} cancels under pinching, hence we see that
\begin{eqnarray*}
&&\C_k^{(d)}(\cP,\rho) = \S^{(d)}_k(\cP(\rho)) -  \S^{(d)}_k(\rho) \\
&&= \Pa{\sum_{\ell=1}^{d-k+1}\mathfrak{C}_{d,k,\ell}\G_\ell(\cP(\rho))-(H_k-1)\Tr{\cP(\rho)}} - \Pa{\sum_{\ell=1}^{d-k+1}\mathfrak{C}_{d,k,\ell}\sfG_\ell(\rho)-(H_k-1)\Tr{\rho}}\\
&&=\sum_{\ell=1}^{d-k+1}\mathfrak{C}_{d,k,\ell}\Br{\G_\ell(\cP(\rho)) - \G_\ell(\rho)} = \sum_{\ell=1}^{d-k+1}\mathfrak{C}_{d,k,\ell}\int_{\op{St}(d,\ell)} \B_+(\bsV^\dagger\rho\bsV|| \bsV^\dagger\cP(\rho)\bsV)\dif\mu_\ell(\bsV),
\end{eqnarray*}
that is,
\begin{eqnarray}\label{eq:B1-Ck-relative}
 \C_k^{(d)}(\cP,\rho) =\sum_{\ell=1}^{d-k+1}\frak{C}_{d,k,\ell} \int_{\op{St}(d,\ell)} \B_+(\bsV^\dagger\rho\bsV|| \bsV^\dagger\cP(\rho)\bsV)\dif\mu_\ell(\bsV).
\end{eqnarray}
For every fixed $\bsV$, both arguments of $\B_+$ are linear in $\rho$.
Joint convexity and positivity of the coefficients therefore imply
\begin{eqnarray}\label{eq:B1-convexity}
 \C_k^{(d)}(\cP,(1-t)\rho_0+t\rho_1) \leqslant(1-t)\C_k^{(d)}(\cP,\rho_0)  + t\C_k^{(d)}(\cP,\rho_1), \qquad 0\leqslant t\leqslant1.
\end{eqnarray}
Nonnegativity follows from the same representation. When $k=1$,
\eqref{eq:B1-compression-gain} with $\ell=d$ gives instead
$\C_1^{(d)}(\cP,\rho)=\B_+(\rho||\cP(\rho))$, which has the same convexity and nonnegativity properties.
One can also obtain the boundary cases by applying the positive definite
inequalities to $(1-\varepsilon)\rho_i+\varepsilon\I_d/d$ and passing
to the limit using continuity of \eqref{eq:kentropy-intrep}.

\subsubsection{Proof of Lemma~\ref{lem:pinching}}\label{subsubsect:4.1.3}

\begin{proof}
The lower bound follows from the relative-entropy representation above.
It also follows from Schur concavity in Corollary~\ref{cor:concavity}, since
$$
 \bdlambda(\cP(\rho))\prec\bdlambda(\rho),\qquad  \S_k^{(d)}(\cP(\rho))\geqslant\S_k^{(d)}(\rho).
$$
For the upper bound, first take a unit vector $\bsv$. Put
$p_j^{(\bsv)}=\norm{\Pi_j\bsv}^2$ and define
$\bsu_j=\Pi_j\bsv/\norm{\Pi_j\bsv}$ only when
$p_j^{(\bsv)}>0$. The vectors so defined are orthonormal, and
$$
 \cP(\proj{\bsv}) =\sum_{j:\,p_j^{(\bsv)}>0}p_j^{(\bsv)}\proj{\bsu_j}.
$$
Thus its nonzero eigenvalues are exactly the positive block masses.
After appending zeros to dimension $d$, its complete spectrum agrees
with that of $\diag(\bsp^{(\bsv)})$. Since pure states have zero
$\S_k^{(d)}$ by Proposition~\ref{prop:extval}, we obtain
\begin{eqnarray}\label{eq:cohinf}
 \C_k^{(d)}(\cP,\proj{\bsv}) = \S_k^{(d)}(\diag(\bsp^{(\bsv)})).
\end{eqnarray}
Every diagonal matrix in this argument is understood in the same ambient
dimension $d$, including all zero masses and any additional zero entries.

Choose a finite pure-state decomposition
$\rho=\sum_iw_i\proj{\bsv_i}$ with $w_i>0$ and $\sum_iw_i=1$.
Convexity \eqref{eq:B1-convexity} and \eqref{eq:cohinf} give
\begin{eqnarray}\label{eq:cohconvex}
 \C_k^{(d)}(\cP,\rho) \leqslant\sum_iw_i\C_k^{(d)}(\cP,\proj{\bsv_i}) =\sum_iw_i\S_k^{(d)}(\diag(\bsp^{(\bsv_i)})).
\end{eqnarray}
By Proposition~\ref{prop:concavity},
\begin{eqnarray}\label{eq:Skconcave}
 \sum_iw_i\S_k^{(d)}(\diag(\bsp^{(\bsv_i)}))\leqslant\S_k^{(d)}\Pa{  \diag\Pa{\sum_iw_i\bsp^{(\bsv_i)}}} = \S_k^{(d)}(\diag(\bsp)),
\end{eqnarray}
because $\sum_iw_ip_j^{(\bsv_i)}=\sum_iw_i\norm{\Pi_j\bsv_i}^2=\Tr{\Pi_j\rho}=p_j$. This proves the upper bound, including zero block masses.
\end{proof}

The lower bound in Lemma~\ref{lem:pinching} is strict unless $\rho=\cP(\rho)$. For $k\geqslant2$, the coefficient of the $\ell=1$ term in \eqref{eq:B1-Ck-relative} is positive. If the gain vanishes, the nonnegative scalar relative entropy therefore vanishes for Haar-almost every unit vector $\bsv$. Its scalar arguments must agree: $\bra{\bsv}\rho\ket{\bsv} =\bra{\bsv}\cP(\rho)\ket{\bsv}$. The difference of these quadratic forms is continuous. Haar measure has full support on the unit sphere, so the equality holds for every unit
vector. Polarization then gives $\rho=\cP(\rho)$. For $k=1$, the same conclusion is the equality condition for ordinary relative entropy. The converse holds directly from the definition of the gain. In particular, at $k=d$ the gain is the scalar-compression formula
\begin{eqnarray*}
 \C_d^{(d)}(\cP,\rho) =d\int_{\op{St}(d,1)} \B_+(\bra{\bsv}\rho\ket{\bsv}||  \bra{\bsv}\cP(\rho)\ket{\bsv})\dif\mu_1(\bsv).
\end{eqnarray*}

\subsection{Subadditivity and its equality cases}\label{proof:subadditivityequality}

We next apply the pinching inequality to block matrices. Two matrices
with the same nonzero eigenvalues need not have the same intermediate
$k$-entropy in their original dimensions. Every spectral comparison
below therefore takes place after padding both matrices to one common
dimension.

\subsubsection{A block inequality at arbitrary trace}\label{subsubsect:4.2.1}

\begin{prop}\label{prop:appsubadditivity}
Let $\bsX_1,\ldots,\bsX_n\geqslant0$ act on $\bbC^m$, and let $t=\sum_{j=1}^n\Tr{\bsX_j}$,
\begin{eqnarray*}
 \bsY=\sum^n_{j=1}\bsX_j\ot\proj{j}=\bigoplus_{j=1}^n\bsX_j,\qquad
 \bsA=\sum_{j=1}^n\bsX_j,\qquad
 \bsp=(\Tr{\bsX_1},\ldots,\Tr{\bsX_n}).
\end{eqnarray*}
For $D\geqslant mn$, $1\leqslant k\leqslant D$, and $t>0$,
\begin{equation}\label{eq:applowerupper}
 0\leqslant\sfS_k^{(D)}(\bsY)    -  \sfS_k^{(D)}(\bsA) \leqslant t \sfS_k^{(D)}(\diag(\bsp/t)).
\end{equation}
Here all matrices are recognized as the embedding obtained by appending zero
eigenvalues to dimension $D$. If $t=0$, all the matrices vanish, and
the two entropy differences and the bound are zero.
\end{prop}

\begin{proof}
Positivity and $t=0$ imply $\bsX_j=0$ for every $j$, proving that case.
Suppose $t>0$. Stack the square roots into
$$
 \bsW=\begin{pmatrix}  \sqrt{\bsX_1}\\ \vdots\\ \sqrt{\bsX_n}   \end{pmatrix}:\bbC^m\longrightarrow\bbC^{mn}.
$$
Then $\bsW^\dagger\bsW=\sum^n_{j=1}\bsX_j=\bsA$, and 
$$
\bsW\bsW^\dagger=\sum^n_{j,j'=1}\bsX^{\frac12}_j\bsX^{\frac12}_{j'}\ot\out{j}{j'}
$$ 
has the same nonzero eigenvalues as $\bsA$. Both have trace $t=\Tr{\bsA}=\Tr{\bsW^\dagger\bsW}=\Tr{\bsW\bsW^\dagger}$, so
$\rho=\frac1t\bsW\bsW^\dagger$ is a density matrix.
Pinch its first $mn$ coordinates into $n$ blocks of size $m$, adding
the orthogonal complement as one further block if $D>mn$.
The pinched state is 
$$
\cP(\rho)=\frac1t\sum^n_{j=1}\bsX_j\ot\proj{j}=\frac1t\bigoplus_{j=1}^n\bsX_j=\frac1t \bsY
$$
and its block masses are
$\frac1t p_j=\frac1t\Tr{\bsX_j}$, together with a possible additional zero mass.
Lemma~\ref{lem:pinching} therefore gives
$$
 0\leqslant\S_k^{(D)}(\bsY/t)    -  \S_k^{(D)}(\bsA/t) \leqslant\S_k^{(D)}(\diag(\bsp/t)).
$$
The normalized positive-cone scaling identity is
\begin{eqnarray}\label{eq:scaling-correct}
 \S_k^{(D)}(c\bsX) =c\S_k^{(D)}(\bsX)+\eta(c)\Tr{\bsX},\qquad c>0.
\end{eqnarray}
It follows by dividing \eqref{eq:Entscalingrule} in
Appendix~\ref{app:scaling} by $\binom{D-1}{k-1}$; its extension at
$c=0$ uses $\eta(0)=0$.
Apply \eqref{eq:scaling-correct} with $c=t$ to the two trace-one
matrices above, i.e., 
\begin{eqnarray*}
&&\sfS_k^{(D)}(\bsY)    -  \sfS_k^{(D)}(\bsA) = \sfS_k^{(D)}(t\bsY/t)    -  \sfS_k^{(D)}(t\bsA/t)\\
&&=\Pa{t\S_k^{(D)}(\bsY/t)+\eta(t)\Tr{\bsY/t}} - \Pa{t\S_k^{(D)}(\bsA/t)+\eta(t)\Tr{\bsA/t}}\\
&&=t \Pa{\S_k^{(D)}(\bsY/t) - \S_k^{(D)}(\bsA/t)}\leqslant t  \S_k^{(D)}(\diag(\bsp/t)),
\end{eqnarray*}
where the two $\eta(t)$ terms cancel, and multiplication by
$t$ proves \eqref{eq:applowerupper}.
\end{proof}

\subsubsection{Proof of Theorem~\ref{th:subadditivity}}\label{subsubsect:4.2.2}

\begin{proof}
Choose an orthonormal eigenbasis $\{\ket{j}\}_{j=1}^n$ of $\rho_B$.
Pinching the second subsystem gives
\begin{eqnarray*}
 \rho'=\sum_{j=1}^n\bsX_j\ot\proj{j},\qquad  \bsX_j=(\I_m\otimes\bra{j})\rho_{AB}    (\I_m\ot\ket{j})\geqslant\zero.
\end{eqnarray*}
The blocks satisfy
$\sum_{j=1}^n\bsX_j=\rho_A$ and $\Tr{\bsX_j}=\lambda_j(\rho_B)$.
Extend this pinching to $\bbC^D$ by retaining the extra $D-mn$
coordinates as a separate block when $D>mn$. The lower bound of
Lemma~\ref{lem:pinching} yields
$$
 \S_k^{(D)}(\rho_{AB}) \leqslant\S_k^{(D)}(\rho').
$$
Under the identification
$\bbC^m\otimes\bbC^n=\bigoplus_{j=1}^n\bbC^m$, form
$\bsW=\sum_{j=1}^n\sqrt{\bsX_j}\otimes\ket{j}$. Then
$$
 \bsW^\dagger\bsW=\rho_A,\qquad  \bsW\bsW^\dagger  =\sum^n_{i,j=1}\sqrt{\bsX_i}\sqrt{\bsX_j}\ot\out{i}{j}.
$$
The two Gram matrices have the same nonzero eigenvalues. We view both
matrices after adjoining zero eigenvalues to the same ambient dimension $D$;
then $\bsW\bsW^\dagger$ and $\rho_A$ have identical complete spectra in
that dimension. The block pinching of the former is
$\rho'$, with block masses $\lambda_j(\rho_B)$ and the possible
extra zero mass. The upper bound of Lemma~\ref{lem:pinching} gives
$$
 \S_k^{(D)}(\rho')-\S_k^{(D)}(\rho_A) \leqslant\S_k^{(D)}(\rho_B)\iff  \S_k^{(D)}(\rho') \leqslant \S_k^{(D)}(\rho_A) + \S_k^{(D)}(\rho_B).
$$
Combining the two inequalities proves subadditivity
$$
 \S_k^{(D)}(\rho_{AB}) \leqslant \S_k^{(D)}(\rho') \leqslant \S_k^{(D)}(\rho_A) + \S_k^{(D)}(\rho_B).
$$
Equivalently, the
second step is the trace-one instance of Proposition~\ref{prop:appsubadditivity},
whose proof also records the normalization for arbitrary total trace.
\end{proof}

\subsubsection{Compressed entropies that vanish on pure states}\label{subsubsect:4.2.3}

The same two-pinch argument gives an exact nonnegative remainder in the
comparison of every intermediate deficit with the subentropy deficit.
We first carry it out in dimension $d=mn$, with $d\geqslant2$.
The value of $\G_\ell$ on a pure state is given by \eqref{eq:value@purestate}, i.e.,
\begin{eqnarray*}
 g_\ell:=\G_\ell(\proj{\psi})  = \frac{\ell}{d}(H_d-H_\ell).
\end{eqnarray*}
To calculate it, for $\ell<d$, observe that the only possible nonzero eigenvalue of $\bsV^\dagger\proj{\psi}\bsV$ is $T=\Innerm{\psi}{\bsV\bsV^\dagger}{\psi}$, where $\bsV\bsV^\dagger$ is the projector of rank $\ell$. Thus $T$ can be realized as $r_1+\cdots+r_\ell$, the sum of the first $\ell$ squared coordinates of a Haar unit vector. Since these squared coordinates are uniform on the simplex, the moment can be calculated as
\begin{eqnarray*}
 \bbE[T^s] = \Gamma(d)\int_{\bbR^d_{\geqslant0}} (r_1+\cdots+r_\ell)^s\delta\Pa{1-\sum^d_{j=1}r_j}[\dif\bsr]= \frac{\Gamma(\ell+s)\Gamma(d)}{\Gamma(\ell)\Gamma(d+s)},\quad s>-\ell,\quad [\dif\bsr]:=\prod^d_{j=1}\dif r_j.
\end{eqnarray*}
Differentiating at $s=1$ gives
\begin{eqnarray*}
\bbE[T\log T]= \frac{\dif}{\dif s}\Big|_{s=1}  \bbE[T^s] = \frac{\dif}{\dif s}\Big|_{s=1} \Pa{\frac{\Gamma(\ell+s)\Gamma(d)}{\Gamma(\ell)\Gamma(d+s)}} = \frac{\ell}{d}(H_\ell-H_d),
\end{eqnarray*}
implying that
\begin{eqnarray*}
\G_\ell (\proj{\psi}) &=& \int_{\op{St}(d,\ell)}\Tr{\eta(\bsV^\dagger \proj{\psi} \bsV)}\dif \mu_\ell(\bsV) \\
&=& -\frac{\dif}{\dif s}\Big|_{s=1} \int_{\op{St}(d,\ell)}\Tr{(\bsV^\dagger \proj{\psi} \bsV)^s}\dif \mu_\ell(\bsV)=- \frac{\dif}{\dif s}\Big|_{s=1}  \bbE[T^s]\\
&=& -\frac{\ell}{d}(H_\ell-H_d) = \frac{\ell}{d}(H_d-H_\ell),
\end{eqnarray*}
proving \eqref{eq:value@purestate}. 

On density matrices in dimension $d$, we know that 
$\widetilde \G_\ell(\rho)=\G_\ell(\rho)-g_\ell$.
This functional is unitarily invariant and concave: compression is
linear and $\bsX\mapsto\Tr{\eta(\bsX)}$ is concave.
It vanishes on every pure state, so concavity also makes it nonnegative
on states. By \eqref{eq:B1-compression-gain}, its gain under a fixed
pinching is convex and nonnegative.
For a pure input $\proj{\bsv}$, that gain equals
\begin{eqnarray*}
\G_\ell(\cP(\proj{\bsv})) -\G_\ell(\proj{\bsv}) = \G_\ell(\cP(\proj{\bsv})) - g_\ell = \widetilde \G_\ell(\diag(\bsp^{(\bsv)}))
\end{eqnarray*}
by the spectral calculation in
the proof of Lemma~\ref{lem:pinching}. For the spectral decomposition $\rho=\sum_i w_i\proj{\bsv_i}$, applying convexity of the gain to a pure-state decomposition and then
concavity of $\widetilde \G_\ell$ gives
\begin{eqnarray*}
 0\leqslant \widetilde \G_\ell(\cP(\rho)) - \widetilde \G_\ell(\rho) &\leqslant& \sum_i w_i\Pa{\G_\ell(\cP(\proj{\bsv_i})) -\G_\ell(\proj{\bsv_i}) }\\
&\leqslant& \sum_i w_i\widetilde \G_\ell(\diag(\bsp^{(\bsv_i)}))\leqslant  \widetilde \G_\ell\Pa{\diag\Pa{\sum_i w_i\bsp^{(\bsv_i)}}} \\
&=&\widetilde \G_\ell(\diag(\bsp)),
\end{eqnarray*}
where $\bsp = \sum_i w_i\bsp^{(\bsv_i)}$, that is,
\begin{eqnarray}\label{eq:F-pinching}
 0\leqslant \widetilde \G_\ell(\cP(\rho)) - \widetilde \G_\ell(\rho) \leqslant \widetilde \G_\ell(\diag(\bsp)).
\end{eqnarray}
All states and diagonal matrices in this inequality have trace one.

Apply \eqref{eq:F-pinching} first to $\rho_{AB}$, pinching $B$ in an
eigenbasis of $\rho_B$. As above, the result is
$\rho'=\bigoplus_j\bsX_j$, where
$\sum_j\bsX_j=\rho_A$ and $\Tr{\bsX_j}=\lambda_j(\rho_B)$.
The matrix $\bsW$ formed from the square roots of these blocks satisfies
$\bsW^\dagger\bsW=\rho_A$. Thus $\bsW\bsW^\dagger$ has the spectrum
of $\rho_A$, and its block pinching is $\rho'$ with masses
$\bdlambda(\rho_B)$. The two uses of \eqref{eq:F-pinching} give
\begin{eqnarray*}
\widetilde \G_\ell(\rho_{AB})\leqslant \widetilde \G_\ell(\rho')  \leqslant \widetilde \G_\ell(\rho_A)+\widetilde \G_\ell(\rho_B).
\end{eqnarray*}
Consequently
\begin{eqnarray*}
 \delta_\ell:=\widetilde \G_\ell(\rho_A)   +  \widetilde \G_\ell(\rho_B) - \widetilde \G_\ell(\rho_{AB})  \geqslant0.
\end{eqnarray*}
Every argument of $\widetilde \G_\ell$ has trace one; in particular, the functional
is never applied to an individual unnormalized block.

Evaluation of \eqref{eq:kentropy-intrep} on a pure state shows that
\begin{eqnarray*}
0=\sfS_k^{(d)}(\proj{\psi})=\sum_{\ell=1}^{d-k+1}\mathfrak{C}_{d,k,\ell}\sfG_\ell(\proj{\psi})-(H_k-1)\Tr{\proj{\psi}},
\end{eqnarray*}
implying that $\sum_{\ell=1}^{d-k+1}\frak{C}_{d,k,\ell}g_\ell=H_k-1$. Therefore, on density matrices,
\begin{eqnarray*}
 \S_k^{(d)}&=& \sum_{\ell=1}^{d-k+1}\mathfrak{C}_{d,k,\ell}\sfG_\ell - (H_k-1) =\sum_{\ell=1}^{d-k+1}\mathfrak{C}_{d,k,\ell}\sfG_\ell -\sum_{\ell=1}^{d-k+1}\frak{C}_{d,k,\ell}g_\ell \\
&=&\sum_{\ell=1}^{d-k+1}\mathfrak{C}_{d,k,\ell}(\sfG_\ell-g_\ell)=\sum_{\ell=1}^{d-k+1}\frak{C}_{d,k,\ell}\widetilde \G_\ell.
\end{eqnarray*}
At $k=d$, the coefficient in \eqref{eq:constantC} is 
$$
\frak{C}_{d,d,1}=\Pa{ \frac{k}{\binom{d-1}{k-1}}\binom{d-\ell-1}{k-2}}_{(k,\ell)=(d,1)} = \frac{d}{\binom{d-1}{d-1}}\binom{d-1-1}{d-2}=  d
$$ 
and
$\S_d^{(d)}=\Q$, so 
\begin{eqnarray*}
\Q = \S_d^{(d)} =\sum_{\ell=1}^{d-d+1}\frak{C}_{d,d,\ell}\widetilde \G_\ell= \frak{C}_{d,d,1}\widetilde \G_1= d\widetilde \G_1\implies  \widetilde \G_1=\frac1d\Q.
\end{eqnarray*}
It follows that
\begin{eqnarray*}
 \widetilde\Delta_k^{(d)} - \frac{k(k-1)}{d(d-1)}\Delta_{\Q} &=&  \Pa{\S_k^{(d)}(\rho_A) + \S_k^{(d)}(\rho_B) -  \S_k^{(d)}(\rho_{AB}) } - \tfrac{k(k-1)}{d-1} \Pa{\tfrac{\Q(\rho_A)}d + \tfrac{\Q(\rho_B)}d - \tfrac{\Q(\rho_{AB})}d} \\
&=& \Pa{\sum_{\ell=1}^{d-k+1}\frak{C}_{d,k,\ell}\widetilde \G_\ell(\rho_A) + \sum_{\ell=1}^{d-k+1}\frak{C}_{d,k,\ell}\widetilde \G_\ell(\rho_B) - \sum_{\ell=1}^{d-k+1}\frak{C}_{d,k,\ell}\widetilde \G_\ell(\rho_{AB})}\\
&&  - \tfrac{k(k-1)}{d-1} \Pa{\widetilde \G_1(\rho_A) + \widetilde \G_1(\rho_B) - \widetilde \G_1(\rho_{AB})}\\
&=& \sum_{\ell=1}^{d-k+1}\frak{C}_{d,k,\ell}\Pa{\widetilde \G_\ell(\rho_A) + \widetilde \G_\ell(\rho_B) - \widetilde \G_\ell(\rho_{AB})} - \tfrac{k(k-1)}{d-1}\delta_1\\
&=&  \sum_{\ell=1}^{d-k+1}\frak{C}_{d,k,\ell}\delta_\ell - \mathfrak{C}_{d,k,1}\delta_1.
\end{eqnarray*}
That is,
\begin{eqnarray}\label{eq:positive-remainder}
 \widetilde\Delta_k^{(d)} - \frac{k(k-1)}{d(d-1)}\Delta_{\Q} =\sum_{\ell=2}^{d-k+1}\frak{C}_{d,k,\ell}\delta_\ell
 \geqslant0,
\end{eqnarray}
where an empty sum is zero. Here we used
\[
 \frac{\frak{C}_{d,k,1}}{d}
 =\frac{k}{d\binom{d-1}{k-1}}\binom{d-2}{k-2}
 =\frac{k(k-1)}{d(d-1)}.
\]

For $D\geqslant mn$ with $D\geqslant2$, repeat the construction in
dimension $D$, extending each pinching by the zero complement.
In this repetition $\G_\ell$, $g_\ell$ and $\widetilde \G_\ell$ are defined in
dimension $D$, and all matrices in their arguments are padded to $D$.
The same Gram construction proves $\delta_\ell\geqslant0$, now with
the coefficients $\frak{C}_{D,k,\ell}$. Its rank-one term is
$\widetilde \G_1=\frac1D \Q$; subentropy is invariant under appending zero eigenvalues.
We obtain
\begin{eqnarray}\label{eq:transfer-main}
 \widetilde\Delta_k^{(D)}(\rho_{AB})
 \geqslant\frac{k(k-1)}{D(D-1)}\Delta_{\Q}(\rho_{AB}),
 \qquad 2\leqslant k\leqslant D,
\end{eqnarray}
together with the exact remainder \eqref{eq:positive-remainder} with
$d$ replaced by $D$. This argument uses the pinching inequality and
concavity, independently of any characterization of equality in
subadditivity.

\subsubsection{Proof of Theorem~\ref{th:equality}}\label{subsubsect:4.2.4}

\begin{proof}
For $k=1$, padding does not change von Neumann entropy, and
\[
 \widetilde\Delta_1^{(D)}(\rho_{AB})
 =\B_+(\rho_{AB}||\rho_A\otimes\rho_B).
\]
To see that the expression is finite, let $\Pi_A$ and $\Pi_B$ be
the support projectors of the marginals. The zero expectation of
$(\I_m-\Pi_A)\otimes\I_n$ in $\rho_{AB}$, and positivity, imply
that $\rho_{AB}$ is supported on $\op{supp}[\rho_A]\ot\bbC^n$.
The corresponding statement for $B$ gives
$\op{supp}[\rho_{AB}]\subseteq
 \op{supp}[\rho_A]\ot\op{supp}[\rho_B]$.
On this support,
$\log(\rho_A\ot\rho_B)=\log\rho_A\ot\I_n+\I_m\ot\log\rho_B$.
Taking partial traces yields the displayed relative-entropy identity.
The trace correction vanishes because both arguments have trace one.
The equality condition for relative entropy is therefore precisely
$\rho_{AB}=\rho_A\otimes\rho_B$.

For $k\geqslant2$, combine \eqref{eq:transfer-main} with the subentropy purity estimate \eqref{eq:native-Q-purity}, proved below without using
Theorem~\ref{th:equality}. If both marginals are mixed, then $ (1-\Tr{\rho_A^2})(1-\Tr{\rho_B^2})>0$, so the subentropy deficit, and hence $\widetilde\Delta_k^{(D)}$, is strictly positive. Thus equality requires at least one pure marginal.
The proof of \eqref{eq:native-Q-purity} uses the pinching inequality and explicit Gaussian and four-sample calculations, so there is no
circular dependence on the equality statement.

Conversely, suppose $\rho_A=\proj{\bsa}$ and put $\bsR=(\I_m-\proj{\bsa})\otimes\I_n$. Then
$0=\Tr{\bsR\rho_{AB}}$. Since $\bsR$ is an orthogonal projection,
\begin{eqnarray*}
0=\Tr{\bsR\rho_{AB}}
=\Tr{\rho_{AB}^{1/2}\bsR\rho_{AB}^{1/2}}
=\norm{\bsR\rho_{AB}^{1/2}}_2^2.
\end{eqnarray*}
Thus $\bsR\rho_{AB}^{1/2}=\zero$, which implies
$\bsR\rho_{AB}=\rho_{AB}\bsR=\zero$. With
$\bsE=\proj{\bsa}\otimes\I_n=\I_{mn}-\bsR$, we obtain
$\rho_{AB}=\bsE\rho_{AB}\bsE$. Define the isometry
$\bsK:\bbC^n\to\bbC^m\otimes\bbC^n$ by
$\bsK\ket{\bsb}=\ket{\bsa}\otimes\ket{\bsb}$.
Then $\bsK\bsK^\dagger=\bsE$, and therefore
\begin{eqnarray*}
\rho_{AB}=\bsK(\bsK^\dagger\rho_{AB}\bsK)\bsK^\dagger
=\proj{\bsa}\otimes(\bsK^\dagger\rho_{AB}\bsK).
\end{eqnarray*}
Taking the $B$ marginal identifies
$\bsK^\dagger\rho_{AB}\bsK=\rho_B$, proving
$\rho_{AB}=\proj{\bsa}\otimes\rho_B$. The spectra of $\rho_{AB}$ and $\rho_B$ agree, whereas
$\S_k^{(D)}(\rho_A)=0$. The deficit is therefore zero. The same support argument applies when $\rho_B$ is pure.
\end{proof}

\subsubsection{Proof of Corollary~\ref{cor:multipartite}}\label{subsubsect:4.2.5}

\begin{proof}
Apply Theorem~\ref{th:subadditivity} to the cut
$1\cdots(r-1):r$. Apply it next to
$1\cdots(r-2):(r-1)$ in the first marginal, and continue.
Every intermediate joint space has dimension at most
$\prod_{i=1}^r d_i\leqslant D$. Thus the same $D$ and $k$ apply at
each step, with every reduced spectrum padded to $D$.
After $r-1$ applications the desired sum of the individual entropies
is obtained.
\end{proof}

\subsubsection{Proof of Corollary~\ref{cor:triangle}}\label{subsubsect:4.2.6}

\begin{proof}
Let $r=\rank[\rho_{AB}]$ and choose a purification $\proj{\Psi}_{ABC}$ with $\dim C=r$.
Complementary reductions of a pure state have the same nonzero
eigenvalues. After padding to $D$, they consequently have the same
$\S_k^{(D)}$. The assumption $D\geqslant mnr$ allows all cuts below
to use this one ambient dimension. Subadditivity on $BC$ gives
 \begin{eqnarray*}
 \S_k^{(D)}(\rho_A)=\S_k^{(D)}(\rho_{BC})\leqslant\S_k^{(D)}(\rho_B)   +\S_k^{(D)}(\rho_C)=\S_k^{(D)}(\rho_B)
          +\S_k^{(D)}(\rho_{AB}).
 \end{eqnarray*}
Interchanging $A$ and $B$ gives the other one-sided inequality and
hence the absolute-value bound.

When $k\geqslant2$, equality in the displayed direction holds exactly
when $\rho_B$ or $\rho_C$ is pure by Theorem~\ref{th:equality}.
The state $\rho_C$ is pure exactly when $\rho_{AB}$ is pure.
The other direction gives the additional possibility that $\rho_A$
is pure. Conversely, if $\rho_{AB}$ is pure, the two marginal spectra
agree and $\S_k^{(D)}(\rho_{AB})=0$.
If one marginal is pure, the support argument above makes $\rho_{AB}$
a product with that pure factor, and its padded spectrum agrees with
the other marginal's spectrum. Each listed case therefore attains
equality. At $k=D$, the functional is the padding-invariant subentropy
$\Q$, yielding its stated triangle inequality.
\end{proof}

\subsubsection{Proof of Corollary~\ref{cor:SSApure}}\label{subsubsect:4.2.7}

\begin{proof}
For the pure state $\rho_{ABC}$ the global entropy is zero.
Complementary spectra, compared in dimension $D$, give
\[
 \S_k^{(D)}(\rho_{AB})=\S_k^{(D)}(\rho_C),\qquad
 \S_k^{(D)}(\rho_{BC})=\S_k^{(D)}(\rho_A),\qquad
 \S_k^{(D)}(\rho_B)=\S_k^{(D)}(\rho_{AC}).
\]
Thus strong subadditivity for this state is exactly
\[
 \S_k^{(D)}(\rho_A)+\S_k^{(D)}(\rho_C)
 \geqslant\S_k^{(D)}(\rho_{AC}),
\]
which is Theorem~\ref{th:subadditivity} on $AC$.
For $k\geqslant2$, Theorem~\ref{th:equality} says that equality holds
exactly when $\rho_A$ or $\rho_C$ is pure.
\end{proof}

\subsection{Proofs for the $k$-product gap}\label{proof:productgap}

The common ambient dimension $D$ is fixed throughout this subsection.
For probability vectors $\bdlambda,\bdmu$, every term of
$\widetilde\Delta_k^{(D),\op{prod}}(\bdlambda,\bdmu)$ is therefore
computed after padding to $D$. In particular, the product gap is the
bipartite deficit of
$\diag(\bdlambda)\ot\diag(\bdmu)$ in that same dimension. Its
nonnegativity follows from Theorem~\ref{th:subadditivity}; for
$k\geqslant2$, Theorem~\ref{th:equality} gives
\begin{eqnarray*}
 \widetilde\Delta_k^{(D),\op{prod}}(\bdlambda,\bdmu)=0
 \quad\iff\quad \bdlambda\text{ or }\bdmu\text{ is pure}.
\end{eqnarray*}
For $k=1$, ordinary entropy additivity makes this gap identically zero.
The separate concavity proof requires more than concavity of the three
individual entropy terms: it uses convexity of the entropy gain of one
fixed mixed-unitary channel.

\subsubsection{Proof of Theorem~\ref{th:sepconcavity}} \label{subsubsect:4.3.1}

\begin{proof}
\begin{itemize}
\item \emph{Convexity of a mixed-unitary gain.} Let $\cE(\bsX)=\sum_i p_i\bsU_i\bsX\bsU_i^\dagger$ be a fixed mixed-unitary channel on $\bbC^D$. The probabilities $p_i$ and the unitaries $\bsU_i$ are fixed independently of the input state; terms with $p_i=0$ may be discarded. Recall the compressed entropy functional
\begin{eqnarray*}
 \G_\ell(\bsX)=\int_{\op{St}(D,\ell)} \Tr{\eta(\bsV^\dagger\bsX\bsV)}\dif\mu_\ell(\bsV), \qquad 1\leqslant\ell\leqslant D,
\end{eqnarray*}
where the compressed matrix is not normalized by its trace.
For a fixed $\bsV\in \op{St}(D,\ell)$, put
$\bsX_{i,\bsV}=\bsV^\dagger\bsU_i\rho\bsU_i^\dagger\bsV$
and
$\bsY_{\bsV}=\bsV^\dagger\cE(\rho)\bsV
=\sum_ip_i\bsX_{i,\bsV}$. Expanding the generalized relative entropy
$\B_+$ gives
\begin{eqnarray*}
 \Tr{\eta(\bsY_{\bsV})} -\sum_ip_i\Tr{\eta(\bsX_{i,\bsV})}=\sum_ip_i\B_+(\bsX_{i,\bsV}||\bsY_{\bsV}).
\end{eqnarray*}
Indeed, the terms linear in the traces cancel, and
$\sum_ip_i\Tr{\bsX_{i,\bsV}\log\bsY_{\bsV}}=\Tr{\bsY_{\bsV}\log\bsY_{\bsV}}=-\Tr{\eta(\bsY_{\bsV})}$.
For each positive $p_i$, the inequality
$\bsX_{i,\bsV}\leqslant p_i^{-1}\bsY_{\bsV}$ implies
$\op{supp}[\bsX_{i,\bsV}]\subseteq\op{supp}[\bsY_{\bsV}]$.
Thus all relative entropies are defined on the relevant support, even
when $\rho$ or a compression is singular. On that support the same
identity follows by restriction, or by a decreasing positive-definite
regularization. The trace entropies on the left are bounded because the
compressed spectra lie in $[0,1]$ and the matrix size is fixed; hence the
nonnegative sum on the right is finite and integrable.

Haar invariance under $\bsV\mapsto\bsU_i^\dagger\bsV$ now yields
\begin{eqnarray}\label{eq:mixed-compression-gain}
 \G_\ell(\cE(\rho)) - \G_\ell(\rho) &=&\sum_ip_i\int_{\op{St}(D,\ell)} \B_+\Pa{\bsV^\dagger\bsU_i\rho\bsU_i^\dagger\bsV ||\bsV^\dagger\cE(\rho)\bsV}\dif\mu_\ell(\bsV).
\end{eqnarray}
Both arguments of every integrand are linear in $\rho$. Joint convexity
of $\B_+$ therefore proves that this gain is convex in $\rho$.
Subtracting the pure-state value
$g_\ell=\frac\ell D(H_D-H_\ell)$ from $G_\ell$ does not change its
gain. Consequently the same conclusion holds for
$\widetilde \G_\ell=\G_\ell-g_\ell$ on density matrices. The positive compression
representation proved earlier reads
\begin{eqnarray*}
 \S_k^{(D)}=\sum_{\ell=1}^{D-k+1} \frak{C}_{D,k,\ell}\widetilde \G_\ell, \qquad \frak{C}_{D,k,\ell} =\frac{k}{\binom{D-1}{k-1}}\binom{D-\ell-1}{k-2} \quad (k\geqslant2).
\end{eqnarray*}
Its coefficients are positive. It follows from
\eqref{eq:mixed-compression-gain} that
$\C_k^{(D)}(\cE,\rho)$ is convex. At $k=1$ the same argument applies
with $\G_D=\S$.

\item \emph{A fixed channel implementing the product spectrum.}  Fix a probability vector $\bdmu$ and write the variable spectrum as
$\bdlambda$, where
$$
 \bdmu=(\mu_1,\ldots,\mu_n),\qquad
 \bdlambda=(\lambda_1,\ldots,\lambda_m),\qquad mn\leqslant D.
$$
Arrange the first $mn$ coordinates of $\bbC^D$ in $n$ consecutive
blocks of size $m$. For $1\leqslant j\leqslant n$, choose the permutation
unitary $\bsU_j$ that shifts the first block to the $j$-th block cyclically
and acts as the identity on the remaining $D-mn$ coordinates. Define
\begin{eqnarray*}
 \cE_{\bdmu}(\bsX)=\sum_{j=1}^n\mu_j\bsU_j\bsX\bsU_j^\dagger,
 \qquad
 \bsX_{\bdlambda}=\diag(\lambda_1,\ldots,\lambda_m,0,\ldots,0).
\end{eqnarray*}
The $j$-th block of $\cE_{\bdmu}(\bsX_{\bdlambda})$ is
$\mu_j\diag(\bdlambda)$. Its nonzero spectrum is therefore
$\bdlambda\ot\bdmu$, up to a permutation. This channel depends on
$\bdmu$, but not on the variable $\bdlambda$. Hence
\begin{eqnarray*}
 \widetilde\Delta_k^{(D),\op{prod}}(\bdlambda,\bdmu)
 &=&\S_k^{(D)}(\bdmu)
 -\C_k^{(D)}(\cE_{\bdmu},\bsX_{\bdlambda}).
\end{eqnarray*}
The first term is constant and the second is convex in $\bdlambda$.
This proves concavity in the first argument. Symmetry of the product
spectrum proves concavity in the second argument. The identical argument
also proves separate concavity of the product gap formed with each
$\widetilde \G_\ell$.

\item \emph{Strictness of the subentropy component.} We give the calculation on the closed simplex, since strictness at
boundary spectra is needed for the rank-constrained maximum.
Let $\gamma$ be Euler's constant. If $\xi_1,\ldots,\xi_D$ are independent exponential random
variables of mean one, then
\begin{eqnarray}\label{eq:product-gaussian-Q}
 \Q(\bdlambda) =1-\gamma + \bbE\Br{\eta\Pa{\sum^D_{i=1}\lambda_i\xi_i}} =1-\gamma + \bbE\Br{\eta\Pa{\Inner{\bdlambda}{\bdxi}}}.
\end{eqnarray}
Here $\bdlambda$ is padded to length $D$ if necessary. To verify this
identity, put $R=\sum^D_{i=1}\xi_i=\Inner{\bse}{\bdxi}$ and $T_i=\xi_i/R$. The vector
$\bsT=(T_1,\ldots,T_D)$ is uniform on $\Delta_{D-1}$ and is independent of
$R$, whose law is Gamma$(D,1)$. Write $v=\sum^D_{i=1}\lambda_iT_i=\Inner{\bdlambda}{\bsT}$, so
$\bbE [v]=\frac1D$. The moment formula
$\bbE [R^s]=\frac{\Gamma(D+s)}{\Gamma(D)}$ gives
\begin{eqnarray*}
 \bbE [R]=D,\qquad   \bbE[R\log R]=D(H_D-\gamma).
\end{eqnarray*}
Using independence and $-\eta(Rv)=Rv\log R+Rv\log v$ gives
\begin{eqnarray*}
 -\bbE [\eta(Rv)] = H_D-\gamma + D\bbE[v\log v].
\end{eqnarray*}
The simplex formula \eqref{eq:Skrho} at $k=D$ gives
$\Q(\bdlambda)=-D\bbE[v\log v]-(H_D-1)$, which is
\eqref{eq:product-gaussian-Q}. Zero eigenvalues are allowed: the
random argument lies between zero and $R$, while $-\eta$ is bounded below
and its positive part is bounded by $R\log_+R$. The latter has finite
expectation, so dominated convergence extends the identity to every
boundary spectrum.

Now fix a mixed $\bdmu$ and distinct probability vectors
$\bdlambda^0,\bdlambda^1$. Set
$\bdlambda(t)=(1-t)\bdlambda^0+t\bdlambda^1$ and
$\bsv=\bdlambda^1-\bdlambda^0$. Discard coordinates that vanish at
both endpoints. On the remaining support $\lambda_i(t)>0$ for
$0<t<1$. With independent mean-one exponentials $\xi_{ij}$, define
\begin{eqnarray*}
 X_j=\sum_i\lambda_i(t)\xi_{ij},\qquad
 Y_j=\sum_i v_i\xi_{ij}.
\end{eqnarray*}
For different $j$ the pairs $(X_j,Y_j)$ are independent and have the
same law. Since $-\eta''(x)=\frac1x$, differentiating
\eqref{eq:product-gaussian-Q} twice yields
\begin{eqnarray}\label{eq:Q-gap-hessian}
 \frac{\dif^2}{\dif t^2}
 \Br{\Q(\bdlambda(t))+\Q(\bdmu)-\Q(\bdlambda(t)\ot\bdmu)}
 =-\bbE\Br{\frac{Y_1^2}{X_1}}
 +\bbE\Br{\frac{\Pa{\sum_j\mu_jY_j}^2}{\sum_j\mu_jX_j}}.
\end{eqnarray}
To justify differentiation, fix a compact interval inside $(0,1)$.
Every retained $\lambda_i(t)$ has a positive lower bound on that
interval, so $|v_i|\leqslant C\lambda_i(t)$ with a common constant
$C$. Thus $|Y_j|\leqslant CX_j$ and
$Y_j^2/X_j\leqslant C^2X_j$. The latter has finite expectation.
The corresponding product term is bounded by
$C^2\sum_j\mu_jX_j$. These bounds justify the second derivative under
the expectation; the first derivative is also integrable because the
exponential variables have finite logarithmic moments.

Weighted Cauchy--Schwarz gives pointwise
\begin{eqnarray*}
 \frac{\Pa{\sum_j\mu_jY_j}^2}{\sum_j\mu_jX_j}
 \leqslant\sum_j\mu_j\frac{Y_j^2}{X_j}.
\end{eqnarray*}
The expectation of the right side equals
$\bbE[Y_1^2/X_1]$ and therefore cancels the first term of
\eqref{eq:Q-gap-hessian}. Equality in Cauchy--Schwarz would require
$Y_j/X_j$ to agree for all $j$ with $\mu_j>0$. There are at least two
such indices because $\bdmu$ is mixed. The ratio $Y_j/X_j$ is
nonconstant: otherwise, for some constant $c$,
\begin{eqnarray*}
 \sum_i\Br{v_i-c\lambda_i(t)}\xi_{ij}=0
 \quad\text{almost surely}.
\end{eqnarray*}
The joint exponential density is positive on the open positive orthant,
so this linear identity forces $v_i=c\lambda_i(t)$ for every retained
coordinate. Summing gives $c=0$ because $\sum_i v_i=0$ and
$\sum_i\lambda_i(t)=1$. This contradicts $\bsv\neq\zero$.
Two independent copies of a nonconstant random variable differ with
positive probability. Thus the Cauchy--Schwarz inequality is strict
with positive probability, and
\begin{eqnarray*}
 \frac{\dif^2}{\dif t^2}
 \Br{\Q(\bdlambda(t))+\Q(\bdmu)-\Q(\bdlambda(t)\ot\bdmu)}<0
 \qquad(0<t<1).
\end{eqnarray*}
The subentropy product gap is therefore strictly concave along every
nonconstant interior segment. Its continuity extends strict concavity
to distinct boundary endpoints as well: equality at an interior point
of a closed segment for a concave function would force that function
to be affine on the whole segment, contradicting the strictly negative
second derivative on its interior.

Finally, in the fixed ambient dimension $D$, the rank-one compression
satisfies $\widetilde \G_1=\frac1D\Q$. Its coefficient in the positive expansion of
$\S_k^{(D)}$ is
\begin{eqnarray*}
 \frac{\frak{C}_{D,k,1}}D =\frac{k(k-1)}{D(D-1)}>0 \qquad(k\geqslant2).
\end{eqnarray*}
The product gaps for all remaining $\widetilde \G_\ell$ are separately concave by
the fixed-channel argument above. Adding them to this strictly concave
subentropy component proves strict concavity of the full product gap
when the fixed other argument is mixed. If that argument is pure, the
product gap vanishes identically, as already noted. This completes the
proof of Theorem~\ref{th:sepconcavity}. \qedhere
\end{itemize}
\end{proof}

\subsubsection{Proof of Theorem~\ref{th:rank-max}}\label{subsubsect:4.3.2}

\begin{proof}
Use $\alpha,\beta$ for the two support-size bounds in the statement.
Permuting coordinates first, each permitted spectrum can be represented
as a probability vector of length $\alpha$ or $\beta$, respectively,
by appending zeros when its support is smaller. This does not change any
of the three spectra after padding to $D$, and
$\alpha\beta\leqslant D$ ensures that every product spectrum fits in
the common ambient space.

Let $\bsu_\alpha=(1/\alpha,\ldots,1/\alpha)$. Averaging all
$\alpha!$ permutations $\pi$ of $\bdlambda$ gives
$\bsu_\alpha$. Separate concavity and permutation invariance imply
\begin{eqnarray*}
 \widetilde\Delta_k^{(D),\op{prod}}(\bdlambda,\bdmu)=\frac1{\alpha!}\sum_\pi
 \widetilde\Delta_k^{(D),\op{prod}}(\pi\bdlambda,\bdmu)\leqslant  \widetilde\Delta_k^{(D),\op{prod}}(\bsu_\alpha,\bdmu).
\end{eqnarray*}
Averaging the $\beta!$ permutations of $\bdmu$ similarly gives
\begin{eqnarray*}
 \widetilde\Delta_k^{(D),\op{prod}}(\bdlambda,\bdmu)
 \leqslant \widetilde\Delta_k^{(D),\op{prod}}(\bsu_\alpha,\bsu_\beta)=T_{D,k}(\alpha)+T_{D,k}(\beta)-T_{D,k}(\alpha\beta)
 =d_{\alpha\beta}^{(D,k)}.
\end{eqnarray*}
The evaluation uses the augmentation identity
\eqref{eq:padding-normalized} and the uniform values from
Proposition~\ref{prop:extval}; equivalently it is the formula
\eqref{eq:padded-uniform}. The maximum is positive because both uniform
spectra are mixed and $k\geqslant2$.

If equality holds at the maximum, neither original spectrum can be
pure, since that would give the zero gap. Thus strict concavity from
Theorem~\ref{th:sepconcavity} applies to both permutation averages.
Equality in the first average requires all permutations of
$\bdlambda$ to coincide, so $\bdlambda=\bsu_\alpha$; equality in the
second requires $\bdmu=\bsu_\beta$. Conversely these uniform spectra
attain the stated value. This also shows that a smaller support cannot
attain the positive maximum. If either support bound is one, one factor
is necessarily pure and the gap is zero for every permitted pair.
At $k=1$ ordinary product additivity gives zero for every pair, so no
uniqueness is asserted. At the subentropy endpoint,
$T_{D,D}(\ell)=\log\ell-H_\ell+1$ gives
$ d_{\alpha\beta}^{(D,D)}
 =H_{\alpha\beta}-H_\alpha-H_\beta+1$, as stated.
\end{proof}

\subsection{Proofs for failure of strong subadditivity}\label{proof:ssafailure}

We retain the sign convention of \eqref{eq:kSSAdefect}: a positive
$\V_k^{(D)}$ violates strong subadditivity. The common dimension $D$
is unchanged when the global state is replaced by any marginal.

\subsubsection{Proof of Proposition~\ref{prop:cexam}}\label{subsubsect:4.4.1}

\begin{proof}
Write the Schmidt decomposition of $\ket\psi$ as in
\eqref{eq:counterexample}. The global state
$\rho_{ABC}=\proj\psi_{AB}\ot\tau_C$ has the same nonzero spectrum
$\bdmu$ as $\tau_C$, because its $AB$ factor is pure. Its $B$ marginal
has nonzero spectrum $\bdlambda$, and its $AB$ marginal is pure.
Finally,
$\rho_{BC}=\rho_B\ot\tau_C$ has nonzero spectrum
$\bdlambda\ot\bdmu$. Padding all four spectra to $D$, and using that
$\S_k^{(D)}$ vanishes on pure states, gives
\begin{eqnarray*}
 \V_k^{(D)}(\rho_{ABC})
 &=&\S_k^{(D)}(\bdmu)+\S_k^{(D)}(\bdlambda)
 -0-\S_k^{(D)}(\bdlambda\ot\bdmu)\\
 &=&\widetilde\Delta_k^{(D),\op{prod}}(\bdlambda,\bdmu).
\end{eqnarray*}
This uses equality of padded nonzero spectra, and does not identify
entropies computed in different ambient dimensions.
\end{proof}

\subsubsection{Proof of Theorem~\ref{th:FSSA}}\label{subsubsect:4.4.2}

\begin{proof}
Proposition~\ref{prop:cexam} reduces the defect to the bipartite deficit
of $\diag(\bdlambda)\ot\diag(\bdmu)$ in dimension $D$.
Theorems~\ref{th:subadditivity} and \ref{th:equality} show that this
quantity is strictly positive for $k\geqslant2$ precisely when both
spectra are mixed. For the Schmidt spectrum $\bdlambda$, mixedness is
equivalent to entanglement of $\ket\psi$; for $\bdmu$ it is precisely
mixedness of $\tau_C$. If either spectrum is pure, the product spectrum
has the same nonzero entries as the other one and the product gap is
zero. For $k=1$, ordinary entropy additivity makes the defect zero for
every state in this family.

For the particular state \eqref{eq:ABotC}, the two spectra are both
$(1/2,1/2)$. Each physical factor has a two-dimensional subspace because
$m,n,r\geqslant2$, so the state embeds in every space stipulated in the
theorem. Its defect is strictly positive for every $2\leqslant k\leqslant D$.
The normalized augmentation formula \eqref{eq:padding-normalized}
gives, for a uniform rank-$\ell$ spectrum,
\begin{eqnarray*}
 \S_k^{(D)}\Pa{\frac{\I_\ell}\ell}
 =\frac1{\binom{D-1}{k-1}}
 \sum_{j=1}^{\min(k,\ell)}
 \binom{D-\ell}{k-j}\binom{\ell-1}{j-1}
 \Pa{\log\ell-H_j+1}
 =T_{D,k}(\ell).
\end{eqnarray*}
Consequently the Bell-family defect is
$2T_{D,k}(2)-T_{D,k}(4)$. Substituting $D=8$ yields the eight entries
listed in Section~\ref{subsect:FSSA}. At $k=D$, zero-padding invariance
of $\Q$ gives independently
\begin{eqnarray*}
 2\Q\Pa{\frac{\I_2}2}-\Q\Pa{\frac{\I_4}4}
 =2\Pa{\log2-\frac12}
 -\Pa{\log4-\frac{13}{12}}=\frac1{12}.
\end{eqnarray*}
More generally, uniform Schmidt rank $\alpha$ and uniform ancilla rank
$\beta$ give the exact value
$T_{D,k}(\alpha)+T_{D,k}(\beta)-T_{D,k}(\alpha\beta)$.

The weak-monotonicity expression follows from the same four spectra.
In the present state, the nonzero spectra of $AB,BC,A,C$ are
$(1),\bdlambda\ot\bdmu,\bdlambda,\bdmu$, respectively. Therefore
\begin{eqnarray*}
 \W_k^{(D)}
 &=&0+\S_k^{(D)}(\bdlambda\ot\bdmu)
 -\S_k^{(D)}(\bdlambda)-\S_k^{(D)}(\bdmu)\\
 &=&-\widetilde\Delta_k^{(D),\op{prod}}(\bdlambda,\bdmu)
 =-\V_k^{(D)}.
\end{eqnarray*}
The ordinary von Neumann entropies, by contrast, satisfy
\begin{eqnarray*}
 \sfI(A:C|B)
 &=&\S(\rho_{AB})+\S(\rho_{BC})-\S(\rho_B)-\S(\rho_{ABC})\\
 &=&0+\Br{\S(\bdlambda)+\S(\bdmu)}
 -\S(\bdlambda)-\S(\bdmu)=0.
\end{eqnarray*}
Thus these strict violations occur on ordinary quantum Markov states.
\end{proof}

\subsubsection{Proof of Corollary~\ref{cor:markov-signs}}\label{subsubsect:4.4.3}

\begin{proof}
Let $\rho^+_{ABC}=\proj\psi_{AB}\ot\tau_C$ and
$\rho^-_{ABC}=\rho_A^+\ot\proj{\bsb}_B\ot\tau_C$ as in the
statement. The preceding calculation gives
$\sfI(A:C|B)_{\rho^+}=0$. The negative witness is a product of its
three factors, with a pure $B$ factor, so ordinary entropy additivity
also gives $\sfI(A:C|B)_{\rho^-}=0$. Its nonzero spectra in the order
$ABC,B,AB,BC$ are
$\bdlambda\ot\bdmu,(1),\bdlambda,\bdmu$. Hence
\begin{eqnarray*}
 \V_k^{(D)}(\rho^-_{ABC})
 &=&\S_k^{(D)}(\bdlambda\ot\bdmu)+0
 -\S_k^{(D)}(\bdlambda)-\S_k^{(D)}(\bdmu)\\
 &=&-\widetilde\Delta_k^{(D),\op{prod}}(\bdlambda,\bdmu)
 =-\V_k^{(D)}(\rho^+_{ABC}).
\end{eqnarray*}
This proves \eqref{eq:markov-opposite}, including cases in which one
or both spectra are pure. If both are mixed, Theorem~\ref{th:equality}
gives strict positivity of the common magnitude. Taking both spectra
uniform of rank two gives both signs in every stipulated physical
dimension and for every $2\leqslant k\leqslant D$; the negative values
are precisely the negatives of the displayed Bell-family table.

For the quantitative estimate, let
$a=\rank(\rho_A^+)$, $c=\rank(\tau_C)$ and $q=ac$. The diagonal
product state with these spectra is supported on
$\bbC^a\ot\bbC^c$, and $q\leqslant mr\leqslant D$.
The comparison \eqref{eq:transfer-main}, in common dimension $D$, gives
\begin{eqnarray*}
 \widetilde\Delta_k^{(D),\op{prod}}(\bdlambda,\bdmu)
 \geqslant\frac{k(k-1)}{D(D-1)}
 \Br{\Q(\bdlambda)+\Q(\bdmu)-\Q(\bdlambda\ot\bdmu)}.
\end{eqnarray*}
Subentropy and this deficit are invariant under all zero padding.
Thus, for $q\geqslant2$, the purity bound
\eqref{eq:native-Q-purity} can be applied in the support dimension $q$:
\begin{eqnarray*}
 \Q(\bdlambda)+\Q(\bdmu)-\Q(\bdlambda\ot\bdmu)
 \geqslant
 \frac{\Pa{1-\Tr{(\rho_A^+)^2}}\Pa{1-\Tr{\tau_C^2}}}
 {8q(q+1)(q+2)(q+3)}.
\end{eqnarray*}
Multiplying the two inequalities gives exactly
\eqref{eq:markov-quantitative}. If $q=1$, both factors are pure and
both sides vanish. If exactly one factor is pure, the product-support
identity makes the gap zero and the corresponding purity factor is zero.
If both are mixed, both purity factors are positive, so the displayed
lower bound is strictly positive. Since $\ket\psi$ is pure bipartite,
this is equivalent to the entanglement and mixedness condition in the
corollary.
\end{proof}

For completeness, the maximum on the pure-bipartite-factor family in
Section~\ref{subsect:FSSA} follows without another optimization.
For Schmidt rank at most $\alpha$ and ancilla rank at most $\beta$,
Proposition~\ref{prop:cexam} identifies the defect with the product gap,
and Theorem~\ref{th:rank-max} bounds it by
$d_{\alpha\beta}^{(D,k)}$. Equality requires uniform Schmidt
probabilities on a rank-$\alpha$ subspace and a uniform rank-$\beta$
ancilla, because these are exactly the equality spectra in the product-gap
maximum. Uniform Schmidt probabilities mean that $\ket\psi$ is maximally
entangled on the chosen support. Local unitary changes and changes of
support do not affect the spectra. If either rank bound is one, one
factor spectrum is pure and every permitted state has zero defect.
This proves the stated maximum and all of its equality cases within
the specified family.

\subsection{Quantitative bounds}\label{proof:quantitativebounds}

We prove the assertions of Section~\ref{subsect:QB} in their stated
order. The first three concern the optimal constants for two qubits.
Their common reduction is a pinching in a product of marginal eigenbases,
followed by an exact comparison of scalar kernels. We give the two
polynomial calculations in full, including coverage of the parameter
region and the equality cases. The general-dimensional estimate is
then proved independently from a Gaussian representation, before we
deduce the trace-distance bound.

\subsubsection{Proof of Theorem~\ref{th:lowerboundfor2-qubit}}\label{subsubsect:4.5.1}

\begin{proof}
Pinch in a product of marginal eigenbases. That is, let $\Pi_{ij}=\proj{ij}$, where $\Pi^A_i=\proj{i}$ and $\Pi^B_j=\proj{j}$ are the eigen-projections of $\rho_A$ and $\rho_B$, respectively, where $i\in[m]$ and $j\in[n]$. Let $\rho'_{AB}=\cP(\rho_{AB})=\sum^{m,n}_{i,j=1}\Pi_{ij}\rho_{AB}\Pi_{ij}$. The marginals are unchanged,
$$
\rho_A = \sum^m_{i=1}\Pi^A_i\rho_A\Pi^A_i=\rho'_A,\qquad \rho_B = \sum^n_{j=1}\Pi^B_j\rho_B\Pi^B_j=\rho'_B
$$
and concavity and unitary invariance give $\Q(\rho_{AB})\leqslant \Q(\rho'_{AB})$, where $m=n=2$. Now
\begin{eqnarray*}
\Delta_{\Q}(\rho_{AB}) = \Q(\rho_A) + \Q(\rho_B) - \Q(\rho_{AB}) \geqslant  \Q(\rho'_A) + \Q(\rho'_B) - \Q(\rho'_{AB})=\Delta_{\Q}(\rho'_{AB}). 
\end{eqnarray*}
Thus
$$
x=\Innerm{00}{\rho_{AB}}{00},\quad y=\Innerm{01}{\rho_{AB}}{01},\quad z= \Innerm{10}{\rho_{AB}}{10},\quad w=\Innerm{11}{\rho_{AB}}{11}.
$$
With respect to the orthonormal basis $\set{\ket{00},\ket{01},\ket{10},\ket{11}}$, the pinched state $\rho'_{AB}=\cP(\rho_{AB})=\diag(x,y,z,w)$ with $\rho'_A=\diag(x+y,z+w)$ and $\rho'_B=\diag(x+z,y+w)$. 
Thus 
$$
\Q(\rho'_{AB})=\Q(\diag(x,y,z,w));
$$ 
the same convention applies to $\S_k^{(D)}$. By \eqref{eq:integrationrep-single}, 
\begin{eqnarray*}
\Q(\rho)&=&\S^{(d)}_d(\rho)=\int_0^\infty\Pa{\op{s}_d(\bsy)-\frac1{1+t}}\dif t=\int_0^\infty\Pa{1-\ee_d(\bsy)-\frac1{1+t}}\dif t\\
&=&  \int_0^\infty\Pa{\frac t{1+t}-\prod^d_{i=1}\frac t{\lambda_i(\rho)+t}}\dif t =  \int_0^\infty\Pa{\frac t{1+t}- \frac1{\det(\I+t^{-1}\rho)}}\dif t,
\end{eqnarray*}
thus
\begin{eqnarray}\label{eq:intrep-of-Q}
\Q(\rho)=  \int_0^\infty\Pa{\frac t{1+t}- \frac1{\det(\I+t^{-1}\rho)}}\dif t,
\end{eqnarray}
leading to the following representation of $\Delta_{\Q}(\rho')$:
\begin{eqnarray*}
\Delta_{\Q}(\rho')&=&  \int_0^\infty\Pa{\frac t{1+t}- \frac1{\det(\I+t^{-1}\rho'_A)}- \frac1{\det(\I+t^{-1}\rho'_B)}+ \frac1{\det(\I+t^{-1}\rho'_{AB})}}\dif t\\
&=&  \int_0^\infty\dif t\Pa{\frac t{1+t}- \frac{t^2}{A_1}- \frac{t^2}{B_1}+ \frac{t^4}{C_1}},
\end{eqnarray*}
where $a'$ and $b'$ are the minimal eigenvalues of $\rho'_A$ and $\rho'_B$, respectively, 
\begin{eqnarray*}
\begin{cases}
A_1=\det(t\I+\rho'_A)=t^2+t+a'(1-a'),\\
B_1=\det(t\I+\rho'_B)=t^2+t+b'(1-b'),\\
C_1=\det(t\I+\rho'_{AB})=(t+x)(t+y)(t+z)(t+w).
\end{cases}
\end{eqnarray*}
For the pointwise comparison, relabel rows and columns so that the
smaller marginal masses occur first; if necessary, exchange the two
subsystems. These operations preserve the deficit and the product of
the two smaller marginal masses. Thus we may assume
\begin{eqnarray*}
0\leqslant a'=x+y\leqslant b'=x+z\leqslant \frac12.
\end{eqnarray*}
Next, we consider this problem in a more general framework. We do not require the normalization of $\rho'$. Set $\kappa:=x+y+z+w=\Tr{\rho'}>0$ and
\begin{eqnarray*}
\begin{cases}
A_\kappa:=\det(t\I+\rho'_A)=t^2+\kappa t+a'(\kappa-a'),\\
B_\kappa:=\det(t\I+\rho'_B)=t^2+\kappa t+b'(\kappa-b'),\\
C_\kappa:=\det(t\I+\rho'_{AB})=(t+x)(t+y)(t+z)(t+w).
\end{cases}
\end{eqnarray*}
By \eqref{appeq:unnormalizedQ}, we also define
\begin{eqnarray}\label{eq:sharp-kernels}
\begin{cases}
 F_{\rho',\kappa}(t)=\frac t{t+\kappa}-\frac{t^2}{A_\kappa}-\frac{t^2}{B_\kappa}+\frac{t^4}{C_\kappa},\\
 f_{\bu,\kappa}(t)=\frac t{t+\kappa}-\frac{2t^2}{(t+\tfrac\kappa2)^2}
 +\frac{t^4}{(t+\frac\kappa4)^4}.
\end{cases}
\end{eqnarray}
Section~\ref{subsubsect:4.5.4} proves that, for every $t>0$, 
\begin{eqnarray}\label{eq:point-sharp}
F_{\rho',\kappa}(t)\geqslant\frac{4a' b' }{\kappa^2}f_{\bu,\kappa}(t).
\end{eqnarray}
It also proves strictness for $a' b' >0$ unless $\rho'$ is completely mixed, i.e., $\rho'=\frac\kappa4\I_4$.

For $\kappa=1$, the $k=d$ case of \eqref{eq:integrationrep-single} gives
$\Delta_{\Q}(\rho')=\int_0^\infty F_{\rho',1}(t)\dif t$. All terms are bounded near zero,
and their constant and $t^{-1}$ terms cancel at infinity, so the integral
converges, including zero entries. From \eqref{eq:extval},
$$
 \int_0^\infty f_{\bu,1}(t)\dif t  =2\Pa{\log2-\frac12} - \Pa{\log4-\frac{13}{12}} = \frac1{12}.
$$
Integration of \eqref{eq:point-sharp} gives 
\begin{eqnarray*}
\Delta_{\Q}(\rho') = \int_0^\infty F_{\rho',1}(t)\dif t \geqslant (4a'b') \int_0^\infty f_{\bu,1}(t)\dif t = \frac{a'b'}3.
\end{eqnarray*}
Since $\rho_A=\rho'_A$ and $\rho_B=\rho'_B$, it follows that 
$$
a=\lambda_{\min}(\rho_A)=\lambda_{\min}(\rho'_A)=a',\qquad b=\lambda_{\min}(\rho_B)=\lambda_{\min}(\rho'_B)=b'.
$$
Therefore
\begin{eqnarray*}
\Delta_{\Q}(\rho_{AB})\geqslant\Delta_{\Q}(\rho'_{AB}) \geqslant \frac{a'b'}3=\frac{ab}3.
\end{eqnarray*}
The arbitrary-mass kernels were used only to homogenize this pointwise
comparison; the entropy deficit is evaluated on normalized states.
At $\frac14\I_4$ the two sides equal $\frac1{12}$, proving optimality.

The certificate proves strictness for every nonuniform table with $ab>0$.
If the pinched table is uniform, equality for the original state requires
$\Q(\rho_{AB})=\Q(\frac14\I_4)$. The Gaussian formula in
Section~\ref{subsubsect:4.5.6} shows that $\Q$ is strictly concave as a function
of its spectrum: $-\eta(u)$ is strictly convex, and a nonzero linear
combination of independent exponential variables is not identically zero.
Symmetry then makes the uniform spectrum its unique maximum.
Thus $\rho_{AB}=\frac14\I_4$. The cases $ab=0$ give equality by the product-support
argument in Theorem~\ref{th:equality} and zero-padding invariance of $\Q$.
\end{proof}

\subsubsection{Proof of Corollary~\ref{cor:twoqubit-allk}}\label{subsubsect:4.5.2}

\begin{proof}
Pinch in a product of marginal eigenbases, which preserves the marginals
and increases the joint $\S_k^{(4)}$. Section~\ref{subsubsect:4.5.5}
proves a sharp pointwise comparison for each resulting classical-table
kernel. For $k=3$ it is the positive multiplier $\frac{4t+1}{3t}$ of
\eqref{eq:point-sharp}; for $k=2$ it follows from a new integer-polynomial
certificate on both complete parameter regions. The uniform kernel
integral is $1/6$ at either index, giving $4ab/6$.
The comparison is strict for $ab>0$ unless the table is uniform. In that
case equality for the original state requires it to maximize $\S_k^{(4)}$,
whose unique maximizer is $\I_4/4$ by Proposition~\ref{prop:extval}.
Pure-marginal states have zero deficit by Theorem~\ref{th:equality}.
Finally,
\[
 \S_k^{(4)}(\I_2/2\oplus0_2)=\log2-\frac{k-1}{6},\qquad
 \S_k^{(4)}(\I_4/4)=\log4+1-H_k,
\]
so the deficit at $\I_4/4$ is $1/6$ for $k=2,3$ and $ab=1/4$.
This proves optimality.
\end{proof}

\subsubsection{Proof of Corollary~\ref{cor:twoqubit-padded-sharp}}\label{subsubsect:4.5.3}

\begin{proof}
Apply \eqref{eq:padding-normalized} with $l=4$ to the joint spectrum and to each marginal spectrum already padded to dimension four:
$$
 \widetilde\Delta_k^{(D)} =\sum_{j=1}^4w_j\widetilde\Delta_j^{(4)},\qquad w_j=\frac{\binom{D-4}{k-j}\binom3{j-1}}{\binom{D-1}{k-1}}.
$$
The weights are nonnegative and sum to one. The lower coefficients are
$0$, $2/3$, $2/3$, and $1/3$ for $j=1,2,3,4$, respectively, by ordinary subadditivity,
Corollary~\ref{cor:twoqubit-allk} and Theorem~\ref{th:lowerboundfor2-qubit}.
Pascal's identity combines the $j=2,3$ terms and gives
\eqref{eq:twoqubit-all-D}. At least one $w_j$ with $j\geqslant2$ is positive
when $k\geqslant2$, so the strict equality characterization passes to the sum.
All four bounds are simultaneously attained at $\I_4/4$, proving
optimality. To make the coefficient explicit, Pascal's identity gives
$$
 \frac{2w_2}3+\frac{2w_3}3+\frac{w_4}3 =\frac{2\binom{D-3}{k-2}+\frac13\binom{D-4}{k-4}} {\binom{D-1}{k-1}}=c_{D,k}.
$$
Cancelling factorials in the two summands gives
$$
c_{D,k}=\frac{2(k-1)(D-k)}{(D-1)(D-2)} +\frac{(k-1)(k-2)(k-3)}{3(D-1)(D-2)(D-3)}.
$$
This identity includes the endpoint indices by the zero-binomial
convention. For fixed $k$, the first term is $\frac{2(k-1)}D+O(D^{-2})$
and the second is $O(D^{-3})$, proving the stated asymptotic.
\end{proof}

\subsubsection{The exact subentropy kernel comparison}\label{subsubsect:4.5.4}

We verify \eqref{eq:point-sharp} over the whole region
$0\leqslant a'=x+y\leqslant b'=x+z\leqslant \frac\kappa2$, where $\kappa=x+y+z+w>0$.
Define
\begin{eqnarray*}
 N_{\kappa}&=&A_{\kappa}B_{\kappa}C_{\kappa} - t(t+\kappa)(A_{\kappa}+B_{\kappa})C_{\kappa}+t^3(t+\kappa)A_{\kappa}B_{\kappa},\\
 U_\kappa &=&(2t+\kappa)^2,\qquad V_\kappa=(4t+\kappa)^4,\\
 E_\kappa&=&128t^4+64\kappa t^3+28\kappa^2t^2+12\kappa^3t+\kappa^4,\\
 \cP_\kappa&=&N_\kappa U_\kappa V_\kappa - (4a'b')E_\kappa A_\kappa B_\kappa C_\kappa.
\end{eqnarray*}
Note that
\begin{eqnarray}\label{eq:certificate-identity}
F_{\rho',\kappa}(t) - \frac{4a'b'}{\kappa^2} f_{\bu,\kappa}(t) = \frac{t\cP_\kappa}{(t+\kappa)A_\kappa B_\kappa C_\kappa U_\kappa V_\kappa}.
\end{eqnarray}
For the completely mixed state $\rho'_{\bu}$, where $a'=b'=\frac{\kappa}2, A_\kappa=B_\kappa=\frac{U_\kappa}4$, and $C_\kappa=\frac{V_\kappa}{256}$, the polynomial $\cP_\kappa$ evaluates to
$$
\cP_\kappa(t) = \frac{U^2_\kappa V_\kappa}{4096}\Pa{U_\kappa V_\kappa - 8t(t+\kappa)V_\kappa + 256t^3(t+\kappa)U_\kappa - \kappa^2 E_\kappa}=0,
$$
which vanishes identically, as confirmed by the algebraic identity
$ U_\kappa V_\kappa - 8t(t+\kappa)V_\kappa + 256t^3(t+\kappa)U_\kappa=\kappa^2E_\kappa$. All denominator factors in \eqref{eq:certificate-identity} are positive
for $t>0$.

Put 
\begin{eqnarray*}
\begin{cases}
q:=z-y=b' - a'\geqslant0,\\
r:=w-x-q=\kappa - 2b'\geqslant0,\\
v:=\min(x,y)\geqslant0,\\
h:=|x-y|\geqslant0.
\end{cases}
\end{eqnarray*}
The parameter region is covered exactly by the following two substitutions:
\begin{center}
\begin{tabular}{c|cccc}
region & $x$ & $y$ & $z$ & $w$\\ \hline
$x\geqslant y$ & $v+h$ & $v$ & $v+q$ & $v+h+q+r$\\
$x\leqslant y$ & $v$ & $v+h$ & $v+h+q$ & $v+q+r$
\end{tabular}
\end{center}
Conversely each substitution with $v,h,q,r\geqslant0$ satisfies the required
ordering. The overlap $h=0$ is harmless. After either substitution,
$\cP_\kappa$ is a homogeneous integer polynomial of total degree $14$,
of degree $12$ in $t$, and every nonzero coefficient is positive:
\begin{eqnarray*}
\cP_\kappa(t) = \zeta_{12}(q,r,v,h)t^{12} + \cdots+\zeta_1(q,r,v,h)t+\zeta_0(q,r,v,h),
\end{eqnarray*}
where $\deg[ \zeta_i(q,r,v,h)]=14-i$ for $i=0,1,\ldots,12$.
Here are the exact coefficient counts:
\begin{center}
\begin{tabular}{c|rr}
region & positive coefficients & negative coefficients\\ \hline
$x\geqslant y$ & $2657$ & $0$\\
$x\leqslant y$ & $2664$ & $0$
\end{tabular}
\end{center}
Appendix~\ref{app:D} gives a complete calculation over $\bbZ$ that
constructs these coefficients and checks the counts, degrees and leading
coefficients. The accompanying sparse coefficient file records the same
expansions. No floating-point arithmetic is used.

To make the implication from the finite coefficient calculation explicit,
write $\cP_\kappa(t)=\sum_{j=0}^{12}\zeta_jt^j$ after either chart
substitution. Each $\zeta_j$ is a polynomial in $v,h,q,r$ with
nonnegative coefficients, so $\zeta_j\geqslant0$ for all
$v,h,q,r\geqslant0$. Equivalently,
\begin{eqnarray*}
\cP_\kappa(t)
=\zeta_0+t\Pa{\zeta_1+t\Pa{\zeta_2+\cdots+t\Pa{\zeta_{11}+t\zeta_{12}}\cdots}}.
\end{eqnarray*}
Successive addition and multiplication of nonnegative real numbers prove
nonnegativity for every $t\geqslant0$. Together with the exhaustive charts
and the positive denominators, this proves the pointwise comparison for
$\cP_\kappa$ and hence
\eqref{eq:point-sharp}. For the equality analysis, the coefficients of
$t^{12}$ in the two regions are respectively
\begin{eqnarray}\label{eq:leading-certificate}
\zeta_{12}(q,r,v,h)=\begin{cases}
 512\Pa{h^2+hq+2hr+2v(q+r)},&x\geqslant y;\\
 512\Pa{h^2+hq+2v(q+r)},&x\leqslant y.
\end{cases}
\end{eqnarray}
If $a'b'>0$, then $a'=2v+h>0$.
A vanishing expression in \eqref{eq:leading-certificate} forces $h=0$,
then $v>0$ and $q=r=0$. This is exactly the completely mixed state.
Every other pinched state $\rho'$ with $a'b'>0$ has a positive $t^{12}$ coefficient;
all remaining coefficients are nonnegative, so $\cP_\kappa(t)>0$ for every
$t>0$. This proves the strictness needed in Theorem~\ref{th:lowerboundfor2-qubit}.

\subsubsection{Sharp kernels for the two intermediate indices}\label{subsubsect:4.5.5}

We prove Corollary~\ref{cor:twoqubit-allk}, keeping all spectra in
dimension four. The variables $\kappa,a',b',A_\kappa,B_\kappa,C_\kappa$ for the unnormalized state $\rho'$ and the ordered region are those above. 
The resolvent formula \eqref{eq:integrationrep-single} gives
\begin{eqnarray*}
\widetilde \Delta^{(4)}_k(\rho') &=& \S^{(4)}_k(\rho'_A) + \S^{(4)}_k(\rho'_B) - \S^{(4)}_k(\rho'_{AB})\\
&=&\kappa\Pa{\S^{(4)}_k(\rho'_A/\kappa) + \S^{(4)}_k(\rho'_B/\kappa) - \S^{(4)}_k(\rho'_{AB}/\kappa)} +\eta(\kappa) \\
&=&\kappa \widetilde \Delta^{(4)}_k(\rho'/\kappa)+\eta(\kappa).
\end{eqnarray*}
Note that $\S^{(4)}_k(\diag(\kappa,0,0,0))=\kappa \S^{(4)}_k(\diag(1,0,0,0)) + \eta(\kappa)=\eta(\kappa)$. That is, 
\begin{eqnarray*}
\kappa \widetilde \Delta^{(4)}_k(\rho'/\kappa) &=& \S^{(4)}_k(\rho'_A) + \S^{(4)}_k(\rho'_B) - \S^{(4)}_k(\rho'_{AB}) -  \S^{(4)}_k(\diag(\kappa,0,0,0))\\
&=& \frac 4k \binom{4}{k}^{-1} \int^\infty_0\dif t \Big[\ee_k\Pa{\tfrac{t}{t+x},\tfrac{t}{t+y},\tfrac{t}{t+z},\tfrac{t}{t+w}} + \ee_k\Pa{\tfrac{t}{t+\kappa},1,1,1} \\
&&~~~~~~~~~- \ee_k\Pa{\tfrac{t}{t+a'},\tfrac{t}{t+(\kappa-a')},1,1} - \ee_k\Pa{\tfrac{t}{t+b'},\tfrac{t}{t+\kappa-b'},1,1}\Big]\\
&=& \int^\infty_0K_k(\rho';t)\dif t 
\end{eqnarray*}
where
\begin{align}\label{eq:intermediate-kernel}
 K_k(\rho';t)=\binom3{k-1}^{-1}& \Big[\ee_k\Pa{\tfrac{t}{t+x},\tfrac{t}{t+y},\tfrac{t}{t+z},\tfrac{t}{t+w}} + \ee_k\Pa{\tfrac{t}{t+\kappa},1,1,1}\notag\\
 &- \ee_k\Pa{\tfrac{t}{t+a'},\tfrac{t}{t+(\kappa-a')},1,1} - \ee_k\Pa{\tfrac{t}{t+b'},\tfrac{t}{t+\kappa-b'},1,1}\Big].
\end{align}
From the above discussion, we see that
\begin{equation}\label{eq:intermediate-kernel-integral}
 \int_0^\infty K_k(\rho';t)\dif t  =\kappa\widetilde\Delta_k^{(4)}(\rho'/\kappa).
\end{equation}
For clarity about the arbitrary-mass normalization, the four-term
integral is the entropy of each marginal vector, minus the entropy of
$\rho'$ and the entropy of $(\kappa,0,0,0)$.
The scaling rule $\S_k(\kappa \rho)=\kappa \S_k(\rho)+\eta(\kappa)$ on normalized $\rho$ shows
that the last term cancels the residual trace correction. For $\kappa=1$
its entropy is zero. Equivalently, substitute $t=\kappa s$ directly in the
kernel. Each $y_i$ lies in $[0,1]$, so the kernel is bounded near zero.
Its constant and $t^{-1}$ terms cancel at infinity, leaving $O(t^{-2})$,
uniformly on fixed-mass compact sets. Thus zero entries are included.
\begin{itemize}
\item \textbf{For $k=3$:} the elementary symmetric sums are
\begin{eqnarray*}
\ee_3\Pa{\tfrac{t}{t+x},\tfrac{t}{t+y},\tfrac{t}{t+z},\tfrac{t}{t+w}} &=&\frac{t^3(4t+\kappa)}{C_\kappa},\qquad \ee_3\Pa{\tfrac{t}{t+\kappa},1,1,1}=\frac{4t+\kappa}{t+\kappa},\\
\ee_3\Pa{\tfrac{t}{t+a'},\tfrac{t}{t+(\kappa-a')},1,1}&=&\frac{t(4t+\kappa)}{A_\kappa} ,\qquad \ee_3\Pa{\tfrac{t}{t+b'},\tfrac{t}{t+\kappa-b'},1,1}=\frac{t(4t+\kappa)}{B_\kappa}.
\end{eqnarray*}
Hence
\begin{eqnarray}\label{eq:k3-Q-kernel}
 K_3(\rho';t)=\frac{4t+\kappa}{3t}F_{\rho',\kappa}(t).
\end{eqnarray}
The same identity holds for the uniform state $\bu_\kappa=\frac{\kappa}4\I_4$.
Its positive multiplier is independent of $\rho'$, so
\eqref{eq:point-sharp} and its strictness give the required sharp
comparison for $K_3$. That is,
\begin{eqnarray*}
 K_3(\rho';t)=\frac{4t+\kappa}{3t}F_{\rho',\kappa}(t)\geqslant \frac{4t+\kappa}{3t}\Pa{ \frac{4a'b'}{\kappa^2}f_{\bu,\kappa}(t)} =  \frac{4a'b'}{\kappa^2} K_3(\bu_\kappa;t).
\end{eqnarray*}

\item \textbf{For $k=2$:} let $\ee_2(\rho')=xy+xz+xw+yz+yw+zw$. Expansion gives
\begin{eqnarray*}
\ee_2\Pa{\tfrac{t}{t+x},\tfrac{t}{t+y},\tfrac{t}{t+z},\tfrac{t}{t+w}} &=&\frac{t^2(6t^2+3\kappa t+\ee_2(\rho'))}{C_\kappa},\qquad \ee_2\Pa{\tfrac{t}{t+\kappa},1,1,1}=\frac{3 t}{\kappa +t}+3,\\
\ee_2\Pa{\tfrac{t}{t+a'},\tfrac{t}{t+(\kappa-a')},1,1}&=&1+\frac{t(5t+2\kappa )}{A_\kappa} ,\qquad \ee_2\Pa{\tfrac{t}{t+b'},\tfrac{t}{t+\kappa-b'},1,1}=1+\frac{t(5t+2\kappa )}{B_\kappa}.
\end{eqnarray*}
Define
\begin{eqnarray*}
 N^{(2)}_\kappa&=&t^2(6t^2+3\kappa t+\ee_2(\rho'))(t+\kappa )A_\kappa B_\kappa \\
       &&- t(5t+2\kappa)(t+\kappa)(A_\kappa+B_\kappa)C_\kappa+(4t+\kappa)A_\kappa B_\kappa C_\kappa,\\
 V^{(2)}_\kappa&=&(4t+\kappa)^2,\qquad E^{(2)}_\kappa=\kappa^3+12\kappa t^2+40t^3.
\end{eqnarray*}
Then
\begin{eqnarray*}
 K_2(\rho';t)=\frac{N^{(2)}_\kappa}{3(t+\kappa)A_\kappa B_\kappa C_\kappa},\qquad
 K_2(\bu_\kappa;t)=\frac{\kappa^2E^{(2)}_\kappa}{3(t+\kappa)U_\kappa V^{(2)}_\kappa}.
\end{eqnarray*}
It follows exactly that
\begin{eqnarray}\label{eq:k2-certificate}
 K_2(\rho';t)-\frac{4a'b'}{\kappa^2}K_2(\bu_\kappa;t) &=& \frac{\cP^{(2)}_\kappa(t)}{3(t+\kappa)A_\kappa B_\kappa C_\kappa U_\kappa V^{(2)}_\kappa }\geqslant0,\\
 \cP^{(2)}_\kappa(t)&=&N^{(2)}_\kappa U_\kappa V^{(2)}_\kappa  - (4a'b')E^{(2)}_\kappa A_\kappa B_\kappa C_\kappa.\notag
\end{eqnarray}
All denominator factors are positive for $t>0$. Use the same two
exhaustive charts as above. In each, $\cP^{(2)}_\kappa$ is a homogeneous
integer polynomial of total degree $13$ and degree $11$ in $t$.
The first chart has $2028$ positive coefficients and the second $2034$;
neither has a negative coefficient. The calculation in
Appendix~\ref{app:D} verifies these assertions over $\bbZ$, together
with the degrees and leading coefficients.

To make the implication from the finite coefficient calculation explicit,
write $\cP^{(2)}_\kappa(t)=\sum_{j=0}^{11}c_jt^j$ after either chart
substitution. Each $c_j$ is a polynomial in $v,h,q,r$ with nonnegative
integer coefficients, so $c_j\geqslant0$ for all
$v,h,q,r\geqslant0$. Equivalently,
\begin{eqnarray*}
\cP^{(2)}_\kappa(t)
=c_0+t\Pa{c_1+t\Pa{c_2+\cdots+t\Pa{c_{10}+tc_{11}}\cdots}}.
\end{eqnarray*}
Successive addition and multiplication of nonnegative real numbers prove
nonnegativity for every $t\geqslant0$. Together with the exhaustive charts
and the positive denominators, this proves the pointwise comparison for
$K_2$.
Its $t^{11}$ coefficients are
\begin{eqnarray*}
c_{11}(q,r,v,h)=\begin{cases}
 160(h^2+hq+2hr+2v(q+r)),&x\geqslant y;\\
 160(h^2+hq+2v(q+r)),&x\leqslant y.
\end{cases}
\end{eqnarray*}
As in \eqref{eq:leading-certificate}, when $a'b'>0$ these vanish
only at the uniform table. Thus every nonuniform table with positive
marginal minima gives a strict comparison for every $t>0$.
\end{itemize}
For either $k=2$ or $k=3$, the rational kernels are continuous for
$t>0$ and absolutely integrable by the bounds above. If $a'b'>0$ and
$\rho'\neq\bu_\kappa$, their difference
$K_k(\rho';t)-\frac{4a'b'}{\kappa^2}K_k(\bu_\kappa;t)$ is strictly
positive for every $t>0$. Continuity gives a positive lower bound on
some compact interval in $(0,\infty)$, so the integral of this difference
is strictly positive. This supplies the strictness used in the quantum
equality characterization.
Since 
$$
K_k(\rho';t)\geqslant \frac{4a'b'}{\kappa^2}K_k(\bu_\kappa;t),\qquad k\in\set{2,3}
$$
it follows that
\begin{eqnarray*}
\kappa\widetilde \Delta^{(4)}_k(\rho'/\kappa)=\int^\infty_0K_k(\rho';t)\dif t\geqslant \frac{4a'b'}{\kappa^2}\int^\infty_0K_k(\bsu_\kappa ;t)\dif t.
\end{eqnarray*}
Now for $\kappa=1$, the uniform integrals of both $K_2$ and $K_3$ are given by the uniform spectral formulas:
\begin{eqnarray*}
\int^\infty_0K_k(\bsu_1 ;t)\dif t=H_k-1-\frac{k-1}3=\frac16.
\end{eqnarray*}
Therefore we get that $\widetilde \Delta^{(4)}_k(\rho)\geqslant \frac23ab$ for $k\in\set{2,3}$. This proves \eqref{eq:sharp-intermediate}, with the strictness used in its quantum equality characterization.

\subsubsection{Proof of Theorem~\ref{th:purity}}\label{subsubsect:4.5.6}

\begin{proof}
The transfer inequality \eqref{eq:transfer-main}, applied with the common
ambient dimension $D$ equal to $d=mn$, i.e.,
\begin{eqnarray*}
 \widetilde\Delta_k^{(D)}(\rho_{AB})
 \geqslant\frac{k(k-1)}{D(D-1)}\Delta_{\Q}(\rho_{AB}),
 \qquad 2\leqslant k\leqslant D,
\end{eqnarray*}
gives the first assertion of
\eqref{eq:purity-main}. We prove the second assertion by a separate
argument on the pinched states and then return to general quantum states.
We now supply the full proof of the subentropy estimate used above:
\begin{eqnarray}\label{eq:native-Q-purity}
 \Delta_{\Q}(\rho_{AB})\geqslant
 \frac{(1-\Tr{\rho_A^2})(1-\Tr{\rho_B^2})}
 {8d(d+1)(d+2)(d+3)},\qquad d=mn.
\end{eqnarray}
This derivation does not assume bipartite subadditivity or its equality
characterization. Consequently it can be used in the strictness argument
of Section~\ref{subsubsect:4.2.4} without a circular dependency.
If one local dimension is one, the state is a product with a pure
one-dimensional factor. Padding invariance of $\Q$ then makes both
sides vanish.

For $m,n\geqslant2$, we first work with a pinched state
$$
\rho'=\sum^{m,n}_{i,j=1}\Pi^A_i\ot\Pi^B_j\rho_{AB}\Pi^A_i\ot\Pi^B_j=\sum^{m,n}_{i,j=1}p_{ij}\out{ij}{ij},\qquad p_{ij}=\Innerm{ij}{\rho}{ij},\sum^{m,n}_{i,j=1}p_{ij}=1,
$$
with row and column sums $r_i=\sum^n_{j=1}p_{ij},c_j=\sum^m_{i=1}p_{ij}$ such that $\rho'_A=\diag(\bsr)$ and $\rho'_B=\diag(\bsc)$.
Write $d=mn$. The symbols $\bsr=(r_1,\ldots,r_m)$ and $\bsc=(c_1,\ldots,c_n)$ denote the marginal probability vectors, and $\bsp=(p_{ij})$ denotes the probability vector of all entries $p_{ij}$.

\paragraph{A Gaussian identity and a classical mutual information.}

Throughout this argument, a standard complex Gaussian $g$ means a random
variable with density $\pi^{-1}e^{-|z|^2}$ on $\bbC$. Consequently,
$|g|^2$ is exponential with mean one,
$\bbE\Br{|g|^2}=1$, and $\bbE\Br{|g|^4}=2$; its argument is uniform on
$[0,2\pi)$ and independent of its modulus. Let
$g_1,\ldots,g_d$ be independent standard complex Gaussians, equivalently
$g_i\sim\bbC\cN(0,1)$ with
$\Re(g_i),\Im(g_i)\sim\cN(0,\frac12)$ (i.i.d.).
For a probability vector $\bdlambda$, we claim
\begin{equation}\label{eq:gaussian-Q}
 \Q(\bdlambda)=1-\gamma + \bbE\Br{\eta\Pa{\sum^d_{i=1}\lambda_i|g_i|^2}},
\end{equation}
where $\gamma$ is Euler's constant and $\eta(x)=-x\log x$. To derive it from
\eqref{eq:kentropy-intrep}, put $R=\sum^d_{i=1}|g_i|^2$ and $\bsu=\bsg/\sqrt R=(g_1,\ldots,g_d)/\sqrt R$.
Then $R\sim\op{Gamma}(d,1)$ is independent of the Haar unit
vector $\bsu$. Thus
one may see this independence directly in polar coordinates. Write
$g_i=\sqrt{x_i}\exp(\mathrm{i}\theta_i)$, with $x_i\geqslant0$ and
$\theta_i\in[0,2\pi)$. The joint law is
\begin{eqnarray*}
 \prod_{i=1}^d e^{-x_i}\dif x_i\,\frac{\dif\theta_i}{2\pi}.
\end{eqnarray*}
With $R=\sum_i x_i$ and $q_i=x_i/R$ for $i<d$, the change of variables
has Jacobian $R^{d-1}$. Hence the radius, the simplex coordinate
$\bsq=(q_1,\ldots,q_{d-1},1-\sum_{i<d}q_i)$, and the phase vector are
independent, with respective laws $\op{Gamma}(d,1)$, the uniform law on
the simplex, and the product uniform phase law. The event $R=0$ has
probability zero. Unitary invariance identifies
$\bsu=\bsg/\sqrt R$ with the Haar unit-vector law used above. Thus the
displayed independence is also valid for
$v=\sum_i\lambda_i|u_i|^2$.
$$
\sum^d_{i=1}\lambda_i|g_i|^2 = R\sum^d_{i=1}\lambda_i|u_i|^2=
Rv,
$$
where $v:=\sum^d_{i=1}\lambda_i|u_i|^2$. Note that $\bbE[|u_1|^2]=\cdots=\bbE[|u_d|^2]$ and $1=|u_1|^2+\cdots+|u_d|^2$. Thus $\bbE[|u_i|^2]=\frac1d$. Then  
\begin{eqnarray*}
\bbE [v]= \sum^d_{i=1}\lambda_i\bbE[|u_i|^2]=\frac1d \sum^d_{i=1}\lambda_i=\frac1d.
\end{eqnarray*}
and, via $\eta(Rv)=\eta(R)v+R\eta(v)$, we see that
\begin{eqnarray*}
 \bbE_{R,v} [\eta(Rv)] = \bbE_R[\eta(R)]\bbE_v[v] + \bbE_R[R]\bbE_v[\eta(v)]  =  d\bbE[\eta(v)] -H_d + \gamma.
\end{eqnarray*}
We used $\bbE [R]=d$ and
$\bbE[\eta(R)]=-d\psi(d+1)=d(\gamma - H_d)$, obtained by differentiating
$\bbE [R^s]=\Gamma(d+s)/\Gamma(d)$ at $s=1$. The $k=d$ compression formula gives
$\Q(\bdlambda)=d\bbE[\eta(v)]-(H_d-1)$, i.e., \eqref{eq:Skrho}, proving
\eqref{eq:gaussian-Q}. The passage to zero entries can be justified with
a bound uniform in the probability vector:
\begin{eqnarray*}
0\leqslant\sum_i\lambda_i|g_i|^2\leqslant R,\qquad
\Abs{\eta\Pa{\sum_i\lambda_i|g_i|^2}}
\leqslant\frac{1+R^2}{e}.
\end{eqnarray*}
The last bound follows from $|\eta(x)|\leqslant(1+x^2)/e$ for
$x\geqslant0$, and is integrable because $R$ has a Gamma law.
Continuity of $\eta$ at zero and dominated convergence therefore include
probability vectors with vanishing entries.

Now take independent standard complex Gaussians $g_{ij}(i\in[m],j\in[n])$, they form an $m\times n$ random matrix $\bsG=(g_{ij})$ and set
$$
t_{ij}=\sqrt{p_{ij}}g_{ij},\qquad s=\sum^{m,n}_{i,j=1}|t_{ij}|^2= \sum^{m,n}_{i,j=1}p_{ij}\abs{g_{ij}}^2.
$$
Since $\sum_{i,j}p_{ij}=1$, there is an entry $p_{ij}>0$.
The corresponding $g_{ij}$ is nonzero almost surely, so $s>0$ almost
surely. On the null event $s=0$, any definition of the normalized density
leaves the expectations unchanged. On the row and column phase tori $\bbT^m\times \bbT^n$, each equipped with
the normalized Haar measure, via $\bdalpha=(\alpha_1,\ldots,\alpha_m)$ and $\bdbeta=(\beta_1,\ldots,\beta_n)$, define
\begin{eqnarray*}
 \Psi(\bdalpha,\bdbeta)=\Abs{\sum_{ij}t_{ij}e^{i\alpha_i+i\beta_j}}^2,   \qquad \widetilde \Psi(\bdalpha,\bdbeta)=\frac1s \Psi(\bdalpha,\bdbeta).
\end{eqnarray*}
Let us show that
$$
\int^{2\pi}_0\cdots\int^{2\pi}_0 \Psi(\bdalpha,\bdbeta)\frac{[\dif \bdalpha]}{(2\pi)^m}\frac{[\dif \bdbeta]}{(2\pi)^n}=s,
$$
so $\widetilde\Psi(\bdalpha,\bdbeta)$ is a probability density. Indeed,
\begin{eqnarray*}
&&\int^{2\pi}_0\cdots\int^{2\pi}_0 \Psi(\bdalpha,\bdbeta)\frac{[\dif \bdalpha]}{(2\pi)^m}\frac{[\dif \bdbeta]}{(2\pi)^n}\\
&&=\int^{2\pi}_0\cdots\int^{2\pi}_0 \Pa{\sum^{m,n}_{i,j=1}t_{ij} e^{\mathrm{i}(\alpha_i+\beta_j)}}\Pa{\sum^{m,n}_{k,l=1}\bar t_{kl} e^{-\mathrm{i}(\alpha_k+\beta_l)}}\frac{[\dif \bdalpha]}{(2\pi)^m}\frac{[\dif \bdbeta]}{(2\pi)^n}\\
&&=\sum^{m,n}_{i,j=1}\sum^{m,n}_{k,l=1}t_{ij}\bar t_{kl} \int^{2\pi}_0\cdots\int^{2\pi}_0e^{\mathrm{i}(\alpha_i-\alpha_k+\beta_j-\beta_l)}\frac{[\dif \bdalpha]}{(2\pi)^m}\frac{[\dif \bdbeta]}{(2\pi)^n}\\
&& = \sum^{m,n}_{i,j=1}\sum^{m,n}_{k,l=1}t_{ij}\bar t_{kl}\delta_{ik}\delta_{jl} = \sum^{m,n}_{i,j=1}\abs{t_{ij}}^2=s.
\end{eqnarray*}
Let $\widetilde\Psi_R,\widetilde\Psi_C$ denote its marginal densities and $\Psi_R=s\widetilde\Psi_R$, $\Psi_C=s\widetilde\Psi_C$.
Here the subscripts $R,C$ identify the phase variables: $\Psi_R$ is the marginal on the row-phase torus and $\Psi_C$ is the marginal on the column-phase torus; their Gaussian expectations involve the column-sum vector $\bsc$ and the row-sum vector $\bsr$, respectively.
Here 
$$
\K\L(f||g)=\int f(\log f-\log g),
$$ 
denotes classical relative entropy
with respect to the product of the normalized Haar measures on the two
tori, with the usual value $+\infty$ if required by the support condition.
The variables $\bdalpha$ and $\bdbeta$ in every integral below carry these
normalized measures, $\frac{[\dif\bdalpha]}{(2\pi)^m}$ and $\frac{[\dif\bdbeta]}{(2\pi)^n}$. We claim
\begin{eqnarray}\label{eq:gaussian-MI}
 \Delta_{\Q}(\rho')=\Q(\bsr)+\Q(\bsc)-\Q(\bsp)
 =\bbE_g\Br{s\K\L(\widetilde \Psi||\widetilde\Psi_R\widetilde\Psi_C)}.
\end{eqnarray}
We first justify the logarithmic expansion, including zero marginal
densities. Fix $\bsg$ with $s>0$. Cauchy--Schwarz gives
$0\leqslant\widetilde\Psi\leqslant d$, so both marginals are
bounded as well. If $\widetilde\Psi_R(\bdalpha)=0$, then
\begin{eqnarray*}
\int^{2\pi}_0\cdots\int^{2\pi}_0
\widetilde\Psi(\bdalpha,\bdbeta)
\frac{[\dif\bdbeta]}{(2\pi)^n}=0.
\end{eqnarray*}
Nonnegativity implies
$\widetilde\Psi(\bdalpha,\bdbeta)=0$ for almost every
$\bdbeta$ on that fiber. The same argument applies to a zero
column marginal. Products with logarithms on these zero fibers are
therefore interpreted as zero. Tonelli's theorem then gives
\begin{eqnarray*}
\int\Abs{\widetilde\Psi\log\widetilde\Psi_R}
&=&\int\widetilde\Psi_R\Abs{\log\widetilde\Psi_R}<\infty,
\end{eqnarray*}
and the analogous statement for $\widetilde\Psi_C$, because
$x\mapsto x|\log x|$ with value zero at $x=0$ is continuous on bounded
intervals. Thus Fubini applies to the weighted logarithms and the
entropy expansion below remains valid when marginal densities vanish.
Indeed, expansion using $\int \Psi=s$ gives
\begin{eqnarray*}
 s\K\L(\widetilde \Psi||\widetilde\Psi_R\widetilde\Psi_C)  &=&\int \eta(\Psi_R) + \int \eta(\Psi_C)-\int \eta(\Psi)- \eta(s)\\
&=& \bbE_{\bdalpha}[\eta(\Psi_R)] +  \bbE_{\bdbeta}[\eta(\Psi_C)] - \bbE_{\bdalpha,\bdbeta}[\eta(\Psi)] - \eta(s),
\end{eqnarray*}
leading to 
\begin{eqnarray*}
 \bbE_{\bsg}\Br{s\K\L(\widetilde \Psi||\widetilde\Psi_R\widetilde\Psi_C)}  &=&\bbE_{\bsg}\Br{\int \eta(\Psi_R) + \int \eta(\Psi_C)-\int \eta(\Psi)- \eta(s)}\\
&=&  \bbE_{\bdalpha}[\bbE_{\bsg}\Br{\eta(\Psi_R)}] +  \bbE_{\bdbeta}[\bbE_{\bsg}\Br{\eta(\Psi_C)}] -  \bbE_{\bdalpha,\bdbeta}[\bbE_{\bsg}\Br{\eta(\Psi)}] - \bbE_{\bsg}[\eta(s)],
\end{eqnarray*}
At fixed phases, the Gaussian linear form defining $\Psi$ has variance one,
so its $\eta$-expectation is $\bbE_{\bsg}[\eta(\Psi)]=\gamma-1$. Also
\begin{eqnarray*}
\Psi_R(\bdalpha)=\sum^n_{j=1}\Abs{\sum^m_{i=1} t_{ij}e^{i\alpha_i}}^2 = \sum^n_{j=1}\Abs{\sum^m_{i=1} \sqrt{p_{ij}}g_{ij}e^{i\alpha_i}}^2=\sum^n_{j=1}\abs{h_j(\bdalpha)}^2,
\end{eqnarray*}
where $h_j(\bdalpha):=\sum^m_{i=1} \sqrt{p_{ij}}g_{ij}e^{i\alpha_i}$. Thus $\Psi_R$ is a sum of independent
exponentials with means $c_j$, and $\Psi_C$ has the analogous distribution
with means $r_i$. Indeed, the $g_{ij}$ are i.i.d. standard complex Gaussians. For fixed phases $\alpha_i$, each $e^{\mathrm{i}\alpha_i}g_{ij}$ is still a standard complex Gaussian. Thus, for fixed $\bdalpha$, 
$$
h_j(\bdalpha)\sim \bbC \cN\Pa{0,\sum^m_{i=1}p_{ij}}=\bbC \cN(0,c_j).
$$
Therefore the law of $\Psi_R(\bdalpha)$ does not depend on $\bdalpha$. Consequently $\bbE_{\bsg}[\eta(\Psi_R(\bdalpha))]$ is constant with respect to $\bdalpha$:
$$
\bbE_{\bdalpha}[\bbE_{\bsg}[\eta(\Psi_R(\bdalpha))]] = \bbE_{\bsg}\Br{\eta\Pa{\sum^n_{j=1}c_j\abs{g^{(1)}_j}^2}}=\Q(\bsc)-(1-\gamma).
$$
Similarly, 
$$
\bbE_{\bdbeta}[\bbE_{\bsg}[\eta(\Psi_C(\bdbeta))]] = \bbE_{\bsg}\Br{\eta\Pa{\sum^m_{i=1}r_i\abs{g^{(2)}_i}^2}}=\Q(\bsr)-(1-\gamma)
$$
and
$$
\bbE_{\bdalpha,\bdbeta}[\bbE_{\bsg}\Br{\eta(\Psi(\bdalpha,\bdbeta))}]  = \gamma-1.
$$
If $F(\bdlambda):=\bbE\Br{\eta\Pa{\sum^d_{i=1}\lambda_i|g_i|^2}}$, then
\begin{eqnarray*}
\bbE_{\bsg}[\eta(s)]=F(\bsp)=\Q(\bsp)-(1-\gamma).
\end{eqnarray*}
The expectation of the displayed expansion is
\begin{eqnarray*}
\bbE_{\bsg}\Br{s\K\L(\widetilde \Psi||\widetilde\Psi_R\widetilde\Psi_C)} &=& F(\bsr) + F(\bsc) - F(\bsp) - (\gamma-1)\\
&=&[\Q(\bsr)-(1-\gamma)] + [\Q(\bsc) -(1-\gamma)]- [\Q(\bsp)-(1-\gamma)] +(1-\gamma)\\
&=& \Q(\bsr) + \Q(\bsc)-\Q(\bsp)=\Delta_{\Q}(\rho')
\end{eqnarray*}
by \eqref{eq:gaussian-Q}, proving \eqref{eq:gaussian-MI}.
For the interchanges, the Cauchy--Schwarz bounds displayed below give
$\Psi\leqslant ds$, $\Psi_R\leqslant ms$, and $\Psi_C\leqslant ns$,
uniformly in the phases. Independence and the moments
$\bbE[|g_{ij}|^2]=1$, $\bbE[|g_{ij}|^4]=2$ imply
\begin{eqnarray*}
\bbE[s^2]
=2\sum_{i,j}p_{ij}^2+
\sum_{(i,j)\neq(k,l)}p_{ij}p_{kl}
=1+\sum_{i,j}p_{ij}^2\leqslant2.
\end{eqnarray*}
Since $m,n\leqslant d$, each entropy integrand in the expansion is
bounded in absolute value by $(1+d^2s^2)/e$. This common bound is
integrable against the Gaussian law and the normalized torus measures,
so Fubini applies term by term. It includes entries $p_{ij}=0$.

\paragraph{A lower bound from four-index Fourier coefficients.}

Cauchy--Schwarz gives $\widetilde\Psi\leqslant d$, $\widetilde\Psi_R\leqslant m$ and $\widetilde\Psi_C\leqslant n$; indeed, 
\begin{eqnarray*}
\Psi(\bdalpha,\bdbeta)&=&\abs{\sum_{i,j}t_{ij}e^{\mathrm{i}(\alpha_i+\beta_j)}}^2\leqslant \Pa{\sum_{i,j}|t_{ij}|}^2\leqslant \Pa{\sum_{i,j}|t_{ij}|^2}\Pa{\sum_{i,j}1}=sd,\\
\Psi_R(\bdalpha)&\leqslant& sm;\qquad \Psi_C(\bdbeta)\leqslant sn,
\end{eqnarray*}
therefore both $\widetilde\Psi$ and $\widetilde\Psi_R\widetilde\Psi_C$ lie in $[0,d]$.
Since $-\eta''(x)\geqslant\frac1d$ on $(0,d]$, the scalar Bregman inequality gives
\begin{equation}\label{eq:MI-quadratic}
 \K\L(\widetilde\Psi||\widetilde\Psi_R\widetilde\Psi_C)\geqslant\frac1{2d}\norm{\widetilde\Psi - \widetilde\Psi_R\widetilde\Psi_C}_2^2.
\end{equation}
For completeness, integrate the scalar version of Bregman divergence inequality
$f(u)-f(v)-f'(v)(u-v)\geqslant \frac1{2d}(u-v)^2$, where $f=-\eta$, with
$u=\widetilde\Psi$, $v=\widetilde\Psi_R\widetilde\Psi_C$. Both marginals integrate to one. Hence Fubini gives
\begin{eqnarray*}
\int\widetilde\Psi_R\widetilde\Psi_C
=\Pa{\int\widetilde\Psi_R}\Pa{\int\widetilde\Psi_C}=1,
\end{eqnarray*}
and the joint density and the product reference have the same total mass and
the same marginals. The other linear term
vanishes because $\log v=\log \widetilde\Psi_R+\log \widetilde\Psi_C$ and $u,v$ have the same
marginals. Zero marginal fibers carry zero joint mass. One may set
$\widetilde\Psi_\varepsilon=(1-\varepsilon)\widetilde\Psi+\varepsilon$ and use the reference density
$\widetilde\Psi_{\varepsilon,R}\widetilde\Psi_{\varepsilon,C}$ formed from its marginals. Apply the
positive-density inequality and take the limit. Since each base measure has
total mass one,
\begin{eqnarray*}
\widetilde\Psi_{\varepsilon,R}
&=&(1-\varepsilon)\widetilde\Psi_R+\varepsilon,\\
\widetilde\Psi_{\varepsilon,C}
&=&(1-\varepsilon)\widetilde\Psi_C+\varepsilon.
\end{eqnarray*}
Thus the regularized densities are positive, have mass one, and satisfy
$\widetilde\Psi_\varepsilon\leqslant d$,
$\widetilde\Psi_{\varepsilon,R}\leqslant m$ and
$\widetilde\Psi_{\varepsilon,C}\leqslant n$. They converge pointwise to
the original densities. The function $\eta$ is continuous and bounded on
$[0,d]$, while the squared difference is bounded by $4d^2$; dominated
convergence therefore gives the desired limit, including zero marginal
fibers.

Note that
\begin{eqnarray*}
&&\widetilde\Psi(\bdalpha,\bdbeta) - \widetilde\Psi_R(\bdalpha)\widetilde\Psi_C(\bdbeta) = \frac1s\abs{\sum_{i,j}t_{ij}e^{\mathrm{i}(\alpha_i+\beta_j)}}^2 -   \frac1{s^2}\Pa{\sum^n_{j=1}\Abs{\sum^m_{i=1} t_{ij}e^{i\alpha_i}}^2 }\Pa{\sum^m_{i=1}\Abs{\sum^n_{j=1} t_{ij}e^{\mathrm i\beta_j}}^2 }\\
&&= \frac1s\sum_{i,j,k,l}t_{ij}\bar t_{kl} e^{\mathrm i(\alpha_i-\alpha_k+\beta_j-\beta_l)} - \frac1{s^2}\Pa{\sum_{i,k}\Pa{\sum_j t_{ij}\bar t_{kj}}e^{\mathrm i(\alpha_i-\alpha_k)}}\Pa{\sum_{j,l}\Pa{\sum_i t_{ij}\bar t_{il}}e^{\mathrm i(\beta_j-\beta_l)}}\\
&&= \frac1s\sum_{i,j,k,l}t_{ij}\bar t_{kl} e^{\mathrm i(\alpha_i-\alpha_k+\beta_j-\beta_l)} - \frac1{s^2}\Pa{\sum_{i,k}\Pa{\sum_b t_{ib}\bar t_{kb}}e^{\mathrm i(\alpha_i-\alpha_k)}}\Pa{\sum_{j,l}\Pa{\sum_a t_{aj}\bar t_{al}}e^{\mathrm i(\beta_j-\beta_l)}}\\
&&=\frac1s\sum_{i,j,k,l}t_{ij}\bar t_{kl} e^{\mathrm i(\alpha_i-\alpha_k+\beta_j-\beta_l)} - \frac1{s^2}\sum_{i,j,k,l}\Pa{\sum_a t_{aj}\bar t_{al}}\Pa{\sum_b t_{ib}\bar t_{kb}}e^{\mathrm i(\alpha_i-\alpha_k+\beta_j-\beta_l)}\\
&&=\sum_{i,j,k,l}\frac1{s^2}\Pa{s t_{ij}\bar t_{kl} - \Pa{\sum_b t_{ib}\bar t_{kb}}\Pa{\sum_a t_{aj}\bar t_{al}}}e^{\mathrm i(\alpha_i-\alpha_k+\beta_j-\beta_l)}=\sum_{i,j,k,l}\frac{C_{ik,jl}}{s^2}e^{\mathrm i(\alpha_i-\alpha_k+\beta_j-\beta_l)},
\end{eqnarray*}
where, for $i\neq k$ and $j\neq l$, the Fourier coefficient of $\widetilde\Psi-\widetilde\Psi_R\widetilde\Psi_C$ at
$\alpha_i-\alpha_k+\beta_j-\beta_l$ is $s^{-2}C_{ik,jl}$, and
\begin{eqnarray}\label{eq:fourier-C}
 C_{ik,jl}:=s t_{ij}\bar t_{kl} - \Pa{\sum_b t_{ib}\bar t_{kb}}\Pa{\sum_a t_{aj}\bar t_{al}}.
\end{eqnarray}
The difference $\widetilde\Psi-\widetilde\Psi_R\widetilde\Psi_C$ has zero
marginal integrals. Hence all Fourier modes depending only on
$\bdalpha$, only on $\bdbeta$, or on neither vanish after terms with the
same frequency have been grouped. The remaining modes have $i\neq k$ and
$j\neq l$. In the frequency lattices $\bbZ^m$ and $\bbZ^n$, the vector
$e_i-e_k$ uniquely determines the ordered pair $(i,k)$, and similarly
$e_j-e_l$ determines $(j,l)$. Thus the remaining frequencies are distinct.
For each rectangle $i<k$, $j<l$, the two orientations and their conjugates
are four distinct modes. Parseval, \eqref{eq:gaussian-MI} and
\eqref{eq:MI-quadratic} imply
\begin{eqnarray*}
 \Delta_{\Q}(\rho') = \bbE_{\bsg}\Br{s\K\L(\widetilde\Psi||\widetilde\Psi_R\widetilde\Psi_C)}
\geqslant \frac1{2d}\bbE_{\bsg}\Br{s\norm{\widetilde\Psi - \widetilde\Psi_R\widetilde\Psi_C}^2_2},
\end{eqnarray*}
where 
\begin{eqnarray*}
\norm{\widetilde\Psi - \widetilde\Psi_R\widetilde\Psi_C}^2_2&=& \int^{2\pi}_0\cdots\int^{2\pi}_0\Abs{\widetilde\Psi(\bdalpha,\bdbeta) - \widetilde\Psi_R(\bdalpha) \widetilde\Psi_C(\bdbeta) }^2\frac{[\dif\bdalpha]}{(2\pi)^m}\frac{[\dif\bdbeta]}{(2\pi)^n}\\
&=&  \int^{2\pi}_0\cdots\int^{2\pi}_0\Abs{\sum_{i,j,k,l}\frac{C_{ik,jl}}{s^2}e^{\mathrm i(\alpha_i-\alpha_k+\beta_j-\beta_l)} }^2\frac{[\dif\bdalpha]}{(2\pi)^m}\frac{[\dif\bdbeta]}{(2\pi)^n}\\
&=&\sum_{\substack{i\ne k\\j\ne l}}\abs{\frac{C_{ik,jl}}{s^2}}^2
 =2 \sum_{i<k,j<l}\Pa{\abs{\frac{C_{ik,jl}}{s^2}}^2
 + \abs{\frac{C_{ik,lj}}{s^2}}^2}.
\end{eqnarray*}
Thus, it holds that
\begin{eqnarray}\label{eq:parseval-C}
 \Delta_{\Q}(\rho') &\geqslant& \frac1{2d}\bbE_{\bsg}\Br{s\norm{\widetilde\Psi - \widetilde\Psi_R\widetilde\Psi_C}^2_2} =\frac1{2d}\bbE_{\bsg}\Br{\sum_{\substack{i\ne k\\j\ne l}}\frac{\abs{C_{ik,jl}}^2}{s^3}} \notag\\ 
&=&\frac1d\sum_{i<k,j<l}
 \bbE_{\bsg}\Br{\frac{|C_{ik,jl}|^2+|C_{ik,lj}|^2}{s^3}}.
\end{eqnarray}
These expectations are finite. Let
$f=\widetilde\Psi-\widetilde\Psi_R\widetilde\Psi_C$.
For $i\ne k$ and $j\ne l$, the coefficient $C_{ik,jl}/s^2$ is the
corresponding Fourier coefficient of $f$.
Since the normalized torus measure has total mass one and both
$\widetilde\Psi$ and $\widetilde\Psi_R\widetilde\Psi_C$ are probability
densities, every such coefficient satisfies
\begin{eqnarray*}
\frac{\abs{C_{ik,jl}}}{s^2}
&\leqslant& \int \abs{f}
 \leqslant \int\left(\widetilde\Psi+\widetilde\Psi_R\widetilde\Psi_C\right)=2.
\end{eqnarray*}
Thus $\abs{C_{ik,jl}}\leqslant 2s^2$ and
$\abs{C_{ik,jl}}^2/s^3\leqslant4s$. Therefore
$$
\bbE\Br{ \frac{|C_{ik,jl}|^2}{s^3}}\leqslant4\bbE [s] = 4\sum_{i,j}p_{ij}\bbE[\abs{g_{ij}}^2]=4\sum_{i,j}p_{ij}=4.
$$
Fix four entries of $\rho'=\diag(p_{ij})$ and write
$(x,y,z,w):=(p_{ij},p_{il},p_{kj},p_{kl})$. The corresponding Gaussian
variables are denoted by
\begin{eqnarray*}
(t_x,t_y,t_z,t_w) &:=& (t_{ij},t_{il},t_{kj},t_{kl}) = \Pa{\sqrt{p_{ij}}g_{ij},\sqrt{p_{il}}g_{il},\sqrt{p_{kj}}g_{kj},\sqrt{p_{kl}}g_{kl}}\\
&=& \Pa{\sqrt{x}g_{ij},\sqrt{y}g_{il},\sqrt{z}g_{kj},\sqrt{w}g_{kl}}.
\end{eqnarray*}
We then rewrite $(C_{ik,jl},C_{ik,lj})$ as $(C_{xw},C_{yz})$. Expand $C_{xw}$ as follows, via 
\begin{eqnarray*}
s &=&\abs{t_x}^2+\abs{t_y}^2+\abs{t_z}^2+\abs{t_w}^2+s',\\
\Pa{\sum_b t_{ib}\bar t_{kb}} &=& t_x \bar t_z + t_y \bar t_w + A',\\
\Pa{\sum_a t_{aj}\bar t_{al}} &=& t_x \bar t_y + t_z \bar t_w + B',
\end{eqnarray*}
Index the Gaussian coordinates by $a$, using $a=x,y,z,w$ for the four
chosen rectangle coordinates, and write $g_a=|g_a|e^{\mathrm i\theta_a}$.
Conditional on all Gaussian absolute values, the phases $\theta_a$ are
independent and uniform on $[0,2\pi)$, so distinct phase characters are
orthogonal. Three full phase sectors of $C_{xw}$ are
\begin{equation}\label{eq:phase-sectors}
 t_x\bar t_w(s-|t_y|^2-|t_z|^2),\qquad
 -t_x^2\bar t_y\bar t_z,\qquad
 -t_yt_z\bar t^2_w.
\end{equation}
Their respective phase characters are
\begin{eqnarray*}
e^{\mathrm i(\theta_x-\theta_w)},\qquad
e^{\mathrm i(2\theta_x-\theta_y-\theta_z)},\qquad
e^{\mathrm i(\theta_y+\theta_z-2\theta_w)}.
\end{eqnarray*}
Conditional on the four selected moduli and on all outside Gaussian
coordinates, these are distinct characters of the four independent phase
circles. The quantities $A',B'$ depend only on the outside coordinates.
The explicit remainder below is
$-[(t_x\bar t_z+t_y\bar t_w)B'+(t_x\bar t_y+t_z\bar t_w)A'+A'B']$,
so its possible characters on the four selected phase circles are
\begin{eqnarray*}
e^{\mathrm i(\theta_x-\theta_z)},\quad
e^{\mathrm i(\theta_y-\theta_w)},\quad
e^{\mathrm i(\theta_x-\theta_y)},\quad
e^{\mathrm i(\theta_z-\theta_w)},\quad 1.
\end{eqnarray*}
None agrees with a character of the three displayed sectors. Hence all
cross terms between these sectors and the remainder vanish after
conditional phase integration.
To check the first sector explicitly, expand the product of sums in
\eqref{eq:fourier-C}. The terms $(a,b)=(i,l)$ and $(k,j)$ subtract
$t_x\overline{t_w}|t_y|^2$ and
$t_x\overline{t_w}|t_z|^2$, respectively. The choices $(i,j)$ and $(k,l)$
give the other two sectors. Any remaining choice contains an outside
coordinate and has a different phase character. More explicitly, in a
general product term $t_{ib}\bar t_{kb}t_{aj}\bar t_{al}$, cancellation
with the first sector can occur only for $(a,b)=(i,l)$ or $(k,j)$; these
are exactly the two terms already displayed. Thus no omitted term can
cancel one of the sectors in \eqref{eq:phase-sectors}, and we get that
\begin{eqnarray*}
C_{xw} &=& C_{ik,jl} = s t_{ij}\bar t_{kl} - \Pa{\sum_b t_{ib}\bar t_{kb}}\Pa{\sum_a t_{aj}\bar t_{al}}\\
&=& \Pa{\abs{t_x}^2+\abs{t_y}^2+\abs{t_z}^2+\abs{t_w}^2+s'} t_x\bar t_w - (t_x \bar t_z + t_y \bar t_w + A')(t_x \bar t_y + t_z \bar t_w + B')\\
&=& \Pa{\abs{t_x}^2+\abs{t_w}^2 + s'} t_x\bar t_w - t_x^2\bar t_y\bar t_z
 -t_yt_z\bar t^2_w  -  (t_x \bar t_z + t_y \bar t_w)B' - (t_x \bar t_y + t_z \bar t_w)A'- A'B'\\
&=& \Pa{s-\abs{t_y}^2-\abs{t_z}^2} t_x\bar t_w - t_x^2\bar t_y\bar t_z
 -t_yt_z\bar t^2_w  -  [(t_x \bar t_z + t_y \bar t_w)B' + (t_x \bar t_y + t_z \bar t_w)A' + A'B']\\
&=& S_1+S_2+S_3+\cR(A',B'),
\end{eqnarray*}
implying that
\begin{eqnarray*}
\bbE[\Abs{C_{xw}}^2] &=& \bbE[\abs{S_1}^2] +  \bbE[\abs{S_2}^2] +  \bbE[\abs{S_3}^2] +\bbE[ \abs{\cR}^2 ]\\ &&+2\Re\Pa{\sum^3_{i=1}\bbE[S_i \overline{\cR(A',B')}]} + 2\Re\Pa{\bbE[S_1\bar S_2]+ \bbE[S_1\bar S_3] + \bbE[S_2\bar S_3]}\\
&=& \bbE[\abs{S_1}^2] +  \bbE[\abs{S_2}^2] +  \bbE[\abs{S_3}^2] +\bbE[ \abs{\cR}^2 ]\\
&\geqslant &\bbE[\abs{S_1}^2] +  \bbE[\abs{S_2}^2] +  \bbE[\abs{S_3}^2] \\
&\geqslant & xw(1-y-z)^2 + 2x^2yz + 2w^2yz.
\end{eqnarray*}
The first sector has squared expectation at least $xw(1-y-z)^2$.
To see this when $xw>0$, weight the independent unit-mean exponentials
$|g_x|^2,|g_w|^2$ by their product, which has expectation one.
Under this probability law they each have mean two; all other means are
unchanged. The nonnegative random variable $s-|t_y|^2-|t_z|^2$ then has
mean $1-y-z+x+w\geqslant1-y-z$. Jensen's inequality proves the assertion.
If $xw=0$, the assertion is immediate. Since
$\bbE[|g|^4]=2$, the second and third sectors contribute
$2x^2yz$ and $2w^2yz$. 

Similarly, 
\begin{eqnarray*}
C_{yz} &=& C_{ik,lj} = s t_{il}\bar t_{kj} - \Pa{\sum_b t_{ib}\bar t_{kb}}\Pa{\sum_a t_{al}\bar t_{aj}}\\
&=& \Pa{\abs{t_x}^2+\abs{t_y}^2+\abs{t_z}^2+\abs{t_w}^2+s'}  t_y\bar t_z - (t_x \bar t_z + t_y \bar t_w + A')(\bar t_x t_y + \bar t_z t_w + \bar B')\\
&=& \Pa{s-\abs{t_x}^2-\abs{t_w}^2} t_y\bar t_z -t^2_y\bar t_x\bar t_w - t_x t_w \bar t_z^2
   -  (t_x \bar t_z + t_y \bar t_w)\bar B' - (\bar t_x  t_y +  \bar t_z t_w)A'- A'\bar B'.
\end{eqnarray*}
Repeating the argument for $C_{yz}$ gives
\begin{eqnarray*}
\bbE[\Abs{C_{yz}}^2] \geqslant  yz(1-x-w)^2 + 2y^2xw + 2z^2xw.
\end{eqnarray*}
Therefore
\begin{eqnarray}\label{eq:rectangle-moment}
 \bbE[|C_{xw}|^2+|C_{yz}|^2]
 &\geqslant& xw((1-y-z)^2+2y^2+2z^2)\notag\\
&&+yz((1-x-w)^2+2x^2+2w^2)\notag\\
 &\geqslant&\frac{xw+yz}{2}.
\end{eqnarray}
The last step is the exact sum-of-squares identity
\begin{eqnarray*}
 (1-u-v)^2+2u^2+2v^2-\frac12
 =2\left(u+v-\frac12\right)^2+(u-v)^2\geqslant0.
\end{eqnarray*}

Let $R=\sum_{ij}|g_{ij}|^2\sim\op{Gamma}(d,1)$.
Since $p_{ij}\leqslant1$, $s\leqslant R$.
The coefficient is homogeneous of degree four in the Gaussian vector:
$$
C_{ik,jl}(\sqrt R\,\bsu)=R^2C_{ik,jl}(\bsu).
$$
Thus, with
$$
J(\bsu):=\abs{C_{ik,jl}(\bsu)}^2,
$$
we have $\abs{C_{ik,jl}(\bsg)}^2=R^4J(\bsu)$.
The function $J$ is continuous on the unit sphere and hence bounded there;
the Gaussian radius $R$ and the direction $\bsu=\bsg/\sqrt R$ are independent.
Consequently,
$$
\bbE[|C_{ik,jl}|^2]
=\bbE[R^4J(\bsu)]
=\bbE[R^4]\bbE[J(\bsu)],
\qquad\text{and hence}\qquad
\bbE[J(\bsu)]
=\frac{\bbE[|C_{ik,jl}|^2]}{\bbE[R^4]}.
$$
and
\begin{eqnarray}\label{eq:radial-C}
 \bbE\Br{\frac{|C_{ik,jl}|^2}{s^3}}
& \geqslant & \bbE\Br{\frac{|C_{ik,jl}|^2}{R^3}} = \bbE[RJ(\bsu)]=\bbE[R]\bbE[J(\bsu)]\notag\\
&=&\bbE[R] \frac{\bbE[|C_{ik,jl}|^2]}{\bbE[R^4]} =\frac{\bbE\Br{|C_{ik,jl}|^2}}{(d+1)(d+2)(d+3)}.
\end{eqnarray}
The factor $d$ cancels explicitly as
\[
\frac{\bbE[R]}{\bbE[R^4]}=
\frac{d}{d(d+1)(d+2)(d+3)}
=\frac1{(d+1)(d+2)(d+3)}.
\]
Here $\bbE[R^k]=d(d+1)\cdots(d+k-1)$ for $k\geqslant1$, in particular
$\bbE[R^4]=d(d+1)(d+2)(d+3)$. Equations \eqref{eq:parseval-C}--\eqref{eq:radial-C} imply
\begin{eqnarray}\label{eq:Q-K-bound}
 \Delta_{\Q}(\rho')&\geqslant& \frac{ \sum_{i<k,j<l} \bbE[|C_{ik,jl}|^2+|C_{ik,lj}|^2]}{d(d+1)(d+2)(d+3)}\notag\\
&\geqslant & \frac1{d(d+1)(d+2)(d+3)}\sum_{i<k,j<l} \frac{p_{ij}p_{kl}+p_{il}p_{kj}}2\notag\\
&=& \frac{\sum_{i<k,j<l} p_{ij}p_{kl}+p_{il}p_{kj}}{2d(d+1)(d+2)(d+3)}= \frac{K(\rho')}{2d(d+1)(d+2)(d+3)},
\end{eqnarray}
where 
\begin{eqnarray}\label{eq:K-table}
 K(\rho') :=\sum_{i<k,j<l}(p_{ij}p_{kl}+p_{il}p_{kj}) =\sum_{i<k}\sum_{j\ne l}p_{ij}p_{kl}.
\end{eqnarray}

\paragraph{The four-sample inequality and return to quantum states.}

We now show
\begin{equation}\label{eq:K-purity}
 K(\rho')\geqslant\frac{(1-\sum_i r_i^2)(1-\sum_j c_j^2)}4.
\end{equation}
Take four independent samples $(I_a,J_a)$, $a=1,2,3,4$, each with distribution
\[
\Pr[(I_a,J_a)=(i,j)]=p_{ij}.
\]
Let $R_0,C_0$ be the sets of edges
on their four positions whose row and column labels, respectively, differ. That is,
\begin{itemize}
\item \textbf{Row mismatch set ($R_0$):} If an edge connects two samples with different row labels, it goes into $R_0$. 
\item \textbf{Column mismatch edge set ($C_0$):} If an edge connects two samples with different column labels, it goes into $C_0$.
\item \textbf{Complementary edge $(e^c)$:} In a $4$-vertex graph, every edge has exactly one "complementary edge" that shares no vertices with it.
\end{itemize}
For an edge $e$, let $e^c$ be the complementary edge and set
$$
 N_A=|\set{e\in R_0:e^c\in C_0}|,\qquad N_B=|R_0\cap C_0|.
$$
For sample positions $a,b$, define the column-mismatch indicator by
\[
\delta_{ab}:=\begin{cases}
1,&J_a\ne J_b,\\
0,&J_a=J_b.
\end{cases}
\]
Thus $\delta_{ab}=\delta_{ba}\in\{0,1\}$ and
\[
\delta_{ac}\leqslant\delta_{ab}+\delta_{bc}
\]
for all sample positions $a,b,c$. All inequalities below involving
$\delta_{ab}$ are pointwise inequalities for these indicators.

We prove $N_A\leqslant2N_B$ for each possible partition of the row labels.
\begin{itemize}
\item If all four rows agree, both counts are zero, i.e., $N_A=N_B=0$. 
\item For type $3+1$, $R_0$ is a
star and its complementary edges form the opposite triangle. Each column
mismatch on that triangle is at most the sum of the two star mismatches
through its endpoints. More explicitly, if the singleton row is at vertex
$4$ and $\{i,j,k\}=\{1,2,3\}$, then
$\delta_{jk}\leqslant\delta_{j4}+\delta_{k4}$ for the complementary edge
of $i4$. Summing over $i$ counts each star mismatch twice and gives
$N_A\leqslant2N_B$. 
\item For type $2+2$,
complementation permutes $R_0$, so $N_A=N_B$. 
\item For type $2+1+1$, put the repeated
positions at $1,2$. Then $R_0$ is all edges except $12$, whence
$N_A=N_B+\delta_{12}-\delta_{34}$. The triangle inequality for mismatch indicators
gives $\delta_{12}\leqslant\delta_{13}+\delta_{23}\leqslant N_B$, proving $N_A\leqslant2N_B$.
\item If all four rows differ, $N_A=N_B$. 
\end{itemize}
These are all five partition types. On complementary edges the samples are independent. Thus
$$
 \bbE [N_A]=6\Pa{1-\sum_i r_i^2}\Pa{1-\sum_j c_j^2},\qquad
 \bbE [N_B]=12K(\rho').
$$
The latter identity follows because two independent samples have both
labels different with probability $2K(\rho')$. Taking expectations in
$N_A\leqslant2N_B$ 
$$
6\Pa{1-\sum_i r_i^2}\Pa{1-\sum_j c_j^2}=\bbE[N_A]\leqslant 2\bbE[N_B]=24K(\rho'),
$$
which proves 
$$
K(\rho')\geqslant \frac14\Pa{1-\sum_i r_i^2}\Pa{1-\sum_j c_j^2}.
$$
This is \eqref{eq:K-purity}; combining it with \eqref{eq:Q-K-bound}
gives the desired version \eqref{eq:native-Q-purity}.

Finally pinch $\rho_{AB}$ in a product of marginal eigenbases. Its
marginals are unchanged, and its diagonal state $\rho'$ has marginal spectra
$\bsr,\bsc$. Concavity and unitary invariance give
$\Q(\rho_{AB})\leqslant \Q(\rho')=\Q(\bsp)$. Hence
$\Delta_{\Q}(\rho_{AB})\geqslant\Delta_{\Q}(\rho')$, proving
\eqref{eq:native-Q-purity} for every quantum state, including singular
states. Strict positivity when both marginals are mixed follows from the
displayed lower bound. If a marginal is pure, the independent support
argument in Theorem~\ref{th:equality} gives a product with that pure
factor, and zero-padding invariance gives equality.
\end{proof}

\subsubsection{Proof of Corollary~\ref{cor:stability}}\label{subsubsect:4.5.7}

\begin{proof}
Write $u_A=1-\Tr{\rho_A^2}$ and $u_B=1-\Tr{\rho_B^2}$.
Combining \eqref{eq:transfer-main} in dimension $D$ with
\eqref{eq:native-Q-purity} in dimension $d$ gives
$$
 \widetilde\Delta_k^{(D)}\geqslant\beta_{D,k,d}u_Au_B.
$$
Thus 
$$
\min(u_A,u_B)\leqslant \Pa{\frac{\widetilde\Delta_k^{(D)}}{\beta_{D,k,d}}}^{\frac12}.
$$
Suppose the smaller impurity is $u_A$, and choose a top eigenvector $\bsa$
of $\rho_A$ with eigenvalue $p$. Since $\lambda_i(\rho_A)\leqslant p$ and
$\sum_i\lambda_i(\rho_A)=1$, we have
$$
 \Tr{\rho_A^2}=\sum_i\lambda_i(\rho_A)^2
 \leqslant p\sum_i\lambda_i(\rho_A)=p,
$$
and hence $u_A\geqslant 1-p$. This argument also applies when the top
eigenvalue is repeated or $\rho_A$ is singular.
Put $\Pi^A=\proj{\bsa}\ot \I_B$ and
$\omega=\Pi^A\rho_{AB}\Pi^A/p=\proj{\bsa}\otimes\sigma_B$.
Here $p>0$ because $\Tr{\rho_A}=1$. More explicitly,
$$
 \sigma_B=\frac1p\Pa{\bra{\bsa}\ot\I_B}\rho_{AB}
 \Pa{\ket{\bsa}\ot\I_B}.
$$
This compression is positive semidefinite and
$\Tr{\sigma_B}=\Tr{\Pi^A\rho_{AB}}/p=1$, so $\sigma_B$ is a state even
when $\rho_{AB}$ is singular.
Let $\ket{\Psi}_{ABR}$ be a purification of $\rho_{AB}$ and define
$\ket{\Phi}_{ABR}=(\Pi^A\ot\I_R)\ket{\Psi}_{ABR}/\sqrt p$.
Its squared norm is $\Tr{\Pi^A\rho_{AB}}/p=1$, $\ket{\Phi}_{ABR}$ purifies $\omega$,
and $\Abs{\inner{\Psi}{\Phi}}=\sqrt p$.
The trace distance between these pure states is $\sqrt{1-p}$.
Contractivity under partial trace therefore gives
\begin{eqnarray*}
 \frac12\norm{\rho_{AB}-\omega}_1&\leqslant& \frac12\norm{\proj{\Psi}-\proj{\Phi}}_1 = \sqrt{1-\abs{\Inner{\Psi}{\Phi}}^2}=\sqrt{1-p}\\
&\leqslant& \sqrt{u_A}  \leqslant \Pa{\frac{\widetilde\Delta_k^{(D)}}{\beta_{D,k,d}}}^{\frac14}.
\end{eqnarray*}
The case of smaller $u_B$ is symmetric. Trace distance between states is
at most one, completing \eqref{eq:stability-distance}. If the deficit
vanishes, this construction has $p=1$ and $\rho_{AB}=\omega$.
\end{proof}

\subsection{Entropy gains of unital channels}\label{proof:entropygains}

We prove Theorem~\ref{th:channel}. Throughout this subsection,
$\cE:M_d(\bbC)\to M_d(\bbC)$ is completely positive, trace-preserving
and unital, and $\rho\in\rD(\bbC^d)$. Since $\cE$ is both trace-preserving and unital, its Hilbert--Schmidt
dual $\cE^\dagger$ has these two properties as well. The pinching $\cP_\rho$ uses the distinct
eigenspaces of $\rho$, including its kernel when present. For a chosen
complementary channel $\widehat\cE$ with output dimension $d_\ee$, write $N:=\max(d,d_\ee)$.

\subsubsection{The adapted divergence and its Bregman identity}\label{subsubsect:4.6.1}

Extend $\S_k^{(d)}$ to positive semidefinite matrices by the spectral
formula of Proposition~\ref{prop:eigfor}, and put
$F_k(\bsX):=-\S_k^{(d)}(\bsX)$. In particular, for $\bsX>0$,
\begin{eqnarray}\label{eq:note-definition}
 F_k(\bsX)=\binom{d-1}{k-1}^{-1} \sum_{|\cI|=k}[\bdlambda_{\cI}](x^k\log x), \qquad (\lambda_1,\ldots,\lambda_d)=\bdlambda(\bsX).
\end{eqnarray}
The affine trace term in the positive compression representation
\eqref{eq:kentropy-intrep} gives
\begin{eqnarray}\label{eq:channel-F-compression}
 F_k(\bsX)=\sum_{\ell=1}^{d-k+1}\frak{C}_{d,k,\ell}  \int_{\op{St}(d,\ell)} \Tr{(\bsV^\dagger\bsX\bsV)\log(\bsV^\dagger\bsX\bsV)}
 \dif\mu_\ell(\bsV)+(H_k-1)\Tr{\bsX}
\end{eqnarray}
for $k\geqslant2$. At $k=1$ we have $F_1(\bsX)=\Tr{\bsX\log\bsX}$.

Let $\sL_k(\bsX):=\nabla F_k(\bsX)$, where the gradient is taken
with respect to the real Hilbert--Schmidt inner product on Hermitian
matrices. The derivative of $\Tr{\bsX\log\bsX}$ in a Hermitian
direction $\bsH$ is 
\begin{eqnarray*}
\frac{\dif}{\dif t}\Big|_{t=0}\Tr{(\bsX+t\bsH)\log(\bsX+t\bsH)}=\Tr{(\log\bsX+\I_d)\bsH}. 
\end{eqnarray*}
Differentiating
\eqref{eq:channel-F-compression} and cancelling its affine trace term
in the Bregman difference therefore gives
\begin{eqnarray}\label{eq:bregman}
 \B_k^{(d)}(\bsX||\bsY)  =F_k(\bsX)-F_k(\bsY)  -\Tr{\sL_k(\bsY)(\bsX-\bsY)},\qquad \bsX,\bsY>0.
\end{eqnarray}
For each compression, the term $+\I_\ell$ in the derivative contributes
$-\Tr{\bsV^\dagger(\bsX-\bsY)\bsV}$. This is exactly the trace
correction in $\B_+$ in \eqref{eq:Dk}. It is necessary because the
compressed matrices generally have different traces.

The generalized relative entropy $\B_+$ is nonnegative, jointly
convex and lower semicontinuous on pairs of positive semidefinite
matrices, with the support convention in Section~\ref{subsect:EG};
joint convexity applies to matrices of arbitrary trace
\cite{Tropp2012}. The compression maps are linear and the coefficients
$\frak{C}_{d,k,\ell}$ are positive. Hence the same properties hold for
$\B_k^{(d)}$. Its simultaneous unitary invariance follows by the
change of variables $\bsV\mapsto\bsU^\dagger\bsV$ in the Haar
integrals:
\begin{eqnarray}\label{eq:channel-divergence-invariance}
 \B_k^{(d)}(\bsU\bsX\bsU^\dagger||\bsU\bsY\bsU^\dagger)
 =\B_k^{(d)}(\bsX||\bsY).
\end{eqnarray}
For the lower semicontinuity assertion at $k\geqslant2$, lower
semicontinuity of $\B_+$ can be integrated using Fatou's lemma, since
every integrand is nonnegative. The case $k=1$ is the corresponding
property of $\B_+$ itself.

\subsubsection{Recovery bounds for mixed-unitary and general unital channels}\label{subsubsect:4.6.2}

First suppose that $\rho>0$ and that
\begin{eqnarray*}
 \cE(\bsX)=\sum_\nu p_\nu\bsU_\nu\bsX\bsU_\nu^\dagger,
 \qquad p_\nu\geqslant0,\quad\sum_\nu p_\nu=1,
\end{eqnarray*}
where all $\bsU_\nu$ are unitary. Unitality implies
$\cE(\rho)>0$ and $\cE^\dagger\circ\cE(\rho)>0$. Summing \eqref{eq:bregman} over the
unitary orbit of $\rho$ yields
\begin{eqnarray*}
\sum_\nu p_\nu  \B_k^{(d)}(\bsU_\nu\rho\bsU_\nu^\dagger||\cE(\rho)) &=& \sum_\nu p_\nu  \Pa{F_k(\bsU_\nu\rho\bsU_\nu^\dagger)-F_k(\cE(\rho))  -\Tr{\sL_k(\cE(\rho))(\bsU_\nu\rho\bsU_\nu^\dagger-\cE(\rho))}}\\
&=& \S^{(d)}_k(\cE(\rho)) - \S^{(d)}_k(\rho) - \Tr{\sL_k(\cE(\rho))\Pa{\sum_\nu p_\nu \bsU_\nu\rho\bsU_\nu^\dagger-\cE(\rho)}}\\
&=&\S^{(d)}_k(\cE(\rho)) - \S^{(d)}_k(\rho).
\end{eqnarray*}
Thus
\begin{eqnarray}\label{eq:mixed-unitary-lower}
 \C_k^{(d)}(\cE,\rho) &=& \S^{(d)}_k(\cE(\rho)) - \S^{(d)}_k(\rho)=\sum_\nu p_\nu
 \B_k^{(d)}(\bsU_\nu\rho\bsU_\nu^\dagger||\cE(\rho))\notag\\
 &=&\sum_\nu p_\nu
 \B_k^{(d)}(\rho||\bsU_\nu^\dagger\cE(\rho)\bsU_\nu)
 \geqslant\B_k^{(d)}(\rho||\cE^\dagger\circ\cE(\rho)).
\end{eqnarray}
Indeed, the first arguments average to $\cE(\rho)$, so the linear
gradient terms cancel, while unitary invariance gives
$F_k(\bsU_\nu\rho\bsU_\nu^\dagger)=F_k(\rho)$.
The second line uses \eqref{eq:channel-divergence-invariance}, joint
convexity, and
$\sum_\nu p_\nu\bsU_\nu^\dagger\cE(\rho)\bsU_\nu
=\cE^\dagger(\cE(\rho))$.

Now let $\cE$ be an \emph{arbitrary} unital channel and still assume
$\rho>0$. Since $\zeta=\cP_\rho(\cE^\dagger\circ\cE(\rho))$ commutes with $\rho$, choose
an orthonormal basis $\{\ket{i}\}_{i=1}^d$ in which
\begin{eqnarray*}
 \rho=\diag(\bsx),\qquad \zeta=\diag(\bsz).
\end{eqnarray*}
Within each degenerate eigenspace of $\rho$, diagonalize the
corresponding block of $\zeta$. Write an eigenvalue decomposition
$\cE(\rho)=\sum^d_{j=1} y_j\proj{\psi_j}$ and define
\begin{eqnarray}\label{eq:channel-bistochastic}
 B_{ji}:=\Innerm{\psi_j}{\cE(\proj{i})}{\psi_j}.
\end{eqnarray}
Positivity of $\cE$ gives $B_{ji}\geqslant0$. Trace preservation and
unitality give the two normalizations
\begin{eqnarray*}
 \sum^d_{j=1}B_{ji}=\Tr{\cE(\proj{i})}=1,
 \qquad
 \sum^d_{i=1}B_{ji}=\Innerm{\psi_j}{\cE(\I_d)}{\psi_j}=1.
\end{eqnarray*}
Thus $\bsB=(B_{ji})_{d\times d}$ is doubly stochastic. It is called the Kraus matrix of $\cE$~\cite{Zhang2011}. By linearity and the definition of the
dual map,
\begin{eqnarray}\label{eq:channel-bistochastic-spectra}
 \bsy=\bsB\bsx,\qquad  z_i=\Innerm{i}{\cE^\dagger\circ\cE(\rho)}{i}
 =\Tr{\cE(\rho)\cE(\proj{i})}
 =\sum_jB_{ji}y_j,
 \qquad \bsz=\bsB^\t\bsy.
\end{eqnarray}
Here $\Innerm{i}{\cP_\rho(\cE^\dagger\circ\cE(\rho))}{i}=\Innerm{i}{\cE^\dagger\circ\cE(\rho)}{i}$ because
$\cP_\rho$ retains the diagonal entries in this basis.

By Birkhoff's theorem \cite[Chapter~2]{Marshall2011} or \cite[Theorem 4.28]{Watrous2018}, write $\bsB$ as a
convex combination of permutation matrices. The associated channel,
defined by the same convex combination of unitary conjugations by
those permutation matrices, sends $\diag(\bsx)$ to $\diag(\bsy)$.
Its dual sends $\diag(\bsy)$ to $\diag(\bsB^\t\bsy)
=\diag(\bsz)$. Applying \eqref{eq:mixed-unitary-lower} to this
channel and using unitary invariance of the entropy and divergence
gives
\begin{eqnarray}\label{eq:channel-pinched-recovery}
 0\leqslant\B_k^{(d)}(\rho||\cP_\rho(\cE^\dagger\circ\cE(\rho)))  \leqslant\S_k^{(d)}(\cE(\rho)) - \S_k^{(d)}(\rho) =\C_k^{(d)}(\cE,\rho).
\end{eqnarray}
If $[\rho,\cE^\dagger\circ\cE(\rho)]=0$, each eigenspace of $\rho$ is invariant under
$\cE^\dagger\circ\cE(\rho)$, so $\cP_\rho(\cE^\dagger\circ\cE(\rho))=\cE^\dagger\circ\cE(\rho)$. In that case the lower bound is
unpinched. The singular case will be treated below.

\subsubsection{Resolvent comparison and the optimal coefficient}\label{subsubsect:4.6.3}

We next prove
$0\leqslant\C_k^{(d)}(\cE,\rho)
\leqslant\alpha_{d,k}\C_1^{(d)}(\cE,\rho)$.
For $\bsX>0$, the logarithm integral and
\eqref{eq:note-definition} give
\begin{eqnarray}\label{eq:resolvent}
 F_k(\bsX)=\int_0^\infty\Br{ \frac{\Tr{\bsX}}{1+t}-\frac dk+ \frac{t^k}{\binom{d-1}{k-1}} \ee_k\Pa{(\bsX+t\I_d)^{-1}}}\dif t.
\end{eqnarray}
To check the constants, use
$$
\log x=\int_0^\infty\Pa{\frac1{1+t}- \frac1{x+t}}\dif t
$$ 
and the divided-difference identities
\begin{equation*}
 [\bdlambda_{\cI}](x^k)=\sum_{i\in\cI}\lambda_i,
 \qquad
 [\bdlambda_{\cI}](\tfrac{x^k}{x+t}) =1-\frac{t^k}{\prod_{i\in\cI}(t+\lambda_i)}.
\end{equation*}
Each index occurs in $\binom{d-1}{k-1}$ subsets, and
$\binom dk/\binom{d-1}{k-1}=d/k$, proving
\eqref{eq:resolvent}. The formula extends to coincident positive
eigenvalues by confluent divided differences. Near zero the
resolvents are bounded on a neighborhood of any fixed positive
definite $\bsX$. At infinity the constant and $t^{-1}$ terms of the
combined integrand cancel, leaving $O(t^{-2})$; its first derivatives
have the same local integrable bound. This justifies differentiation
under the integral on the positive definite cone.

Regard $\S_k^{(d)}$ as a symmetric function of a positive eigenvalue
vector $\bdlambda$. Fix $i\ne j$, set $a=\lambda_i$, $b=\lambda_j$, denote $\partial_a\equiv \partial_i$ and $\partial_b\equiv \partial_j$
and put
\begin{eqnarray*}
 z_l=\frac{t}{t+\lambda_l},\qquad
 \sE_t=\ee_{k-1}\Pa{z_l:l\in[d]\setminus \set{i,j}},\qquad
 \sF_t=\ee_{k-2}\Pa{z_l:l\in[d]\setminus \set{i,j}}.
\end{eqnarray*}
Elementary symmetric polynomials outside their degree range are zero,
and $\ee_0=1$. Differentiating \eqref{eq:resolvent} gives
\begin{eqnarray*}
 &&-\frac{\partial_i\S_k^{(d)}-\partial_j\S_k^{(d)}}{a-b}  = \frac{\partial_i F_k - \partial_j F_k}{a-b} \\
&&= \binom{d-1}{k-1}^{-1}\int_0^\infty \frac{\partial_i \ee_k\Pa{z_1,\ldots,z_d} - \partial_j\ee_k\Pa{z_1,\ldots,z_d}}{a-b}\dif t. 
\end{eqnarray*}
For example, write
$$
\ee_k(z_1,\ldots,z_d)=\ee_k\Pa{z_l:l\in[d]\setminus \set{i,j}}  +(z_i+z_j)\sE_t+z_iz_j\sF_t
$$ 
and use $\partial_a z_i=- \frac{t}{(t+a)^2}$ to get 
\begin{eqnarray*}
\partial_i \ee_k(z_1,\ldots,z_d) &=&0+ \partial_i[(z_i+z_j)\sE_t] + \partial_i[z_iz_j\sF_t] = (\partial_iz_i)\sE_t + (\partial_iz_i) z_j\sF_t\\
&=& (\partial_iz_i)(\sE_t + z_j\sF_t)=- \frac{t}{(t+a)^2}(\sE_t + z_j\sF_t) \\
&=& -\frac1t(z^2_i\sE_t + z^2_iz_j\sF_t),
\end{eqnarray*}
analogously, 
\begin{eqnarray*}
\partial_j \ee_k(z_1,\ldots,z_d) = -\frac1t(z^2_j\sE_t + z_iz^2_j\sF_t),
\end{eqnarray*}
both of which lead to 
\begin{eqnarray*}
\frac{\partial_i \ee_k(z_1,\ldots,z_d)-\partial_j \ee_k(z_1,\ldots,z_d)}{a-b} = -\frac{(z_i+z_j)\sE_t+z_iz_j\sF_t}t \frac{z_i-z_j}{a-b} = \frac{(z_i+z_j)\sE_t+z_iz_j\sF_t}{(t+a)(t+b)}.
\end{eqnarray*}
Thus we obtain 
\begin{eqnarray}\label{eq:pair-derivative}
-\frac{\partial_i\S_k^{(d)}-\partial_j\S_k^{(d)}}{a-b} & =&\binom{d-1}{k-1}^{-1}\int_0^\infty
 \frac{(z_i+z_j)\sE_t+z_iz_j\sF_t}{(t+a)(t+b)}\dif t\\
&\leqslant & \binom{d-1}{k-1}^{-1}\Br{\binom{d-2}{k-1} \int_0^\infty\frac{z_i+z_j}{(t+a)(t+b)}\dif t + \binom{d-2}{k-2} \int_0^\infty\frac{z_iz_j}{(t+a)(t+b)}\dif t},\notag
\end{eqnarray}
where we used the fact that,
\begin{eqnarray*}
 0\leqslant \sE_t\leqslant\binom{d-2}{k-1},\qquad
 0\leqslant \sF_t\leqslant\binom{d-2}{k-2},\qquad (0\leqslant z_l\leqslant1).
\end{eqnarray*}
Let
\begin{eqnarray*}
 I_0:=\int_0^\infty\frac{\dif t}{(t+a)(t+b)}
 =\frac{\log a-\log b}{a-b}.
\end{eqnarray*}
The boundary values of $\frac t{(t+a)(t+b)}$ vanish at zero and infinity.
Integrating its derivative therefore yields
\begin{eqnarray*}
0&=&\frac t{(t+a)(t+b)}\Big|^\infty_0 =  \int_0^\infty\Pa{\frac t{(t+a)(t+b)}}'\dif t\\
&=&  \int_0^\infty\frac{1-z_i-z_j}{(t+a)(t+b)}\dif t = I_0 - \int_0^\infty\frac{z_i+z_j}{(t+a)(t+b)}\dif t,
\end{eqnarray*}
that is,
\begin{eqnarray*}
 \int_0^\infty\frac{z_i+z_j}{(t+a)(t+b)}\dif t=I_0.
\end{eqnarray*}
For the remaining term, partial fractions give
\begin{eqnarray}\label{eq:I2}
 I_2:=\int_0^\infty\frac{z_iz_j}{(t+a)(t+b)}\dif t=\int_0^\infty\frac{t^2}{(t+a)^2(t+b)^2}\dif t
 =\frac{a+b}{(a-b)^2}
 -\frac{2ab(\log a-\log b)}{(a-b)^3}
 \leqslant\frac{I_0}{3}.
\end{eqnarray}
To prove the last inequality, assume that $a\geqslant b$ by symmetry, and it permits $\frac ab=\exp (x)\geqslant1$ with
$x\geqslant0$. After multiplication by the positive denominators,
it is equivalent to $3\sinh x\leqslant x(\cosh x+2)$ for $x\geqslant0$. The function
$g(x)=x(\cosh x+2)-3\sinh x$ satisfies
$g(0)=g'(0)=g''(0)=0$ and $g'''(x)=x\sinh x\geqslant0$;
three integrations prove the inequality. When $a=b$, the continuous
values are 
\begin{eqnarray*}
I_0=\int^\infty_0 \frac{\dif t}{(t+a)^2}=\frac1a,\qquad
I_2= \int^\infty_0 \frac{t^2\dif t}{(t+a)^4} = \frac1{3a}.
\end{eqnarray*}

Combining these estimates with \eqref{eq:pair-derivative} gives
\begin{eqnarray}\label{eq:channel-schur-comparison}
 0&\leqslant&  -\frac{\partial_i\S_k^{(d)}-\partial_j\S_k^{(d)}}{a-b}\leqslant \frac{\binom{d-2}{k-1}+\frac13\binom{d-2}{k-2}}
 {\binom{d-1}{k-1}}I_0
 =\left(1-\frac{2(k-1)}{3(d-1)}\right)I_0
 =\alpha_{d,k}I_0.
\end{eqnarray}
For $\S=\S^{(d)}_1$, the corresponding derivative quotient is exactly
$I_0$. The Schur criterion \cite[Chapter~3]{Marshall2011} therefore
shows that both $\S_k^{(d)}$ and
$\alpha_{d,k}\S-\S_k^{(d)}$ are Schur concave. The construction
\eqref{eq:channel-bistochastic} shows directly that
$\bdlambda(\cE(\rho))\prec\bdlambda(\rho)$. Applying these two Schur
concavity statements gives
\begin{equation}\label{eq:channel-sharp-comparison}
 0\leqslant\C_k^{(d)}(\cE,\rho)
 \leqslant\alpha_{d,k}\C_1^{(d)}(\cE,\rho).
\end{equation}
Continuity of the entropies extends this comparison to singular states.

For the last inequality in \eqref{eq:channel-chain}, let
$\bsW:\bbC^d\to\bbC^d\otimes\bbC^{d_\ee}$ be a Stinespring isometry
for the chosen complement. The bipartite state
$\bsW\rho\bsW^\dagger$, with marginals $\cE(\rho)$ and $\widehat\cE(\rho)$, has the
same nonzero spectrum as $\rho$. The von Neumann triangle inequality
thus gives
\begin{equation*}
 \S(\cE(\rho))-\S(\rho)\leqslant\S(\widehat\cE(\rho)),
\end{equation*}
which proves
$\C_k^{(d)}(\cE,\rho)\leqslant\alpha_{d,k}\S(\widehat\cE(\rho))$.

To establish optimality of $\alpha_{d,k}$, take
\begin{equation*}
 \rho_\varepsilon=\diag\left(\frac12+\varepsilon,
 \frac12-\varepsilon,0,\ldots,0\right),
 \qquad 0<|\varepsilon|<\frac12,
\end{equation*}
and let $\cE$ average the identity conjugation and conjugation by the
unitary that swaps the first two coordinates. Then
$\cE(\rho_\varepsilon)=\rho_0$. Applying
\eqref{eq:padding-normalized} to a two-dimensional state $\varrho$
gives, for every $1\leqslant k\leqslant d$,
\begin{equation}\label{eq:rank-two-padding}
 \S_k^{(d)}(\varrho)
 =\frac{d-k}{d-1}\S(\varrho)
 +\frac{k-1}{d-1}\Q(\varrho).
\end{equation}
For the two-dimensional restriction of $\rho_\varepsilon$, Taylor
expansion at $\varepsilon=0$ gives
\begin{equation*}
 \S(\rho_0)-\S(\rho_\varepsilon)
 =2\varepsilon^2+O(\varepsilon^4),\qquad
 \Q(\rho_0)-\Q(\rho_\varepsilon)
 =\frac23\varepsilon^2+O(\varepsilon^4).
\end{equation*}
The second expansion follows, for example, by substituting
$\frac12\pm\varepsilon$ into the two-eigenvalue formula
\eqref{eq:Phi12}. Consequently
\begin{equation*}
 \frac{\C_k^{(d)}(\cE,\rho_\varepsilon)}
 {\C_1^{(d)}(\cE,\rho_\varepsilon)}
 \longrightarrow\frac{d-k}{d-1}
 +\frac{k-1}{3(d-1)}=\alpha_{d,k}.
\end{equation*}
No smaller coefficient can hold uniformly over states and unital
channels, including at $k=1$.

\subsubsection{Unpinched recovery at the endpoint indices}\label{subsubsect:4.6.4}

We prove the unpinched lower bound for $k\in\set{1,d-1,d}$.
Differentiating \eqref{eq:resolvent} gives
\begin{eqnarray}\label{eq:gradient}
 \sL_k(\bsX)&=& \nabla F_k(\bsX) = \nabla\int_0^\infty\Br{ \frac{ \Tr{\bsX}}{1+t} -\frac dk+  \frac{t^k}{\binom{d-1}{k-1}} \ee_k\Pa{(\bsX+t\I_d)^{-1}}}\dif t \notag\\
&=& \int_0^\infty\Br{ \frac{ \nabla\Tr{\bsX}}{1+t} +  \frac{t^k}{\binom{d-1}{k-1}} \nabla \ee_k\Pa{(\bsX+t\I_d)^{-1}}}\dif t \notag\\
&=& \int_0^\infty\left[ \frac{\I_d}{1+t} -\frac{t^k}{\binom{d-1}{k-1}}\bsR_k(\bsX+t\I_d) \right]\dif t,
 \qquad
 \bsR_k(\bsB):=-\nabla_{\bsB}\ee_k(\bsB^{-1}).
\end{eqnarray}
The integral is interpreted as a combined, convergent integrand.
For $\bsB>0$, the identities
$\ee_d(\bsB^{-1})={\det}^{-1}[\bsB]$ and
$\ee_{d-1}(\bsB^{-1})=\frac{\Tr{\bsB}}{\det[\bsB]}$ imply
\begin{equation}\label{eq:channel-endpoint-R}
 \bsR_d(\bsB)=\frac{\bsB^{-1}}{\det[\bsB]},\qquad
 \bsR_{d-1}(\bsB) =\frac{(\Tr{\bsB})\bsB^{-1}-\I_d}{\det[\bsB]}.
\end{equation}

Let $\Phi$ be completely positive, trace preserving and unital, and
put $\bsC=\Phi(\bsB)>0$. Operator Jensen's inequality
\cite{Hansen2003}, applied to the operator convex function $x^{-1}$
and the operator concave function $\log x$, gives
\begin{equation*}
 \bsC^{-1}\leqslant\Phi(\bsB^{-1}),\qquad
 \log\bsC\geqslant\Phi(\log\bsB).
\end{equation*}
Trace preservation in the second inequality yields
$$
\Tr{\log\bsC}\geqslant\Tr{\Phi(\log\bsB)} \iff  \log\det\bsC\geqslant\log\det\bsB \implies \det\bsC\geqslant \det\bsB.
$$
Also $\Tr{\bsC}=\Tr{\bsB}=:t$ and
$t\bsC^{-1}-\I_d\geqslant0$, since every eigenvalue of $\bsC$ is
at most its trace. Thus
\begin{eqnarray*}
 \frac{\bsC^{-1}}{\det\bsC} \leqslant\frac{\Phi(\bsB^{-1})}{\det\bsB},\qquad
 \frac{t\bsC^{-1}-\I_d}{\det\bsC} \leqslant\frac{t\bsC^{-1}-\I_d}{\det\bsB}
 \leqslant\frac{t\Phi(\bsB^{-1})-\I_d}{\det\bsB}.
\end{eqnarray*}
Using \eqref{eq:channel-endpoint-R} and $\Phi(\I_d)=\I_d$, these
are precisely
\begin{equation*}
 \bsR_k(\Phi(\bsB))\leqslant\Phi(\bsR_k(\bsB)),
 \qquad k=d-1,d.
\end{equation*}
Because $\Phi(\bsX+t\I_d)=\Phi(\bsX)+t\I_d$, the minus sign
in \eqref{eq:gradient} reverses this comparison and gives
\begin{equation}\label{eq:channel-gradient-Jensen}
 \sL_k(\Phi(\bsX))\geqslant\Phi(\sL_k(\bsX)),
 \qquad k=d-1,d.
\end{equation}
One may first integrate over a compact interval in $t$ and then pass
to the convergent combined integrals. For $k=1$,
$\sL_1(\bsX)=\log\bsX+\I_d$, and
\eqref{eq:channel-gradient-Jensen} follows directly from logarithmic
Jensen. This also covers the overlap $d=2$, $k=d-1=1$.

The positive-cone scaling law \eqref{eq:Entscalingrule}, after
normalization, is
\begin{equation*}
 F_k(s\bsX)=sF_k(\bsX) - \eta(s)\Tr{\bsX},\qquad s>0.
\end{equation*}
Differentiating at $s=1$ gives $\Tr{\bsX\sL_k(\bsX)}=F_k(\bsX)+\Tr{\bsX}$, and thus, for $\bsX=\cE^\dagger\circ\cE(\rho)$, we have 
\begin{equation*}
 \Tr{\cE^\dagger\circ\cE(\rho)\sL_k(\cE^\dagger\circ\cE(\rho))}=F_k(\cE^\dagger\circ\cE(\rho))+\Tr{\cE^\dagger\circ\cE(\rho)}=F_k(\cE^\dagger\circ\cE(\rho))+1.
\end{equation*}
Hence \eqref{eq:bregman}, evaluated on positive definite states,
becomes
\begin{eqnarray}\label{eq:channel-state-bregman}
 \B_k^{(d)}(\rho||\cE^\dagger\circ\cE(\rho))  &=& F_k(\rho) - F_k(\cE^\dagger\circ\cE(\rho)) - \Tr{\sL_k(\cE^\dagger\circ\cE(\rho))(\rho-\cE^\dagger\circ\cE(\rho))} \notag\\
&=&F_k(\rho)-\Tr{\rho\sL_k(\cE^\dagger\circ\cE(\rho))}+1.
\end{eqnarray}
Apply \eqref{eq:channel-gradient-Jensen} with
$\Phi=\cE^\dagger$ and $\bsX=\cE(\rho)$, we get that
$$
 \sL_k(\cE^\dagger(\cE(\rho)))\geqslant \cE^\dagger(\sL_k(\cE(\rho))),   \qquad k=d-1,d.
$$
Taking the trace against
$\rho\geqslant0$ gives
\begin{eqnarray*}
 \Tr{\rho\sL_k(\cE^\dagger\circ\cE(\rho))} \geqslant\Tr{\rho\cE^\dagger(\sL_k(\cE(\rho)))}
 =\Tr{\cE(\rho)\sL_k(\cE(\rho))}=F_k(\cE(\rho))+1.
\end{eqnarray*}
Substitution in \eqref{eq:channel-state-bregman} proves
\begin{equation}\label{eq:channel-unpinched-endpoints}
 \B_k^{(d)}(\rho||\cE^\dagger\circ\cE(\rho))
 \leqslant F_k(\rho)-F_k(\cE(\rho))
 =\C_k^{(d)}(\cE,\rho),\qquad k\in\set{1,d-1,d}.
\end{equation}

\subsubsection{Singular states and the lower-bound limits}\label{subsubsect:4.6.5}

For arbitrary $\rho$, let
\begin{equation*}
 \rho_\varepsilon=(1-\varepsilon)\rho+\varepsilon\frac{\I_d}{d},
 \qquad 0<\varepsilon<1.
\end{equation*}
Both $\cE$ and $\cE^\dagger$ are unital. Therefore
\begin{equation*}
\cE(\rho_\varepsilon)
 =(1-\varepsilon)\cE(\rho)+\varepsilon\frac{\I_d}{d},\qquad  \cE^\dagger(\cE(\rho_\varepsilon))  =(1-\varepsilon)\cE^\dagger(\cE(\rho))+\varepsilon\frac{\I_d}{d}.
\end{equation*}
All three states are positive definite. The affine transformation of
the eigenvalues of $\rho$ is injective, so it preserves the distinct
eigenspaces, including the former kernel:
$\cP_{\rho_\varepsilon}=\cP_\rho$. Consequently
\begin{equation*}
\cP_{\rho_\varepsilon}(\cE^\dagger(\cE(\rho_\varepsilon))) =(1-\varepsilon)\cP_\rho(\cE^\dagger\circ\cE(\rho))+\varepsilon\frac{\I_d}{d}.
\end{equation*}
The positive definite bounds already proved, lower semicontinuity of
$\B_k^{(d)}$ and continuity of $\S_k^{(d)}$ give
\begin{eqnarray*}
 \B_k^{(d)}(\rho||\cP_\rho(\cE^\dagger\circ\cE(\rho))) \leqslant\liminf_{\varepsilon\downarrow0} \B_k^{(d)}(\rho_\varepsilon||\cP_{\rho_\varepsilon}(\cE^\dagger(\cE(\rho_\varepsilon))))
 \leqslant\C_k^{(d)}(\cE,\rho).
\end{eqnarray*}
The same argument applies to the unpinched bounds whenever $\cE$ is
mixed-unitary or $k\in\set{1,d-1,d}$. If $[\rho,\cE^\dagger\circ\cE(\rho)]=0$, the
pinched bound itself is already unpinched. This proves every
lower-bound assertion of Theorem~\ref{th:channel}, including its
support and boundary cases.

\subsubsection{Entropy exchange for mixed-unitary channels and for subentropy}\label{subsubsect:4.6.6}

For a fixed mixed-unitary channel, the first line of
\eqref{eq:mixed-unitary-lower} writes
$\C_k^{(d)}(\cE,\rho)$ as a positive sum of jointly convex
divergences, with both arguments linear in $\rho$. It is therefore
convex for positive definite inputs, and continuity extends this
convexity to the whole state space. For a pure-state decomposition
$\rho=\sum_iw_i\proj{\bsv_i}$,
\begin{equation}\label{eq:channel-pure-decomposition}
 \C_k^{(d)}(\cE,\rho)  \leqslant\sum_iw_i\C_k^{(d)}(\cE,\proj{\bsv_i})  =\sum_iw_i\S_k^{(d)}(\cE(\proj{\bsv_i})).
\end{equation}
The last equality uses $\S_k^{(d)}(\proj{\bsv_i})=0$.

For each pure input, $\bsW\ket{\bsv_i}$ is a pure state of the system
and environment. Thus $\cE(\proj{\bsv_i})$ and
$\widehat\cE(\proj{\bsv_i})$ have identical nonzero
spectra. Since $k\leqslant d\leqslant N$, the padding comparison
\eqref{eq:padding-monotone} gives
\begin{equation*}
 \S_k^{(d)}(\cE(\proj{\bsv_i})) \leqslant\S_k^{(N)}(\cE(\proj{\bsv_i})) =\S_k^{(N)}(\widehat\cE(\proj{\bsv_i})).
\end{equation*}
Concavity in the fixed dimension $N$, together with
$\sum_iw_i\widehat\cE(\proj{\bsv_i}) =\widehat\cE(\rho)$, now yields
\begin{equation*}
 \C_k^{(d)}(\cE,\rho) \leqslant\sum_iw_i\S_k^{(N)}(\widehat\cE(\proj{\bsv_i})) \leqslant\S_k^{(N)}(\widehat\cE(\rho)).
\end{equation*}
This is \eqref{eq:exchange-Sk}. The argument applies to every chosen
complement, including a minimal one; $d_\ee$ need not equal the number
of unitaries in the mixed-unitary decomposition.

For subentropy, a mixed-unitary representation is unnecessary.
Apply the subentropy triangle inequality in Corollary~\ref{cor:triangle}
to $\bsW\rho\bsW^\dagger$. Its marginals are $\cE(\rho)$ and
$\widehat\cE(\rho)$, and its nonzero spectrum is that of $\rho$. Padding
invariance at the subentropy endpoint gives
\begin{equation*}
\Q(\rho)\leqslant \Q(\cE(\rho))\leqslant\Q(\rho)+\Q(\widehat\cE(\rho)).
\end{equation*}
Equivalently, purify $\rho$ with a reference system and apply
subentropy subadditivity to the joint marginal of the reference and
environment in the resulting pure Stinespring state. The complementary reductions
have the required nonzero spectra. Combining this exchange bound
with \eqref{eq:channel-unpinched-endpoints} at $k=d$ proves
\eqref{eq:exchange-Q} for every unital channel. The exchange upper
bound itself holds for every channel. This completes the proof of
Theorem~\ref{th:channel}.

\subsubsection{Why the divergence and the common environment dimension matter}\label{subsubsect:4.6.7}

For computational dephasing of a qubit, let
$\rho=\proj{+}$ with $\ket{+}=\frac{\ket{0}+\ket{1}}{\sqrt2}$.
Then $\cE(\rho)=\widehat\cE(\rho)=\frac12\I_2$. The ordinary relative entropy is
\begin{equation*}
 \B_+(\rho||\cE^\dagger\circ \cE(\rho))=\log2,
\end{equation*}
whereas the subentropy gain is
$\Q(\frac12\I_2)-\Q(\rho)=\log2-\frac12$.
Thus replacing the adapted divergence by ordinary relative entropy
does not preserve the lower bound throughout the family.

The dimensional adjustment in \eqref{eq:exchange-Sk} is also needed.
Fix $d>k\geqslant2$, choose a pinching into $k$ nonzero orthogonal
blocks, and take unit vectors $\ket{\bsv_1}$ and $\ket{\bsv_2}$ in its
first two blocks. For
\begin{equation*}
 \ket{\psi_\varepsilon} =\sqrt{1-\varepsilon}\ket{\bsv_1} +\sqrt{\varepsilon}\ket{\bsv_2},\qquad 0<\varepsilon<\frac12,
\end{equation*}
the system output and complementary output have nonzero spectra
$(1-\varepsilon,\varepsilon)$. The $k$ nonzero block projectors are
linearly independent Kraus operators, so a minimal complement has
dimension $k$. Define
\begin{equation*}
 h(\varepsilon)=-(1-\varepsilon)\log(1-\varepsilon)
 -\varepsilon\log\varepsilon,\qquad
 q(\varepsilon)=\Q(\diag(1-\varepsilon,\varepsilon)).
\end{equation*}
By \eqref{eq:rank-two-padding}, the system gain equals
\begin{equation*}
 \frac{d-k}{d-1}h(\varepsilon)
 +\frac{k-1}{d-1}q(\varepsilon),
\end{equation*}
whereas the entropy in the native minimal environment dimension is
$\S_k^{(k)}(\widehat\cE(\proj{\psi_\varepsilon}))
=q(\varepsilon)$. The two-eigenvalue formula gives
\begin{equation*}
 h(\varepsilon)=\eta(\varepsilon)+O(\varepsilon),
 \qquad
 q(\varepsilon)=\varepsilon
 +O(\varepsilon^2|\log\varepsilon|).
\end{equation*}
The ratio of the system gain to this native-dimension environment
entropy is therefore unbounded as $\varepsilon\downarrow0$.
No finite constant can replace the padding in this comparison.
The remaining questions are the unpinched recovery lower bound for
arbitrary unital channels at $d\geqslant4$,
$2\leqslant k\leqslant d-2$, and the padded exchange bound
\eqref{eq:exchange-Sk} for arbitrary unital channels at $1<k<d$.

\subsection{Proof of the operational application}\label{proof:operational}

We prove Theorem~\ref{th:operational} by interpreting the positive
compression formula as an output-entropy identity, then establishing a
coding converse and achievability at the actual input mean. All
entropies involving the Haar setting will be defined by finite-dimensional
conditional states.

\paragraph{The output-entropy identity.}
For a fixed $\bsU$, let
$\mathcal H_{d,k}:=\bigoplus_{C,j}\bbC^{m_j(C)}$.
The output retaining $C,j$ and the residual is the density matrix
\begin{eqnarray}\label{eq:operational-fibre}
 \Omega_\omega(\bsU):=\frac1{L_{d,k}}
                 \bigoplus_{C,j}\bsA_{\bsU,C,j}(\omega),\qquad
 D_{\rm out}:=\dim\mathcal H_{d,k}=L_{d,k}d.
\end{eqnarray}
It has trace one. The complete output is the measurable field
$\bsU\mapsto\Omega_\omega(\bsU)$ together with the classical setting
$\bsU$ distributed according to normalized Haar measure $\mu_{\haar}$.
This specifies the classical--quantum channel: the probability of a
measurable setting set $E$ is $\mu_{\haar}(E)$, independently of $\omega$, and
the conditional state at $\bsU$ is \eqref{eq:operational-fibre}. Define
\begin{eqnarray}\label{eq:operational-reference-entropy}
 \mathfrak h(\cN_{d,k}(\omega))
 :=\int_{\op{U}(d)}\S(\Omega_\omega(\bsU))\dif\mu_{\haar}(\bsU).
\end{eqnarray}
The integrand lies between zero and $\log D_{\rm out}$. For an ensemble
$\{p_a,\omega_a\}$ with mean $\bar\omega$, its output Holevo information
is, by definition, the integral of the conditional Holevo information:
\begin{eqnarray}\label{eq:operational-chi-definition}
 \chi(\{p_a,\cN_{d,k}(\omega_a)\})
 =\mathfrak h(\cN_{d,k}(\bar\omega))
       -\sum_a p_a\mathfrak h(\cN_{d,k}(\omega_a)).
\end{eqnarray}
The setting $\bsU$ is independent of $a$, so it supplies no additional
message information. This definition avoids assigning a discrete
entropy to the continuous Haar record. For $n$ uses, the same definitions
use product Haar measure and the finite output space
$\mathcal H_{d,k}^{\ot n}$ of dimension $D_{\rm out}^n$.
More explicitly, let $\Phi_{\bsU}$ be the channel
$\omega\mapsto\Omega_\omega(\bsU)$.
For an arbitrary, possibly entangled $n$-site input $\omega$, its
conditional block output is
\begin{eqnarray*}
 \Omega_\omega^{(n)}(\bsU^n)
 :=(\Phi_{\bsU_1}\ot\cdots\ot\Phi_{\bsU_n})(\omega).
\end{eqnarray*}
A collective decoder is a field of POVMs
$\{\bsD_w(\bsU^n)\}_{w=1}^{M_n}$ whose matrix entries are measurable
functions of $\bsU^n$, with
\begin{eqnarray*}
 \bsD_w(\bsU^n)\geqslant\zero,\qquad
 \sum_{w=1}^{M_n}\bsD_w(\bsU^n)=\I_{D_{\rm out}^n}
 \quad\text{for almost every }\bsU^n.
\end{eqnarray*}
For the codewords in Section~\ref{sect:application}, the actual average
decoding error is therefore
\begin{eqnarray}\label{eq:operational-decoder-error}
 \epsilon_n
 =1-\frac1{M_n}\sum_{w=1}^{M_n}\int
 \Tr{\bsD_w(\bsU^n)\Omega_{\omega_w^{(n)}}^{(n)}(\bsU^n)}
 \dif\mu^n_{\haar}(\bsU^n).
\end{eqnarray}
The conditional output is block diagonal in the classical flags
$(C,j)^n$. Pinching every POVM effect in these same blocks preserves
positivity, completeness and all decoding probabilities. Thus reading
the classical flags before a conditional collective measurement on the
residual systems realizes the same decoding statistics.

For $k\geqslant2$, cut sets are in bijection with ordered positive
compositions $(m_1,\ldots,m_k)$ of $d$. Fixing one part to be $\ell$
leaves $\binom{d-\ell-1}{k-2}$ compositions of the remaining parts.
Consequently, for $1\leqslant\ell\leqslant d-k+1$,
\begin{eqnarray}\label{eq:operational-block-count}
 \bbE_C\#\{j:m_j(C)=\ell\}
 =\frac{k}{L_{d,k}}\binom{d-\ell-1}{k-2}
 =\mathfrak C_{d,k,\ell}.
\end{eqnarray}
For a fixed interval of length $\ell$, $\bsU\bsV_{C,j}$ has the
normalized Haar distribution on $\op{St}(d,\ell)$. Expanding the
direct-sum entropy in \eqref{eq:operational-fibre} and applying
Proposition~\ref{prop:posrep} therefore gives
\begin{eqnarray}\label{eq:operational-entropy-identity}
 \mathfrak h(\cN_{d,k}(\omega))
 &=&\log L_{d,k}
       +\sum_{\ell=1}^{d-k+1}\mathfrak C_{d,k,\ell}\G_\ell(\omega)
 \nonumber\\
 &=&h_{d,k}+\S_k^{(d)}(\omega),\qquad
 h_{d,k}:=\log L_{d,k}+H_k-1.
\end{eqnarray}
For $k=1$, the identity follows directly from unitary invariance, with
$h_{d,1}=0$. Every pure input has $\S_k^{(d)}=0$, so
\eqref{eq:operational-chi-definition} proves
\eqref{eq:operational-holevo}. Continuity and concavity of the right-hand
side also follow directly from the finite-dimensional entropy integral.

\paragraph{Converse, including entangled codewords.}
Write $Z=(\bsU,C)$ for the setting and $J$ for the outcome, and denote
Shannon entropy by $\H$. The function
\begin{eqnarray*}
 f(\sigma):=\bbE_Z\Br{\H\Pa{ (\Tr{\bsA_{\bsU,C,j}(\sigma)})_{j=1}^k}}
\end{eqnarray*}
is concave in $\sigma$. For a pure input every nonzero residual is pure.
The direct-sum entropy then gives
$\mathfrak h(\cN_{d,k}(\sigma))=\log L_{d,k}+f(\sigma)$, and hence
$f(\sigma)=H_k-1$. A pure-state decomposition yields
$f(\sigma)\geqslant H_k-1$ for every state.

For an arbitrary $n$-site input $\omega$, condition on the earlier
settings and outcomes $h=(z_{<t},j_{<t})$ and let $\sigma_{t|h}$ be the
normalized conditional state of input site $t$. More explicitly, apply
the earlier branch operators
$\bsK_{z_s,j_s}=\bsV_{C_s,j_s}^\dagger\bsU_s^\dagger$ on sites $s<t$,
leave all other sites unchanged, normalize, and take the $t$-th marginal.
These conditional states are defined for almost every history with
nonzero conditional branch probability; null histories may be assigned
arbitrary states. Summing all future outcomes is trace preserving, so
neither this conditional state nor the probability of the past depends
on future settings. With
$\Pi_{z,j}=\bsU\bsV_{C,j}\bsV_{C,j}^\dagger\bsU^\dagger$, it follows
almost everywhere that
\begin{eqnarray*}
 \Pr\Set{J_t=j\mid J_{<t},Z^n}  =\Tr{\Pi_{Z_t,j}\sigma_{t|h}}.
\end{eqnarray*}
The fresh setting $Z_t$ retains its independent prescribed distribution
conditional on $h$. The conditional entropy chain rule therefore gives
\begin{eqnarray}\label{eq:operational-history-bound}
 \H(J^n|Z^n)=\sum_{t=1}^n\bbE_h [f(\sigma_{t|h})]   \geqslant n(H_k-1).
\end{eqnarray}
If $\omega_{Z^n,J^n}$ denotes the normalized final residual, expanding
the block output entropy yields
\begin{eqnarray*}
 \mathfrak h(\cN_{d,k}^{\ot n}(\omega))
 =n\log L_{d,k}+\H(J^n|Z^n)
       +\bbE_{Z^n,J^n}[\S(\omega_{Z^n,J^n})]
 \geqslant nh_{d,k}.
\end{eqnarray*}
Entropy subadditivity at fixed $\bsU^n$, followed by integration, gives
the complementary bound
\begin{eqnarray*}
 \mathfrak h(\cN_{d,k}^{\ot n}(\omega))
 \leqslant\sum_{t=1}^n\mathfrak h(\cN_{d,k}(\omega_t)).
\end{eqnarray*}
For a code, put $\bar\omega_n=M_n^{-1}\sum_w\omega_w^{(n)}$ and
$\rho_{n,t}=(\bar\omega_n)_t$. Apply the last upper bound to
$\bar\omega_n$ and the lower bound to each codeword. Equations
\eqref{eq:operational-chi-definition} and
\eqref{eq:operational-entropy-identity}, followed by concavity, imply
\begin{eqnarray}\label{eq:operational-block-converse}
 \chi_n\leqslant\sum_{t=1}^n\S_k^{(d)}(\rho_{n,t})
       \leqslant n\S_k^{(d)}(\tau_n).
\end{eqnarray}
Conditioned on each $\bsU^n$, the ordinary finite-dimensional Holevo
bound applies to any collective decoder. If $W$ is the uniform message
and $\widehat W$ is the decoded message, independence of $W$ and
$\bsU^n$, followed by integration of the conditional Holevo bound, gives
\begin{eqnarray*}
 I(W;\widehat W)
 \leqslant I(W;\bsU^n,\widehat W)
 =I(W;\widehat W\mid\bsU^n)\leqslant\chi_n.
\end{eqnarray*}
Fano's inequality, with the error in
\eqref{eq:operational-decoder-error}, now gives
\begin{eqnarray}\label{eq:operational-fano}
 (1-\epsilon_n)\log M_n
 \leqslant n\S_k^{(d)}(\tau_n)+h_2(\epsilon_n),
\end{eqnarray}
where $\epsilon_n$ is average error and
$h_2(x)=-x\log x-(1-x)\log(1-x)$. Continuity and
$\norm{\tau_n-\rho}_1\to0$ prove the capacity upper bound.

\paragraph{Achievability at the prescribed mean.}
Take a finite pure-state ensemble $\{p_a,\proj{\psi_a}\}$ of mean
$\rho$, for example its spectral ensemble. Its output Holevo information
is $\S_k^{(d)}(\rho)$. To apply finite-dimensional coding, partition
the compact group $\op{U}(d)$ into finitely many measurable cells $E$
of operator-norm diameter at most $\delta$. Discard null cells and put
\begin{eqnarray*}
 q_E=\mu_{\haar}(E),\qquad
 \bar\Omega_{\omega,E}=q_E^{-1}\int_E
                    \Omega_\omega(\bsU)\dif\mu_{\haar}(\bsU),\qquad
 \cN_\delta(\omega)=\bigoplus_Eq_E\bar\Omega_{\omega,E}.
\end{eqnarray*}
The receiver obtains this finite-dimensional channel by keeping the
cell label instead of the exact setting. Trace-norm contractivity of
the fixed cut-and-outcome channel gives, uniformly for input states,
\begin{eqnarray*}
 \norm{\Omega_\omega(\bsU)-\Omega_\omega(\bsW)}_1
 &\leqslant&\norm{\bsU^\dagger\omega\bsU
                       -\bsW^\dagger\omega\bsW}_1
 \leqslant2\norm{\bsU-\bsW}_{\rm op}.
\end{eqnarray*}
Within a cell the half trace distance to its average is therefore at
most $\delta$. Uniform continuity of entropy on
$\rD(\bbC^{D_{\rm out}})$ shows that the ensemble Holevo information
of $\cN_\delta$ tends to $\S_k^{(d)}(\rho)$ as $\delta\downarrow0$:
apply continuity to the mean output and every signal output within
each cell, then average with weights $q_E$. The input-independent
Shannon entropy of $(q_E)_E$ cancels from the Holevo difference.
Only the fixed dimension $D_{\rm out}$ is used in this continuity
argument. If $d=1$, the channel and the zero capacity are trivial.

For any $0<R<\S_k^{(d)}(\rho)$, first fix a partition whose ensemble
Holevo information exceeds $R$. The finite-dimensional
Holevo--Schumacher--Westmoreland random-coding theorem
\cite{Holevo1998,Schumacher1997} gives iid product codebooks with
$M_n=\lfloor\exp(nR)\rfloor$ and expected average decoding error
tending to zero. Let $\widehat p_a$ be the empirical letter frequency
over all $nM_n$ positions. The realized aggregate mean is
$\tau_n=\sum_a\widehat p_a\proj{\psi_a}$, so
\begin{eqnarray*}
 \bbE\norm{\tau_n-\rho}_1
 \leqslant\sum_a\bbE|\widehat p_a-p_a|
 \leqslant\sum_a\sqrt{\frac{p_a(1-p_a)}{nM_n}}
 \longrightarrow0.
\end{eqnarray*}
Select a deterministic codebook and its decoder for which the sum of
decoding error and mean-state deviation is at most twice its expectation.
Both required limits hold simultaneously. The decoder for $\cN_\delta$
is implemented on the original output by the same cell coarse graining;
its classical flags can be read before the conditional collective
measurement on the residuals. If $\S_k^{(d)}(\rho)=0$, the one-message
input $\rho^{\ot n}$ attains rate zero with exact mean $\rho$. This
proves \eqref{eq:operational-capacity}.

\paragraph{Refinement of the measurement and the capacity differences.}
For $k<d$, given $(\bsU,C,j)$ and its residual, choose one of the
$d-k$ uncut gaps uniformly. If it lies inside the occupied interval,
measure its two subintervals; otherwise retain the residual. Report
the enlarged cut set $C'$ and the final interval and erase the old
history. For a final interval $I'\subseteq I$, the coordinate restriction
$\bsV_{I'}^\dagger\bsV_I$ composed with the old branch operator is
$\bsV_{I'}^\dagger\bsU^\dagger$. Thus each parent produces exactly
the required final quantum branch.

For completeness, this operation has a Kraus representation on the
whole finite output space $\mathcal H_{d,k}$. If $I$ is an interval
determined by $C$, let $\iota_{C,I}:\bbC^{|I|}\to\mathcal H_{d,k}$
be its block inclusion, and write
$\bsV_I:\bbC^{|I|}\to\bbC^d$ for the coordinate inclusion.
For $g\notin C$, set $C'=C\cup\{g\}$. For each interval $I'$ of
$C'$ contained in $I$, let $\iota_{C',I'}$ denote its block inclusion
into $\mathcal H_{d,k+1}$, and define
\begin{eqnarray}\label{eq:operational-refinement-kraus}
 \bsW_{C,I,g,I'}
 :=\frac1{\sqrt{d-k}}\,
 \iota_{C',I'}(\bsV_{I'}^\dagger\bsV_I)\iota_{C,I}^\dagger.
\end{eqnarray}
For fixed $C,I,g$, the final intervals contained in $I$ partition $I$.
Consequently,
\begin{eqnarray*}
 \sum_{I'\subseteq I}
 (\bsV_{I'}^\dagger\bsV_I)^\dagger
 (\bsV_{I'}^\dagger\bsV_I)=\I_{|I|},
\end{eqnarray*}
where the sum is over intervals of $C'$. Summing over all indicated
indices gives
\begin{eqnarray}\label{eq:operational-refinement-complete}
 \sum_{C,I,g,I'}\bsW_{C,I,g,I'}^\dagger\bsW_{C,I,g,I'}
 =\bigoplus_{C,I}\frac1{d-k}\sum_{g\notin C}\I_{|I|}
 =\I_{\mathcal H_{d,k}}.
\end{eqnarray}
Thus
$\bsX\mapsto\sum_{C,I,g,I'}\bsW_{C,I,g,I'}\bsX
\bsW_{C,I,g,I'}^\dagger$
is completely positive and trace preserving on every matrix on
$\mathcal H_{d,k}$. Applying this fixed map to each conditional fibre
and retaining $\bsU$ defines the receiver channel $\cT_k$.

Every size-$k$ cut set $C'$ has
$k$ parents, and its total weight is
\begin{eqnarray*}
 \frac{k}{\binom{d-1}{k-1}(d-k)}
 =\frac1{\binom{d-1}{k}}.
\end{eqnarray*}
This is the uniform law required for $\cN_{d,k+1}$ and proves
\eqref{eq:operational-degradation}. At a maximally mixed mean,
\eqref{eq:operational-capacity} and the uniform entropy value give the
stated loss $H_{k+1}-H_k=1/(k+1)$. For the qutrit example, the only
rank multiset is $(1,2)$. The reported setting identifies the residual
with the corresponding laboratory subspace; conditional on this ray
and plane, the remaining basis choice is input-independent randomness.
It may be discarded or regenerated, yielding the stated equivalent
description.

Finally, for every padded state,
$C_{\sigma^{[D]}}(\cN_{D,k})=\S_k^{(D)}(\sigma)$ by
\eqref{eq:operational-capacity}. Substitute this identity in
\eqref{eq:gap-definition} and \eqref{eq:kSSAdefect}. Theorems
\ref{th:subadditivity} and \ref{th:equality}, the bounds in
Section~\ref{subsect:QB}, and Corollary~\ref{cor:markov-signs}
give every assertion of Corollary~\ref{cor:operational-gap}.

\section{Concluding remarks}\label{sect:5}

In a common ambient dimension, the Nichols--Wootters family preserves
bipartite subadditivity throughout the interpolation from entropy to subentropy, whereas strong
subadditivity is confined to the von Neumann endpoint. At subentropy,
the results settle the general subadditivity conjecture of
Datta \emph{et al.}~\cite{Datta2014} and answer the strong-subadditivity
question of Zhang \emph{et al.}~\cite{Zhang2017a} negatively.
For every $k\geqslant2$, the equality condition and the positive quantitative
gap distinguish this family from von Neumann mutual information even
on mixed product states. Near saturation forces trace-distance proximity
to a product state with a pure marginal, and the SSA defect takes both
signs even at zero conditional mutual information. The positive compression representation
accounts for this distinction and connects it to entropy gains under
quantum channels.

The operational application identifies the same entropy family with the
classical communication capacity retained by a specified isotropic partial
measurement, from the reversible channel at $k=1$ to the random-basis
measurement at $k=d$. Its common-dimension capacity differences are exactly
the bipartite gap and the tripartite defect studied here. Thus the entropy
inequalities also describe quantitative constraints on information retained
by this detector at different prescribed input means.

Several questions remain. First, the optimal constants multiplying the
marginal purity product are open in general dimensions. For two-qubit
states, the sharp coefficient $c_{D,k}$ in
\eqref{eq:twoqubit-all-D} concerns $ab$, whereas the purity product is
$4ab(1-a)(1-b)$. It places the optimal purity-product coefficient
between $\frac14c_{D,k}$ and $c_{D,k}$ but does not determine it.
Second, it remains to decide whether the pinching in the recovery
lower bound of Theorem~\ref{th:channel} can be removed for arbitrary
unital channels when $d\geqslant4$ and $2\leqslant k\leqslant d-2$, and whether the
padded exchange bound \eqref{eq:exchange-Sk} extends to every unital
channel for $1<k<d$.
Finally, the exact maximum of the SSA defect over arbitrary tripartite
states remains open, as do a complete classification of its sign and
its extrema within the quantum Markov class $\sfI(A:C|B)=0$. Theorems~\ref{th:FSSA} and
\ref{th:rank-max} give the exact defect and rank-constrained extrema on
the pure-bipartite-factor family.

\subsubsection*{Acknowledgments}

The authors used AI tools, namely ChatGPT Pro (GPT-6 Pro), Codex (gpt-6.1-sol) and Claude Opus 5.5 (via Claude Code), for literature search, computation and double-checking, including formal verification of the results in Lean. All proofs can be checked by hand, and the computations can be reproduced from the verification package. All outputs were manually verified, and the authors take full responsibility for the work.

\newpage
\appendix
\appendixpage
\addappheadtotoc

\section{Structural identities and their proofs}\label{app:A}

\subsection{Eigenvalue and divided-difference formula}\label{app:eig}

Because $\ee_k(*)$ is multilinear in the eigenvalues of its argument, and the eigenvalues of $\bsR_\rho(z)$ are $\frac{z}{z-\lambda_i}$, one has
\begin{eqnarray}\label{eq:ekRrhoz}
\ee_k[\bsR_\rho(z)]=\sum_{\abs{\cI}=k} \prod_{i\in\cI}\frac{z}{z-\lambda_i} = \sum_{\abs{\cI}=k} \frac{z^k}{\prod_{i\in\cI}(z-\lambda_i)} ,
\end{eqnarray}
the sum running over $k$-element subsets $\cI\subseteq [d]:=\set{1,\ldots,d}$. Inserting Eq.~\eqref{eq:ekRrhoz} into Eq.~\eqref{eq:EntkSk} and applying the residue theorem at $z=\lambda_i$ gives the eigenvalue formula.
\begin{prop}[Eigenvalue formula]\label{prop:eigfor}
For distinct nonzero eigenvalues $\bsx=(x_1,\ldots,x_k)$,
\begin{eqnarray}\label{eq:Phik1}
\Phi_k(x_1,\ldots,x_k):=-\sum^{k}_{i=1}p_i(\bsx) \log x_i,\qquad p_i(\bsx)\equiv\frac{x^k_i}{\prod_{j\in \hat i}(x_i-x_j)},
\end{eqnarray}
where $\hat i:=[k]\backslash\set{i}$ for $1\leqslant i\leqslant k$. Then
\begin{eqnarray}\label{eq:Entkrho}
\op{Ent}_k(\rho)=\sum_{\abs{\cI}=k}\Phi_k(\bdlambda_{\cI}),
\end{eqnarray}
and repeated or zero eigenvalues are handled by continuous extension. Equivalently, $\Phi_k$ is the divided difference
\begin{eqnarray}\label{eq:divdiff}
\Phi_k(x_1,\ldots,x_k)=-[x_1,\ldots,x_k]\Pa{x^k\ln x}.
\end{eqnarray}
In particular,
\begin{eqnarray}\label{eq:Phi12}
\Phi_1(x)=-x\log x,\qquad \Phi_2(x,y)=-\frac{x^2\log x-y^2\log y}{x-y}.
\end{eqnarray}
\end{prop}

\begin{proof}
By multilinearity of $\ee_k$ in the eigenvalues,
\begin{eqnarray*}
\ee_k[\bsR_\rho(z)]=\sum_{|\cI|=k}\prod_{i\in\cI}\frac{z}{z-\lambda_i}=\sum_{|\cI|=k}\frac{z^{k}}{\prod_{i\in\cI}(z-\lambda_i)} .
\end{eqnarray*}
Each term is a rational function of $z$ whose only poles are at the eigenvalues $\lambda_i, i\in\cI$. Thus $\ee_k[\bsR_\rho(z)]$ has poles only at the nonzero eigenvalues of $\rho$ (a zero eigenvalue $\lambda_i=0$ gives $\frac{z}{z-0}=1$, no pole).

For a fixed subset $\cI=\set{i_1,\dots,i_k}$ with distinct nonzero eigenvalues, the partial-fraction (Lagrange interpolation) identity is
\begin{eqnarray*}
\frac{z^{k}}{\prod_{i\in\cI}(z-\lambda_i)}
=1+\sum_{i\in\cI}\frac{\lambda_i^{k}}{\prod_{j\in\cI\setminus\set{i}}(\lambda_i-\lambda_j)}\cdot\frac{1}{z-\lambda_i}.
\end{eqnarray*}
Indeed, let $p(z)=\prod_{i\in\cI}(z-\lambda_i)$. Since both the numerator $z^k$ and the denominator $p(z)$ have the same degree, write
$\frac{z^{k}}{\prod_{i\in\cI}(z-\lambda_i)}
=1+\frac{q(z)}{p(z)}$, where $\deg (q)\leqslant k-1$. Decompose the proper rational part:
$\frac{q(z)}{p(z)} = \sum_{i\in\cI} \frac{A_i}{z-\lambda_i}$. The coefficients are given by the residue formula:
$$
A_i = \lim_{z\to \lambda_i}(z-\lambda_i)\frac{z^k}{p(z)} =\frac{\lambda^k_i}{ \prod_{j\in\cI\setminus\set{i}}(\lambda_i-\lambda_j)}.
$$
Based on this observation, we see that
\begin{eqnarray}\label{eq:equivalentform2.5}
\ee_k[\bsR_\rho(z)]=\binom{d}{k} + \sum_{|\cI|=k}\ \sum_{i\in\cI}\frac{p_i^{(\cI)}}{z-\lambda_i},\qquad p_i^{(\mathcal I)}=\frac{\lambda_i^{k}}{\prod_{j\in\cI\setminus\set{i}}(\lambda_i-\lambda_j)} .
\end{eqnarray}
Insert \eqref{eq:equivalentform2.5} into the contour integral:
\begin{eqnarray*}
\op{Ent}_k(\rho)&=&-\frac{1}{2\pi \mathrm i}\oint \Pa{\binom{d}{k} + \sum_{|\cI|=k}\sum_{i\in\cI}\frac{p_i^{(\cI)}}{z-\lambda_i}}(\log z)\dif z\\
&=& -\frac{1}{2\pi \mathrm i}\binom{d}{k} \oint \log z\dif z -\frac{1}{2\pi \mathrm i}\oint \Pa{\sum_{|\cI|=k}\sum_{i\in\cI}\frac{p_i^{(\cI)}}{z-\lambda_i}}(\log z)\dif z\\
&=&0 - \sum_{|\cI|=k}\sum_{i\in\cI} p_i^{(\cI)} \Br{\frac{1}{2\pi \mathrm i}\oint \frac{\log z}{z-\lambda_i}\dif z},
\end{eqnarray*}
where the first term vanishes because $\oint \log z\dif z=0$. Indeed, $\log z$ possesses an antiderivative on any domain that excludes $0$ and does not cross the branch cut; choosing the cut outside the contour (e.g. along the negative real axis, with the contour lying in the right half-plane, so that it encloses only the nonzero eigenvalues $\lambda_i$) makes $\log z$ analytic on and inside the contour. Hence its integral over the closed contour vanishes by Cauchy's theorem. For the remaining term, the integrand is analytic inside the contour except for the simple poles at $z=\lambda_i$. Applying the residue theorem, and using that the contour is positively oriented,
\begin{eqnarray*}
\frac{1}{2\pi \mathrm i}\oint \frac{\log z}{z-\lambda_i}\dif z=\lim_{z\to \lambda_i}(z-\lambda_i) \frac{\log z}{z-\lambda_i}=\log\lambda_i,
\end{eqnarray*}
by the residue theorem. Summing,
\begin{eqnarray*}
\op{Ent}_k(\rho)=\sum_{|\cI|=k}\Pa{-\sum_{i\in\cI}p_i^{(\cI)}\log\lambda_i}=\sum_{|\cI|=k}\Phi_k(\bdlambda_{\cI}),
\end{eqnarray*}
where we defined
\begin{eqnarray*}
\Phi_k(x_1,\dots,x_k):=-\sum_{i=1}^k\frac{x_i^{k}}{\prod_{j\neq i}(x_i-x_j)} \log x_i.
\end{eqnarray*}
Recall that the divided difference of a function $f$ at distinct points $x_1,\dots,x_k$ is
$$
[x_1,\dots,x_k]f=\sum_{i=1}^k\frac{f(x_i)}{\prod_{j\neq i}(x_i-x_j)} .
$$
Taking $f(x)=x^k\log x$, we get
$$
[x_1,\dots,x_k](x^k\log x)=\sum_{i=1}^k\frac{x_i^{k}\log x_i}{\prod_{j\ne i}(x_i-x_j)}=-\Phi_k(x_1,\dots,x_k).
$$
Hence this is the claimed divided-difference formula \eqref{eq:divdiff}. The explicit expressions of $\Phi_k(k=1,2)$ follow directly.

Finally, consider repeated and zero eigenvalues. The right-hand side of \eqref{eq:Entkrho} is a finite sum of $\Phi_k$ evaluated on $k$-subsets of the spectrum. The function $x^k\log x$, extended by zero at $x=0$, belongs to $C^{k-1}([0,\infty))$ and is smooth on $(0,\infty)$. Divided differences of order $k-1$ of a $C^{k-1}$ function extend continuously to coincident arguments. Concretely, for repeated eigenvalues one takes limits; e.g.
$$
\lim_{y\to x}\Phi_2(x,y)=-\frac{\dif}{\dif x}(x^2\log x)=-(2x\log x+x),
$$
which is the derivative form of the divided difference. For a zero eigenvalue, $\frac{z}{z-0}=1$, so the corresponding factor contributes no pole and no $\log 0$ term arises in the residue computation; equivalently $\eta(0)=0$ and the continuous extension of $\Phi_k$ gives the correct limit. Thus \eqref{eq:Entkrho} holds for arbitrary spectra by continuity, and the contour integral \eqref{eq:EntkSk} is well defined because $\ee_k[\bsR_\rho(z)]$ has no pole at $z=0$.

For clarity, the continuity of the genuine contour integral at positive
repeated nodes can be seen on a fixed contour. Given positive nodes, put
$m=\min_i x_i$ and $M=\max_i x_i$. The positively oriented circle with
center $M$ and radius $M-m/2$ encloses every node, and its closed disc
lies in $\Re [z]\geqslant m/2>0$. For all sufficiently close positive node
tuples, every pole remains separated from this circle by a common
positive distance. The rational integrands, multiplied by $\log z$,
therefore converge uniformly on the circle, so their contour integrals
converge as the nodes coalesce. At an exactly zero node, cancel the
factor $z/(z-0)=1$ before integrating. This reduces the block to its
positive support; the corresponding identity for divided differences is
proved in Lemma~\ref{applem:0padding}. For an all-zero block, the
rational factor is identically one; choosing the contour in the right
half-plane gives $\oint\log z\,\dif z=0$ by Cauchy's theorem.
Thus the support reduction and the continuous divided-difference extension
give the same value also at zero nodes; no continuity of $\log z$ at zero
is required.

The present proposition as stated concerns $\op{Ent}_k$; the normalized $\S_k$ is obtained by multiplying by $\binom{d-1}{k-1}^{-1}$, which does not affect the eigenvalue formula. For $k=1$,
\begin{eqnarray*}
\S_1(\rho)=\binom{d-1}{0}^{-1}\op{Ent}_1(\rho)=\sum^d_{i=1}\Phi_1(\lambda_i)=-\sum_{i=1}^d\lambda_i\log\lambda_i=\S(\rho),
\end{eqnarray*}
recovering the von Neumann entropy. For $k=d$, $\Phi_d(\lambda_1,\ldots,\lambda_d)$ is the subentropy summand, and summing over the single subset $[d]:=\set{1,\dots,d}$ gives the subentropy $\Q(\rho)$ (in the contour form of Jozsa--Mitchison \cite{Jozsa2015}). This confirms that the formula interpolates correctly between the two endpoints.
\end{proof}

Formula \eqref{eq:Entkrho} is the form used in the quantitative arguments: it exhibits $\S_k$ as a finite sum of divided differences of the single-variable function $x^k\log x$, each of which is continuous on the closed nonnegative simplex, smooth in its positive interior, and has an explicit integral representation.

\subsection{Simplex representation}\label{app:simrep}

The Hermite--Genocchi formula \cite{Boor2005} for divided differences, together with
\begin{eqnarray}\label{eq:k-1derivative}
\frac{\dif^{k-1}}{\dif x^{k-1}}\Pa{x^k\log x}=k! x(\log x+H_k-1),
\end{eqnarray}
yields a representation of $\Phi_k$ as an average over the probability simplex.

\begin{prop}[Simplex representation]\label{prop:simplexrep}
Let $\bsT\equiv(T_1,\ldots,T_k)$ be uniformly distributed random vector on the $(k-1)$-dimensional probability simplex
$$
\Delta_{k-1}:=\Set{\bst\equiv(t_1,\ldots,t_k)\in\bbR^k_{\geqslant0}:\Inner{\bse}{\bst}=1},\quad\bse=(1,\ldots,1)\in\bbR^k,
$$
equipped with the normalized probability measure $\Gamma(k)\delta(1-\Inner{\bse}{\bst})[\dif \bst]$, where $[\dif\bst]$ is the Lebesgue volume element in $\bbR^k$, $\delta$ is Dirac delta function, and $\Gamma$ is the Gamma function. Then
\begin{eqnarray}\label{eq:Phik2}
\Phi_k(\bsx)=k\cdot \bbE_{\bsT}\Br{\eta\Pa{\Inner{\bsT}{\bsx}}} - (H_k-1)\Inner{\bse}{\bsx},
\end{eqnarray}
where $\bsx=(x_1,\ldots,x_k)\in \bbR^k_{\geqslant0}$. Choose a uniformly random $k$-element subset
$\cI=\set{i_1,\ldots,i_k}\subseteq[d]$, independently of
$\bsT=(T_1,\ldots,T_k)$, and set
$X=\sum_{j=1}^kT_j\lambda_{i_j}=\Inner{\bsT}{\bdlambda_{\cI}}$.
Then
\begin{eqnarray}\label{eq:Skrho}
S_k(\rho)=d\cdot \bbE_{\cI,\bsT}[\eta(\Inner{\bsT}{\bdlambda_{\cI}})]-(H_k-1)=d\cdot \bbE[\eta(X)]-(H_k-1).
\end{eqnarray}
\end{prop}

\begin{proof}
Recall from Proposition~\ref{prop:eigfor} that
\begin{eqnarray*}
\Phi_k(x_1,\ldots,x_k)=-[x_1,\ldots,x_k](x^k\log x) = -\sum^k_{i=1} \frac{x^k_i\log x_i}{\prod_{j\neq i}(x_i-x_j)}
\end{eqnarray*}
for distinct $x_i$; the general case follows by continuity.

The Hermite-Genocchi formula for divided differences states that for $f\in C^{k-1}$ on an interval containing the points $x_1,\ldots,x_k$,
\begin{eqnarray*}
[x_1,\ldots,x_k]f&=&\int_{\bbR^k_{\geqslant0}} f^{(k-1)}\Pa{\sum^k_{i=1}t_ix_i}\delta\Pa{1-\sum^k_{i=1}t_i}[\dif \bst]\\
&=&\frac1{\Gamma(k)}\bbE_{\bsT}\Br{f^{(k-1)}\Pa{\Inner{\bsT}{\bsx}}},
\end{eqnarray*}
where the expectation uses the normalized probability measure $\Gamma(k)\delta\Pa{1-\Inner{\bse}{\bst}}[\dif\bst]$ supported on $\Delta_{k-1}$.

Applying the Hermite--Genocchi formula to $f(z)=z^k\log z$, together with \eqref{eq:k-1derivative}, we get that
\begin{eqnarray*} 
&&-[x_1,\ldots,x_k]\Pa{z^k\log z} =- \frac1{\Gamma(k)}\bbE_{\bsT}\Br{\Pa{\frac{\dif^{k-1}}{\dif z^{k-1}}\Pa{z^k\log z}}\Big|_{z=\Inner{\bsT}{\bsx}}}\\
 &&=-\frac1{\Gamma(k)}\bbE_{\bsT}\Br{k! z(\log z+H_k-1)\Big|_{z=\Inner{\bsT}{\bsx}}}\\
 &&= k\bbE_{\bsT}\Br{\eta\Pa{\Inner{\bsT}{\bsx}}} - (H_k-1)k\bbE_{\bsT}\Br{\Inner{\bsT}{\bsx}} \\
 &&=  k\bbE_{\bsT}\Br{\eta\Pa{\Inner{\bsT}{\bsx}}} - (H_k-1)\Inner{\bse}{\bsx},
\end{eqnarray*}
where we used the fact that $\bbE_{\bsT}[T_i]=\frac1k$ for each $i=1,\ldots,k$. This is exactly \eqref{eq:Phik2}.

Next we derive the state-level formula. Now let $\rho$ have spectrum $\bdlambda=(\lambda_1,\ldots,\lambda_d)$. By Proposition~\ref{prop:eigfor},
$$
\S_k(\rho) = \binom{d-1}{k-1}^{-1}\op{Ent}_k(\rho)=\binom{d-1}{k-1}^{-1}\sum_{|\cI|=k}\Phi_k(\bdlambda_{\cI}),
$$
where $\bdlambda_{\cI}=(\lambda_i)_{i\in\cI}$. Insert \eqref{eq:Phik2} for each subset:
\begin{eqnarray*}
\sum_{|\cI|=k}\Phi_k(\bdlambda_{\cI}) = k\sum_{|\cI|=k} \bbE_{\bsT}\Br{\eta\Pa{\Inner{\bsT}{\bdlambda_{\cI}}}} - (H_k-1)\sum_{|\cI|=k}\Inner{\bse}{\bdlambda_{\cI}}.
\end{eqnarray*}
For the second term: each index $i$ belongs to exactly $\binom{d-1}{k-1}$ of the $k$-element subsets, so
$$
\sum_{|\cI|=k}\Inner{\bse}{\bdlambda_{\cI}}=\binom{d-1}{k-1}\sum_{i=1}^d\lambda_i=\binom{d-1}{k-1}.
$$
For the first term: choose $\cI$ uniformly at random among the $\binom{d}{k}$ subsets, independent of $\bsT=(T_1,\ldots,T_k)$. Then
\begin{eqnarray*}
\binom{d}{k}^{-1} \sum_{|\cI|=k} \bbE_{\bsT}\Br{\eta\Pa{\Inner{\bsT}{\bdlambda_{\cI}}}} =\bbE_{\cI,\bsT}\Br{\eta\Pa{\Inner{\bsT}{\bdlambda_{\cI}}}}\equiv \bbE[\eta(X)],
\end{eqnarray*}
where $X=\Inner{\bsT}{\bdlambda_{\cI}}$ with $\cI$ uniform. Thus
\begin{eqnarray*}
\sum_{|\cI|=k} \bbE_{\bsT}\Br{\eta\Pa{\Inner{\bsT}{\bdlambda_{\cI}}}}  =\binom{d}{k}\bbE[\eta(X)],
\end{eqnarray*}
implying that
\begin{eqnarray*}
\S_k(\rho) &=& \binom{d-1}{k-1}^{-1}\op{Ent}_k(\rho) = \binom{d-1}{k-1}^{-1} \sum_{|\cI|=k}\Phi_k(\bdlambda_{\cI})\\
&=& \binom{d-1}{k-1}^{-1}\Br{k\sum_{|\cI|=k} \bbE_{\bsT}\Br{\eta\Pa{\Inner{\bsT}{\bdlambda_{\cI}}}} - (H_k-1)\sum_{|\cI|=k}\Inner{\bse}{\bdlambda_{\cI}}}\\
&=& \binom{d-1}{k-1}^{-1}\Br{k\binom{d}{k}\bbE[\eta(X)] - (H_k-1) \binom{d-1}{k-1}}\\
&=& d\cdot \bbE[\eta(X)] - (H_k - 1),
\end{eqnarray*}
completing the proof.
\end{proof}

Two consequences are used repeatedly. First, \eqref{eq:Skrho} exhibits $\S_k$ as a concave functional of the spectrum, since $\eta$ is concave and $X$ is linear in $\lambda_i$. Second, it gives the uniform value in Subsection~\ref{app:minmaxhier} since $X$ is deterministic when the spectrum is uniform.

\subsection{The normalized spectral restriction formula}\label{app:specres}

\paragraph{Boundary convention} The normalized restriction below is defined only when $p_{\cI}>0$.
For $p_{\cI}=0$, the entire summand is zero. The calculation for distinct
positive nodes extends to all nonnegative nodes: $\Q$ is continuous and
bounded on the compact $k$-component simplex, so
$$
|p_{\cI}\Q(\bdsigma_{\cI})|\leqslant p_{\cI}\sup_{\Delta_{k-1}}|\Q|\to0,
$$
while $-p_{\cI}\log p_{\cI}\to0$. Continuity of $\Phi_k$ then gives the identity
at every boundary point.

Let $\bdlambda = (\lambda_1, \ldots, \lambda_d)\in\Delta_{d-1}$ be a probability vector. Let $\cI$ be a $k$-element subset of $[d]$, and let $p_{\cI} = \sum_{i \in \cI} \lambda_i$. Define the normalized restriction $\bdsigma_{\cI}$ with components $(\bdsigma_{\cI})_i = \lambda_i / p_{\cI}$ for $i \in \cI$.
Then for distinct, strictly positive eigenvalues in $\cI$:
\begin{eqnarray}\label{appeq:PhivsSubq}
 \Phi_k(\bdlambda_{\cI}) = -\sum_{i \in \cI} \frac{\lambda_i^k}{\prod_{j \in \mathcal{I} \setminus \set{i}} (\lambda_i - \lambda_j)} \log \lambda_i = p_{\cI} \Q(\bdsigma_{\cI}) - p_{\cI} \log p_{\cI},
\end{eqnarray}
where $\Q(\cdot)$ is the subentropy of the normalized $k$-component state $\bdsigma_{\cI}$. The proof of \eqref{appeq:PhivsSubq} can be described as follows:
\begin{itemize}
\item \textbf{Step 1: Write the subentropy in terms of the restricted spectrum.} By definition, the subentropy $\Q$ of a $k$-dimensional state with normalized spectrum $\bdsigma = (\sigma_1, \dots, \sigma_k)\in\Delta_{k-1}$ is given by the eigenvalue formula:
\begin{equation*}
\Q(\bdsigma) = -\sum_{i\in\cI} \frac{\sigma_i^k}{\prod_{j\in\cI\setminus\set{i}} (\sigma_i - \sigma_j)} \log \sigma_i.
\end{equation*}
Substitute $\sigma_i = \lambda_i /p_{\cI}$ into this formula. Note that:
\begin{enumerate}[(1)]
\item The term $\log \sigma_i = \log(\lambda_i / p_{\cI}) = \log \lambda_i - \log p_{\cI}$.
\item The denominator transforms as:
\begin{eqnarray*}
 \prod_{j \in \cI \setminus \set{i}} (\sigma_i - \sigma_j) = \prod_{j \in \cI \setminus \set{i}} \Pa{\frac{\lambda_i}{p_{\cI}} - \frac{\lambda_j}{p_{\cI}} } = \frac{1}{p_{\cI}^{k-1}} \prod_{j \in \cI\setminus \set{i}} (\lambda_i - \lambda_j).
\end{eqnarray*}
\item The numerator transforms as $\sigma_i^k = (\lambda_i / p_{\cI})^k = \lambda_i^k / p_{\cI}^k$.
\end{enumerate}
Substituting these into the subentropy formula gives:
\begin{eqnarray*}
\Q(\bdsigma_{\cI}) = -\sum_{i \in \cI} \frac{\lambda_i^k / p_{\cI}^k}{\frac{1}{p_{\mathcal{I}}^{k-1}} \prod_{j \in \cI \setminus \set{i}} (\lambda_i - \lambda_j)} (\log \lambda_i - \log p_{\cI}).
\end{eqnarray*}
Simplify the fraction: $p_{\cI}^{k-1} / p_{\cI}^k = 1/p_{\cI}$. Thus:
\begin{eqnarray*}
\Q(\bdsigma_{\cI}) = -\frac{1}{p_{\cI}} \sum_{i \in \cI} \frac{\lambda_i^k}{\prod_{j \in \cI \setminus \set{i}} (\lambda_i - \lambda_j)} (\log \lambda_i - \log p_{\cI}).
\end{eqnarray*}
\item \textbf{Step 2: Multiply by $p_{\cI}$.} Multiply both sides by $p_{\cI}$:
\begin{eqnarray*}
 p_{\cI} \Q(\bdsigma_{\cI}) = -\sum_{i \in \cI} \frac{\lambda_i^k}{\prod_{j \in \cI \setminus \set{i}} (\lambda_i - \lambda_j)} \log \lambda_i + \log p_{\cI} \sum_{i \in \cI} \frac{\lambda_i^k}{\prod_{j \in \cI \setminus \set{i}} (\lambda_i - \lambda_j)}.
\end{eqnarray*}
Notice that the first term on the right-hand side is exactly $\Phi_k(\bdlambda_{\cI})$. So we have:
\begin{eqnarray}\label{eq:QvsPhik}
 p_{\cI} \Q(\bdsigma_{\cI}) = \Phi_k(\bdlambda_{\cI}) + \log p_{\cI} \sum_{i \in \cI} \frac{\lambda_i^k}{\prod_{j \in \cI \setminus \set{i}} (\lambda_i - \lambda_j)}.
\end{eqnarray}
\item \textbf{Step 3: The divided difference identity.} We need to evaluate the sum appearing in the second term of Eq.~\eqref{eq:QvsPhik}:
\begin{eqnarray*}
 \sum_{i \in \cI} \frac{\lambda_i^k}{\prod_{j \in  \cI \setminus \set{i}} (\lambda_i - \lambda_j)}
\end{eqnarray*}
This sum is precisely the $(k-1)$-th divided difference of the function $f(x) = x^k$ evaluated at the points $\set{\lambda_i}_{i \in \cI}$, denoted as $[\bdlambda_{\cI}](x^k)$. By the fundamental property of divided differences for polynomials, if $f(x) = x^k$, its $(k-1)$-th divided difference is a symmetric polynomial of degree $k - (k-1) = 1$. Specifically, it is the sum of the nodes:
\begin{eqnarray*}
 \sum_{i \in \cI} \frac{\lambda_i^k}{\prod_{j \in \cI \setminus \set{i}} (\lambda_i - \lambda_j)} = \sum_{i \in \cI} \lambda_i = p_{\cI}.
\end{eqnarray*}
\item \textbf{Step 4: Final substitution.}
Substitute the result from Step 3 back into Eq.~\eqref{eq:QvsPhik}:
\begin{eqnarray*}
 p_{\cI} \Q(\bdsigma_{\cI}) = \Phi_k(\lambda_{\cI}) + \log p_{\cI} \cdot p_{\cI}.
\end{eqnarray*}
Rearranging the terms to solve for $\Phi_k(\bdlambda_{\cI})$:
$ \Phi_k(\bdlambda_{\cI}) = p_{\cI} \Q(\bdsigma_{\cI}) - p_{\cI} \log p_{\cI}$.
\end{itemize}
This concludes the proof.\qed

\subsection{Scaling and connection with lower-dimensional subentropy}\label{app:scaling}

The definition extends to positive semidefinite matrices $\bsA$ that are not normalized.
\begin{prop}
For a positive semidefinite matrix $\bsA$ acting on $\bbC^d$, we have
\begin{eqnarray}\label{eq:Entscalingrule}
\op{Ent}_k(c\bsA)  = c\op{Ent}_k(\bsA) +\eta(c)\binom{d-1}{k-1}\Tr{\bsA}, \qquad c>0.
\end{eqnarray}
\end{prop}

\begin{proof}
First suppose that the eigenvalues $\alpha_1,\ldots,\alpha_d$ of
$\bsA$ are positive and pairwise distinct. By
Proposition~\ref{prop:eigfor}, its unnormalized $k$-entropy is:
$$
\op{Ent}_k(\bsA) = \sum_{\abs{\cI}=k} \Phi_k(\bdalpha_{\cI}),\qquad \bdalpha_{\cI}=(\alpha_i)_{i\in\cI}
$$
where
$$
\Phi_k(\bdalpha_{\cI}) = -\sum_{i\in\cI}p^{(\cI)}_i\log\alpha_i,\qquad p^{(\cI)}_i = \frac{\alpha^k_i}{\prod_{j\in\cI\setminus\set{i}}(\alpha_i-\alpha_j)}.
$$
Consider the scaled matrix $c\bsA(c>0)$. Its eigenvalues are $c\alpha_1,\ldots,c\alpha_d$. Applying the eigenvalue formula:
\begin{eqnarray*}
\Phi_k(c\bdalpha_{\cI}) &=& - \sum_{i\in\cI}  \frac{(c\alpha_i)^k}{\prod_{j\in\cI\setminus\set{i}}(c\alpha_i-c\alpha_j)} \log(c\alpha_i) = - c\sum_{i\in\cI}  p^{(\cI)}_i (\log c+\log\alpha_i)\\
&=& \eta(c)\sum_{i\in\cI}  p^{(\cI)}_i - c\sum_{i\in\cI}  p^{(\cI)}_i\log\alpha_i=c\Phi_k(\bdalpha_{\cI}) + \eta(c)\sum_{i\in\cI}  p^{(\cI)}_i.
\end{eqnarray*}
For the second term, we use the divided difference property. The weights $p^{(\cI)}_i$ arise from the partial fraction decomposition of the divided difference. Specifically, the sum of the weights is the $(k-1)$-th divided difference of the polynomial $f(x)=x^k$:
\begin{eqnarray*}
\sum_{i\in\cI}p^{(\cI)}_i = \sum_{i\in\cI} \frac{\alpha^k_i}{\prod_{j\in\cI\setminus\set{i}}(\alpha_i-\alpha_j)} = [\bdalpha_{\cI}](x^k).
\end{eqnarray*}
For the monomial $f(x)=x^n$ and $0\leqslant m\leqslant n$, its $m$-th divided difference is a symmetric homogeneous polynomial of degree $n-m$ in the nodes. For $f(x)=x^k$ and $m=k-1$, the divided difference is the sum of the nodes: $[\bdalpha_{\cI}](x^k) = \sum_{i\in\cI}\alpha_i =: p_{\cI}$.
Thus $\sum_{i\in\cI}p^{(\cI)}_i = p_{\cI}$.

Substitute $p_{\cI}$ back into the scaling equation:
$$
\Phi_k(c\bdalpha_{\cI}) = c\Phi_k(\bdalpha_{\cI}) +\eta(c)p_{\cI}
$$
Summing over all $k$-element subsets $\cI$:
\begin{eqnarray*}
\op{Ent}_k(c\bsA) &=& \sum_{\abs{\cI}=k}\Phi_k(c\bdalpha_{\cI}) = c\sum_{\abs{\cI}=k} \Phi_k(\bdalpha_{\cI}) + \eta(c)\sum_{\abs{\cI}=k}p_{\cI}\\
&=& c\op{Ent}_k(\bsA) + \eta(c)\sum_{\abs{\cI}=k}\sum_{i\in\cI}\alpha_i.
\end{eqnarray*}
Each index $i\in[d]$ appears in exactly $\binom{d-1}{k-1}$ subsets of size $k$. Therefore
$$
\sum_{\abs{\cI}=k}\sum_{i\in\cI}\alpha_i = \binom{d-1}{k-1}\sum^d_{i=1}\alpha_i = \binom{d-1}{k-1}\Tr{\bsA}.
$$
Substituting this yields \eqref{eq:Entscalingrule} for positive distinct
spectra. Both sides are continuous in the nonnegative eigenvalues by
Proposition~\ref{prop:eigfor}. Since positive distinct tuples are dense
in the nonnegative orthant, the identity extends to every positive
semidefinite $\bsA$.
\end{proof}

For a $k$-element subset $\cI$ with $p_{\cI}=\sum_{i\in\cI}\lambda_i>0$, let
\begin{eqnarray}
\bdsigma_{\cI}=\diag\Pa{\frac{\lambda_i}{p_{\cI}}:i\in \cI}
\end{eqnarray}
be the normalized $k$-component restriction. Then
\begin{eqnarray}\label{eq:PQentropy}
\Phi_k(\bdlambda_{\cI})=- \sum_{i\in\cI}  \frac{\lambda_i^k}{\prod_{j\in\cI\setminus\set{i}}(\lambda_i-\lambda_j)} \log\lambda_i=p_{\cI} \Q(\bdsigma_{\cI}) - p_{\cI} \log p_{\cI},
\end{eqnarray}
where the displayed fractions are evaluated first at positive distinct
entries and otherwise by their continuous extension from
Proposition~\ref{prop:eigfor}; see Appendix~\ref{app:specres}.
Consequently,
\begin{eqnarray} \label{eq:SkQ}
\op{Ent}_k(\rho)=\sum_{\abs{\cI}=k} p_{\cI}\Pa{\Q(\bdsigma_{\cI})-\log p_{\cI}}.
\end{eqnarray}
Terms with $p_{\cI}=0$ are interpreted as zero. Equivalently, since $\Q=\S_{\abs{\cI}}$ on the $k$-dimensional restriction,
\begin{eqnarray}\label{eq:Sdkrho}
\S^{(d)}_k(\rho) = \binom{d-1}{k-1}^{-1}
\left[\sum_{\abs{\cI}=k}p_{\cI} \S^{(k)}_k(\bdsigma_{\cI}) - \sum_{\abs{\cI}=k}p_{\cI}\log p_{\cI}\right].
\end{eqnarray}
Eqs.~\eqref{eq:SkQ} and \eqref{eq:Sdkrho} express $\S_k$ as a binomially normalized sum of subentropies of the conditional restrictions and the mass terms $-p_{\cI}\log p_{\cI}$. These formulas complement the simplex representation used below.

\subsection{Concavity, Schur concavity, and mixing}\label{app:concavity}

A key result of majorization is that if $\bsy\prec \bsx$ and $f$ is symmetric and concave, then $f(\bsy)\geqslant f(\bsx)$ \cite{Marshall2011}. This gives the following concavity statement:
\begin{prop}[Concavity]\label{prop:concavity}
For every $k\in[d]$, both $\op{Ent}_k$ and $\sfS_k$ are concave on $\rD(\bbC^d)$:
\begin{eqnarray}\label{eq:Skrhoconcave}
\sfS_k(t\rho+(1-t)\sigma)\geqslant t\sfS_k(\rho)+(1-t)\sfS_k(\sigma), \qquad t\in[0,1].
\end{eqnarray}
Moreover they are strictly concave: for $t\in(0,1)$, equality in \eqref{eq:Skrhoconcave} holds only if $\rho=\sigma$.
\end{prop}

\begin{proof}
(1) \emph{Concavity.}  By the simplex representation (Proposition~\ref{prop:simplexrep}, Eq.~\eqref{eq:Skrho}), the $k$-entropy of a state $\rho$ with spectrum $\bdlambda = (\lambda_1, \ldots, \lambda_d)$ is given by: $\S_k(\rho)=d\cdot\bbE[\eta(X)] - (H_k-1)$, where the random variable $X$ is defined as $X = \Inner{\bsT}{\bdlambda_{\cI}}$, which depends on $(\cI,\bsT)$ for given $\bdlambda$. Here, $\cI$ is a uniformly random $k$-element subset of $[d]$, and $\bsT\equiv(T_1, \ldots, T_k)$ is uniformly distributed on the probability simplex $\Delta_{k-1}$, independent of $\cI$. This can be rewritten as
\begin{eqnarray*}
\S_k(\bdlambda) = d \cdot \bbE_{\cI,\bsT}[\eta(\Inner{\bsT}{\bdlambda_{\cI}})] - (H_k - 1),
\end{eqnarray*}
The function $\eta(x) = -x \ln x$ (with $\eta(0)=0$) is strictly concave on $[0, \infty)$. Write $g(\bdlambda)=\bbE_{\cI,\bsT}[\eta(\Inner{\bsT}{\bdlambda_{\cI}})]$. We verify symmetry and concavity separately.
\begin{itemize}
\item Let $\pi$ be any permutation of all $d$ coordinates.
It permutes the uniformly chosen $k$-subset $\cI$ into another uniformly
chosen subset $\pi(\cI)$. Within that subset, the induced reordering of
coordinates preserves the uniform simplex law of $\bsT$. Relabelling
both variables in the expectation therefore gives
$g(\pi\bdlambda)=g(\bdlambda)$.
\item The map $\bdlambda \mapsto g(\bdlambda) = \bbE_{\cI,\bsT}[\eta(\Inner{\bsT}{\bdlambda_{\cI}})]$ is concave on the region where $\Inner{\bsT}{\bdlambda_{\cI}}\geqslant0$ a.s. We show concavity of $\bdlambda \mapsto \eta(\Inner{\bsT}{\bdlambda_{\cI}})$ for fixed $\bsT$. Indeed, fix a realization $\bsT = \bst$. The map $\bdlambda \mapsto\Inner{\bst}{\bdlambda_{\cI}}$ is linear (hence both convex and concave) in $\bdlambda$. Composing the concave function $\eta$ with an affine map preserves concavity: for any $\bdlambda, \bdmu \in \bbR^d_{\geqslant0}$ and $t\in[0,1]$,
$$
\Inner{\bsT}{(t\bdlambda + (1-t)\bdmu)_{\cI}} =t \Inner{\bsT}{\bdlambda_{\cI}} + (1-t)\Inner{\bsT}{\bdmu_{\cI}},
$$
so by the concavity of $\eta$,
$$
\eta(\Inner{\bsT}{(t\bdlambda + (1-t)\bdmu)_{\cI}}) \geqslant t \eta(\Inner{\bsT}{\bdlambda_{\cI}}) + (1-t)\eta(\Inner{\bsT}{\bdmu_{\cI}}).
$$
This holds pointwise for every realization $\bsT$. From preservation of concavity under expectation, taking expectations gives the desired inequality.
Hence $g$ is concave.
\end{itemize}
Let $\rho, \sigma \in \rD(\bbC^d)$ with spectra $\bdlambda(\rho)$ and $\bdlambda(\sigma)$, and let $t \in [0,1]$. Because $\S_k$ is a unitarily invariant spectral function, we can work directly with the spectra. The spectrum of the convex combination $t\rho + (1-t)\sigma$ is not simply the convex combination of the spectra. However, the map from a density matrix to its $k$-entropy is a symmetric concave function of its eigenvalues (and by definition, a spectral function).

To show concavity over the state space $\rD(\bbC^d)$, we use the variational representation. By the spectral theorem, $\bdlambda(t\rho + (1-t)\sigma) \prec t\bdlambda(\rho) + (1-t)\bdlambda(\sigma)$ (majorization) \cite[Theorem~4.3.47]{Horn2015}. Since $\S_k$ is Schur-concave, we have:
\begin{eqnarray*}
\S_k(t\rho + (1-t)\sigma) = \S_k(\bdlambda(t\rho + (1-t)\sigma))  \geqslant \S_k(\diag(t\bdlambda(\rho) + (1-t)\bdlambda(\sigma))).
\end{eqnarray*}
Now, applying the simplex representation to the diagonal state with spectrum $\bdlambda_t = t\bdlambda(\rho) + (1-t)\bdlambda(\sigma)$:
$$
\S_k(\bdlambda_t) = d \cdot \bbE_{\cI,\bsT}\Br{\eta\Pa{\Inner{\bsT}{(\bdlambda_t)_{\cI}}}} - (H_k - 1).
$$
Because $\eta$ is concave and the map $\bdlambda_{\cI} \mapsto \Inner{\bsT}{\bdlambda_{\cI}}$ is linear in $\bdlambda_{\cI}$ for each $\cI$, the function inside the expectation is concave in $\bdlambda_{\cI}$. Thus:
\begin{eqnarray*}
 \eta\Pa{\Inner{\bsT}{t\bdlambda_{\cI}(\rho) + (1-t)\bdlambda_{\cI}(\sigma)}} \geqslant t \eta\Pa{\Inner{\bsT}{\bdlambda_{\cI}(\rho)}} +
(1-t) \eta\Pa{\Inner{\bsT}{\bdlambda_{\cI}(\sigma)}}.
\end{eqnarray*}
Taking the expectation over $(\cI,\bsT)$ preserves the inequality, yielding:
\begin{eqnarray*}
\S_k(\bdlambda_t) \geqslant t \S_k(\bdlambda(\rho)) + (1-t) \S_k(\bdlambda(\sigma)).
\end{eqnarray*}
Since $\S_k(t\rho + (1-t)\sigma) \geqslant \S_k(\bdlambda_t)$, concavity of $\S_k$ on $\rD(\bbC^d)$ follows. The unnormalized $\op{Ent}_k$ is just a positive scalar multiple $\binom{d-1}{k-1} \S_k$, so it is also concave.

(2) \emph{Strict concavity.} To prove strict concavity for $t \in (0,1)$, we analyze the Hessian of $\S_k$ with respect to the eigenvalues. Since the function is smooth on the interior of the simplex, we can compute the second directional derivative.

Let $\bdlambda$ be a strictly positive probability vector ($\lambda_i > 0$ for all $i$). Let $\bsv \neq \zero$ be a real vector such that $\sum_{i=1}^d v_i = 0$ (to remain in the tangent space of the simplex). Consider the perturbation $\bdlambda(\epsilon) = \bdlambda + \epsilon \bsv$. From the representation $\S_k(\bdlambda) = d \cdot \bbE_{\cI,\bsT}[\eta\Pa{\Inner{\bsT}{\bdlambda_{\cI}}}] - (H_k - 1)$, the second derivative with respect to $\epsilon$ is:
\begin{eqnarray*}
 \frac{\dif^2}{\dif\epsilon^2} S_k(\bdlambda(\epsilon)) = d \cdot \bbE_{\cI,\bsT}\Br{\eta''\Pa{\Inner{\bsT}{\bdlambda_{\cI}(\epsilon)}} \Inner{\bsT}{\bsv_{\cI}}^2}.
\end{eqnarray*}
The second derivative of $\eta(x) = -x \ln x$ is $\eta''(x) = -\frac1x$. Since $\lambda_i > 0$ and $T_i \geqslant 0$ (with $\sum^k_{i=1} T_i=1$), the argument of $\eta''$ is strictly positive on a set of positive measure. Thus:
\begin{eqnarray*}
\frac{\dif^2}{\dif\epsilon^2} S_k(\bdlambda(\epsilon)) \Big|_{\epsilon=0} = -d \cdot \bbE_{\cI,\bsT}\Br{\frac{\Inner{\bsT}{\bsv_{\cI}}^2}{\Inner{\bsT}{\bdlambda_{\cI}}}}.
\end{eqnarray*}
Because the denominator is strictly positive almost surely, the expectation is strictly positive as long as the numerator $\Inner{\bsT}{\bsv_{\cI}}^2$ is not identically zero. Since the random variables $T_i$ span the simplex and $\bsv \neq \zero$, the linear form $\Inner{\bsT}{\bsv_{\cI}}$ cannot vanish for almost every $\bsT$. Therefore, the second derivative is strictly negative:
\begin{eqnarray*}
\D^2 \S_k(\bdlambda)[\bsv,\bsv] = -d \cdot \bbE_{\cI,\bsT}\Br{\frac{\Inner{\bsT}{\bsv_{\cI}}^2}{\Inner{\bsT}{\bdlambda_{\cI}}}} < 0
\end{eqnarray*}
for any non-zero tangent vector $\bsv$.

This proves strict spectral concavity in the positive interior. On the
closed simplex, take distinct probability vectors $\bsx,\bsy$ and choose a
coordinate where $\bsx-\bsy$ is nonzero, together with a $k$-subset containing
it. The linear form $\Inner{\bsT}{(\bsx-\bsy)_{\cI}}$ is nonzero on a set
of positive simplex measure. Strict scalar concavity of $\eta$ therefore
makes the averaged Jensen inequality strict, including boundary vectors.
A symmetric strictly concave function is strictly Schur concave between
sorted vectors that are not equal: express a majorized vector as a convex
combination of permutations and apply strict Jensen.

For matrices with distinct sorted spectra, combine the strict spectral
inequality with
$\bdlambda(t\rho+(1-t)\sigma)\prec t\bdlambda(\rho)+(1-t)\bdlambda(\sigma)$.
If the sorted spectra coincide but $\rho\ne\sigma$, then
\[
 \Tr{(t\rho+(1-t)\sigma)^2}
 =\Tr{\rho^2}-t(1-t)\Tr{(\rho-\sigma)^2}<\Tr{\rho^2}.
\]
Here spectral majorization also gives
$\bdlambda(t\rho+(1-t)\sigma)\prec\bdlambda(\rho)$, since
$\bdlambda(\rho)=\bdlambda(\sigma)$. The strict purity inequality
excludes equality of these sorted spectra. Thus the mixture has a different
sorted spectrum, strictly majorized by that of $\rho$. Strict Schur concavity proves strictness in this remaining
case. Consequently equality in \eqref{eq:Skrhoconcave} for $0<t<1$
holds if and only if $\rho=\sigma$.
\end{proof}

\begin{cor}[Schur concavity]\label{cor:concavity}
It holds that $\sfS_k$ is symmetric and concave,
\begin{eqnarray}\label{eq:majorization}
\bdlambda(\sigma)\prec\bdlambda(\rho) \quad\Longrightarrow\quad  \sfS_k(\sigma)\geqslant \sfS_k(\rho).
\end{eqnarray}
Consequently, every unital quantum channel $\cE$ on a fixed $d$-dimensional space satisfies
$\sfS_k(\cE(\rho))\geqslant \sfS_k(\rho)$, and $\sfS_k$ is continuous and unitarily invariant:
\begin{eqnarray*}
\sfS_k(\bsU\rho \bsU^\dagger)=\sfS_k(\rho).
\end{eqnarray*}
\end{cor}

Schur concavity gives the lower bound in the pinching lemma; its upper
bound also uses convexity of the pinching gain. Schur concavity also justifies the local pinching step in
the quantitative bounds of Section~\ref{sect:proofs}.

\subsection{Minimum, maximum, and the hierarchy}\label{app:minmaxhier}

\begin{prop}[Extremal values]\label{prop:extval}
For a pure state, $\op{Ent}_k(\rho)=\sfS_k(\rho)=0$ for all $k$. At the maximally mixed state,
\begin{eqnarray}\label{eq:extval}
\sfS_k(\tfrac{\I_d}d)=\ln d-H_k+1,\qquad \op{Ent}_k(\tfrac{\I_d}d)=\binom{d-1}{k-1}(\ln d-H_k+1).
\end{eqnarray}
Hence, by Schur concavity,
\begin{eqnarray*}
0\leqslant \sfS_k(\rho)\leqslant \log d-H_k+1,
\end{eqnarray*}
with the lower bound attained exactly at pure states and the upper bound exactly at $\frac{\I_d}d$.
\end{prop}

\begin{proof}
For a pure spectrum $(1,0,\ldots,0)$,
$$
\ee_k[\bsR_\rho(z)]=\binom{d-1}{k} +\binom{d-1}{k-1}\frac{z}{z-1}.
$$
After multiplication by $\ln z$, the residue at $z=1$ is
$\binom{d-1}{k-1}\log1=0$. Thus $\op{Ent}_k(\rho)=\S_k(\rho)=0$.
For the maximally mixed state, $X\equiv\frac1d$, so \eqref{eq:Skrho} gives
\begin{eqnarray*}
d\cdot \eta\Pa{\tfrac1d} - (H_k-1)=\ln d - H_k + 1.\qedhere
\end{eqnarray*}
\end{proof}

\begin{lem}[The elementary inequality for the entropy hierarchy]\label{applem:monotone}
Let $d\geqslant2$, $1\leqslant k<d$ and $\bsy\in[0,1]^d$, and put
$\op{s}_k(\bsy)=1-\binom{d}{k}^{-1}\ee_k(\bsy)$. Then
\begin{eqnarray}\label{appeq:decreasing}
 \frac{\op{s}_k(\bsy)}{k}\geqslant\frac{\op{s}_{k+1}(\bsy)}{k+1}.
\end{eqnarray}
In particular, this applies to $y_i=\frac t{t+\lambda_i}$ for $t>0$.
\end{lem}

\begin{proof}
Fix a $(k+1)$-subset $\cJ\subseteq[d]$, and take independent Bernoulli
random variables $Y_i$ with success probabilities $y_i(i\in\cJ)$. Let $N=\sum_{i\in\cJ}Y_i$, which is the number of successes. The probability of at most one failure (equivalently described as "at least $k$ successes" when there are $(k+1)$ independent trials) is
\begin{eqnarray*}
\Pr\set{N\geqslant k} &=& \Pr\Set{N=k+1} + \Pr\Set{N=k}=\prod_{j\in\cJ}y_j+   \sum_{i\in\cJ}(1-y_i)\prod_{j\in\cJ\setminus\{i\}}y_j\\
 &=&\prod_{j\in\cJ}y_j + \Pa{\sum_{i\in\cJ}\prod_{j\in\cJ\setminus\set{i}}y_j - (k+1)\prod_{j\in\cJ}y_j}\\
&=&\sum_{i\in\cJ}\prod_{j\in\cJ\setminus\set{i}}y_j  -  k\prod_{j\in\cJ}y_j\leqslant1.
\end{eqnarray*}
Each $k$-subset occurs in the first sum for exactly $d-k$ choices of
$\cJ$. Averaging it over all $(k+1)$-subsets therefore gives
\begin{eqnarray*}
\binom{d}{k+1}^{-1}\sum_{\abs{\cJ}=k+1}\Pa{\sum_{i\in\cJ}\prod_{j\in\cJ\setminus\set{i}}y_j  -  k\prod_{j\in\cJ}y_j}\leqslant1.
\end{eqnarray*}
For a fixed $k$-subset $\cI$, $\abs{\set{\cJ=\cI\cup\set{i}:i\in[d]\setminus\cI}}=d-k$. For each such $\cJ=\cI\cup\set{i}$ for $i\in[d]\setminus\cI$, the monomial $\prod_{j\in\cI}y_j$ appears once (with $i$ being the omitted index). Hence
\begin{eqnarray*}
\sum_{\abs{\cJ}=k+1}\sum_{i\in\cJ}\prod_{j\in\cJ\setminus\set{i}}y_j &=& (d-k)\sum_{\abs{\cI}=k}\prod_{j\in\cI}y_j = (d-k)\ee_k(\bsy),\\
\sum_{\abs{\cJ}=k+1} k\prod_{j\in\cJ}y_j &=& k\ee_{k+1}(\bsy).
\end{eqnarray*}
Note that  $(d-k)\binom{d}{k+1}^{-1}=(k+1)\binom{d}{k}^{-1}$. We get that
\begin{eqnarray}\label{appeq:ekineq}
\binom{d}{k+1}^{-1} \Br{(d-k)\ee_k(\bsy) - k\ee_{k+1}(\bsy)}= \frac{(k+1)\ee_k(\bsy)}{\binom{d}{k}}
 -\frac{k\ee_{k+1}(\bsy)}{\binom{d}{k+1}}\leqslant1,
\end{eqnarray}
Rearranging \eqref{appeq:ekineq} gives
$(k+1)\op{s}_k(\bsy)-k\op{s}_{k+1}(\bsy)\geqslant0$, which is
\eqref{appeq:decreasing}.

The strictness used in the hierarchy also follows from this event.
If two coordinates satisfy $y_i,y_j<1$, a $(k+1)$-subset containing
both has probability of at least two failures at least
$(1-y_i)(1-y_j)>0$. Its contribution makes the averaged inequality
strict. For a mixed spectrum, two eigenvalues are positive, so their
$y_i,y_j$ are less than one for every $t>0$. This includes spectra
with zero eigenvalues; no interior approximation is needed for strictness.
\end{proof}

\begin{thrm}[Entropy-subentropy hierarchy]
For every $\rho\in\rD(\bbC^d)$, we have
\begin{equation}\label{eq:chainofineq}
\sfS(\rho)=\sfS_1(\rho)\geqslant \sfS_2(\rho)\geqslant\cdots\geqslant \sfS_d(\rho)=\sfQ(\rho).
\end{equation}
If $\rho$ is mixed, every adjacent inequality is strict:
\begin{equation}\label{eq:chainofineqstrict}
\sfS_1(\rho)>\sfS_2(\rho)>\cdots>\sfS_d(\rho).
\end{equation}
For pure states all terms vanish, i.e., $\sfS_1(\rho)=\cdots=\sfS_d(\rho)=0$.
\end{thrm}

\begin{proof}
Our proof can be described as follows. Recall from \eqref{eq:divdiff} that
\begin{equation*}
\Phi_k(x_1,\ldots,x_k)=-[x_1,\ldots,x_k]\Pa{x^k\log x}.
\end{equation*}
Use the elementary Frullani-type identity, valid for $x>0$:
$$
-\log x = \int^\infty_0 \Pa{\frac1{t+x} - \frac1{1+t}}\dif t\implies -x^k\log x = \int^\infty_0 \Pa{\frac{x^k}{t+x} - \frac{x^k}{1+t}}\dif t.
$$
Using linearity of divided differences,
\begin{align*} &\Phi_k(x_1,\ldots,x_k) = - [x_1,\ldots,x_k]\Pa{x^k\log x}\\
 &= \int^\infty_0 \Pa{[x_1,\ldots,x_k]g_t-\frac1{1+t} [x_1,\ldots,x_k](x^k)}\dif t,
\end{align*}
where $g_t(x):=\frac{x^k}{t+x}$.
\begin{enumerate}[(i)]
\item \textbf{The first term.} Since $x^k$ is a polynomial of degree $k$, its $(k-1)$-th divided difference is a symmetric homogeneous polynomial of degree $k-(k-1)=1$, namely the sum of the nodes:
$$
[x_1,\ldots,x_k](x^k)=\sum^k_{i=1}\frac{x^k_i}{\prod_{j\neq i}(x_i-x_j)}=\sum^k_{i=1}x_i.
$$
Hence
$$
\frac{[x_1,\ldots,x_k](x^k)}{1+t} = \frac{\sum^k_{i=1}x_i}{1+t}.
$$
\item \textbf{The second term.} Write
$$
g_t(x) = q_t(x) + (-t)^k h_t(x) ,\quad q_t(x):=\sum^{k-1}_{i=0} (-t)^ix^{k-1-i},\quad h_t(x):=\frac1{t+x}.
$$
Note that $q_t$ is a polynomial of degree $k-1$ (with $q_t(x)\equiv1$ when $k=1$). By linearity of divided differences and the same monomial property (the $(k-1)$-th divided difference of a polynomial of degree $\leqslant k-1$ is its coefficient of $x^{k-1}$),
$$
[x_1,\ldots,x_k]g_t = [x_1,\ldots,x_k]q_t + (-t)^k[x_1,\ldots,x_k] h_t.
$$
The leading coefficient of $q_t$ is $1$, so from Hermite--Genocchi integral formula, we get that
$$
[x_1,\ldots,x_k]q_t=\int_{\bbR^k_{\geqslant0}}q^{(k-1)}_t(\Inner{\bsT}{\bsx})\delta\Pa{1-\Inner{\bse}{\bsT}}[\dif \bsT]=1.
$$
Thus $[x_1,\ldots,x_k]g_t = 1 + (-t)^k[x_1,\ldots,x_k] h_t$. For the remaining term, the $(k-1)$-th divided difference of $h_t(x)$ is
\begin{equation*}
[x_1,\ldots,x_k]h_t = \int_{\bbR^k_{\geqslant0}}h^{(k-1)}_t(\Inner{\bsT}{\bsx})\delta\Pa{1-\Inner{\bse}{\bsT}}[\dif \bsT] = \frac{(-1)^{k-1}}{\prod^k_{i=1}(t+x_i)}.
\end{equation*}
The integral expression follows from Hermite--Genocchi and
$h^{(k-1)}_t(x)=(-1)^{k-1}\Gamma(k)/(t+x)^k$. To verify its product
value directly, use the divided-difference recursion. The case $k=1$
is immediate. For distinct nodes and $k\geqslant2$, induction gives
\begin{eqnarray*}
[x_1,\ldots,x_k]h_t
&=&\frac{[x_2,\ldots,x_k]h_t-[x_1,\ldots,x_{k-1}]h_t}{x_k-x_1}\\
&=&\frac{(-1)^{k-2}}{x_k-x_1}
\Pa{\frac1{\prod_{i=2}^k(t+x_i)}
-\frac1{\prod_{i=1}^{k-1}(t+x_i)}}\\
&=&\frac{(-1)^{k-1}}{\prod_{i=1}^k(t+x_i)}.
\end{eqnarray*}
For fixed $t>0$, continuity includes repeated and zero nonnegative nodes.
Therefore
$$
[x_1,\ldots,x_k]g_t = 1+ \frac{(-t)^k(-1)^{k-1}}{\prod^k_{i=1}(t+x_i)} = 1 - \frac{t^k}{\prod^k_{i=1}(t+x_i)}.
$$
\end{enumerate}
In summary, we have the following integral representation:
\begin{equation}
\Phi_k(x_1,\ldots,x_k)=\int_0^\infty\Br{1-\prod_{i=1}^k\frac{t}{t+x_i}-\frac{\sum^k_{i=1} x_i}{1+t}}\dif t.
\end{equation}
This representation also holds at zero nodes. To justify passage to the
boundary under the improper integral, fix $L<\infty$ and take
$0\leqslant x_i\leqslant L$. The absolute value of the integrand is at most $1+kL$
for $0<t\leqslant1$. For large $t$,
$$
 \prod^k_{i=1}\frac{t}{t+x_i}=1-\frac{\sum^k_{i=1} x_i}{t}+O(t^{-2}),\qquad
 \frac{\sum^k_{i=1} x_i}{1+t}=\frac{\sum^k_{i=1} x_i}{t}+O(t^{-2}),
$$
with both remainders uniform on the compact node cube. Thus a common
integrable bound exists, and dominated convergence, together with
continuity of $\Phi_k$, includes repeated and zero nodes.
It follows, with $y_i=\frac{t}{t+\lambda_i}$, that
\begin{eqnarray*}
\S_k(\rho)&=&\binom{d-1}{k-1}^{-1}\op{Ent}_k(\rho) = \binom{d-1}{k-1}^{-1} \sum_{|\cI|=k}\Phi_k(\bdlambda_{\cI})\notag\\
&=&\int_0^\infty \binom{d-1}{k-1}^{-1} \sum_{|\cI|=k} \Br{1-\prod_{i\in\cI}\frac{t}{t+\lambda_i}-\frac{\sum_{i\in\cI} \lambda_i}{1+t}}\dif t \\
&=&\int_0^\infty\Br{\frac{d}{k}\Pa{1-\binom{d}{k}^{-1}\ee_k(\bsy)}-\frac1{1+t}}\dif t.
\end{eqnarray*}
That is,
\begin{eqnarray}\label{eq:integrationrep-single}
\S_k(\rho)=\int_0^\infty\Pa{d \frac{\op{s}_k(\bsy)}k-\frac1{1+t}}\dif t.
\end{eqnarray}
where $\op{s}_k(\bsy):=1-\binom{d}{k}^{-1}\ee_k(\bsy)$ for $0\leqslant y_i\leqslant1$. Furthermore,
\begin{eqnarray}\label{eq:integrationrep}
\S_k(\rho)-\S_{k+1}(\rho)=d\int_0^\infty\Pa{ \frac{\op{s}_k(\bsy)}k - \frac{\op{s}_{k+1}(\bsy)}{k+1} }\dif t.
\end{eqnarray}
Thus $\S_k(\rho)\geqslant \S_{k+1}(\rho)$ can be obtained by using the elementary inequality \eqref{appeq:decreasing} in Lemma~\ref{applem:monotone}, where $y_i=\frac t{t+\lambda_i}$. The left-hand side of \eqref{appeq:ekineq} is the probability of at most one failure in $k+1$ independent trials with success probabilities $y_i(t)$, averaged over $(k+1)$-subsets. Substituting \eqref{appeq:decreasing} into \eqref{eq:integrationrep} proves \eqref{eq:chainofineq}. If $\rho$ is mixed, choose two indices $i,j$ with positive eigenvalues.
For every $t>0$, their failure probabilities
$1-y_i(t)$ and $1-y_j(t)$ are positive.
For each $1\leqslant k<d$, a $(k+1)$-subset containing both has probability
of at least two failures at least
$(1-y_i(t))(1-y_j(t))>0$.
Its contribution makes the averaged inequality strict. Integration
proves strictness even when other eigenvalues are zero. 
\end{proof}

If $\bsX\geqslant0$ with $\Tr{\bsX}=\kappa>0$, assume that $\bsx=(x_1,\ldots,x_d)$ are the spectrum of $\bsX$. Then from \eqref{eq:integrationrep-single}, we see that
\begin{eqnarray*}
\S_k(\bsX) &=& \S_k(\kappa \bsX/\kappa) = \kappa \S_k(\bsX/\kappa) + \eta(\kappa)\\
&=&\kappa \int_0^\infty\Pa{\frac dk \Pa{1-\binom{d}{k}^{-1}\ee_k\Pa{\frac{t}{t+x_1/\kappa},\ldots,\frac{t}{t+x_d/\kappa}}} - \frac1{1+t}}\dif t +  \eta(\kappa)\\
&=& \int_0^\infty\Pa{\frac dk \Pa{1-\binom{d}{k}^{-1}\ee_k\Pa{\frac{\kappa t}{\kappa t+x_1},\ldots,\frac{\kappa t}{\kappa t+x_d}}} - \frac{\kappa}{\kappa+\kappa t}}\dif (\kappa t) +  \eta(\kappa).
\end{eqnarray*}
Replace $\kappa t$ by $t$, we get that
\begin{eqnarray}\label{appeq:unnormalizedSk}
\S_k(\bsX) = \int_0^\infty\Pa{\frac dk \Pa{1-\binom{d}{k}^{-1}\ee_k\Pa{\frac{t}{t+x_1},\ldots,\frac{t}{t+x_d}}} - \frac{\kappa}{\kappa+t}}\dif t +  \eta(\kappa).
\end{eqnarray}
For $k=d$, we get that
\begin{eqnarray}\label{appeq:unnormalizedQ}
\Q(\bsX) = \int_0^\infty\Pa{ \frac{t}{t+\kappa} -\frac{t^d}{ \prod^d_{i=1}(t+x_i)}}\dif t +  \eta(\kappa),\qquad \kappa=x_1+\cdots+x_d.
\end{eqnarray}

\subsection{Dependence on ambient dimension and zero padding}\label{app:depenambient}

Unlike $\S_1$ and $\S_d$, the intermediate quantities $\S_k$ depend on the ambient dimension, even when one merely appends a zero eigenvalue. Writing $\S^{(d)}_k$ and $\S^{(d+1)}_k$ for the two ambient dimensions, and using $\ee_k(\bsB\oplus 1)=\ee_k(\bsB)+\ee_{k-1}(\bsB)$, one finds the following result:
\begin{prop}\label{prop:embeddingvspadding}
Let $\rho\in\rD(\bbC^d)$ and let $\rho\oplus 0$ denote its embedding into $\bbC^{d+1}$ by appending one zero eigenvalue. Then, for $1\leqslant k\leqslant d+1$, we have
\begin{equation}\label{eq:embedding}
\op{Ent}^{(d+1)}_k(\rho\oplus0)=\op{Ent}^{(d)}_k(\rho)+\op{Ent}^{(d)}_{k-1}(\rho),
\end{equation}
with $\op{Ent}_0^{(d)}\equiv0$ and $\op{Ent}_k^{(d)}=0$ whenever $k>d$ (out-of-range terms vanish). For $1\leqslant k\leqslant d$, the normalized version is
\begin{equation}\label{eq:Sd1kpluszero}
\sfS^{(d+1)}_k(\rho\oplus0)=\frac{d-k+1}{d}\sfS^{(d)}_k(\rho) + \frac{k-1}{d}\sfS^{(d)}_{k-1}(\rho).
\end{equation}
Here the term with coefficient $k-1=0$ is omitted when $k=1$.
\end{prop}
For example, if $\rho=\frac12\I_2$,
\begin{equation}
\sfS^{(2)}_2(\rho)=\ln2-\frac12,\qquad
\sfS^{(3)}_2(\rho\oplus0)=\ln2-\frac14.
\end{equation}

\begin{proof}
Let $\bsR_{\rho\oplus 0}(z) = (\I_{d+1} - z^{-1}(\rho\oplus 0))^{-1}$. In block form,
$$
\bsR_{\rho\oplus 0}(z) = \begin{pmatrix} \bsR_\rho(z) & 0 \\ 0 & 1 \end{pmatrix},
$$
because the appended zero eigenvalue gives $1 - z^{-1}\cdot 0 = 1$ in the extra diagonal entry. Applying the elementary symmetric polynomial identity $\ee_k(\bsB\oplus 1) = \ee_k(\bsB) + \ee_{k-1}(\bsB)$ (which follows from $\prod_{i=1}^{d+1}(t - \mu_i) = (t-1)\prod_{i=1}^d(t-\mu_i)$ with $\mu_{d+1}=1$, or directly from the definition $\det(t\I - \bsM) = \sum^D_{k=0} (-1)^k \ee_k(\bsM)t^{D-k}$) to $\bsB = \bsR_\rho(z)$, we get
$$
\ee_k[\bsR_{\rho\oplus 0}(z)] = \ee_k[\bsR_\rho(z)] + \ee_{k-1}[\bsR_\rho(z)].
$$
Inserting into the contour definition \eqref{eq:EntkSk},
\begin{align*}
\op{Ent}_k^{(d+1)}(\rho\oplus 0)&= -\frac{1}{2\pi \mathrm i}\oint \ee_k[\bsR_{\rho\oplus 0}(z)](\log z)\dif z \\
&= -\frac{1}{2\pi \mathrm i}\oint \ee_k[\bsR_\rho(z)](\log z)\dif z - \frac{1}{2\pi \mathrm i}\oint \ee_{k-1}[\bsR_\rho(z)](\log z)\dif z \\
&= \op{Ent}_k^{(d)}(\rho) + \op{Ent}_{k-1}^{(d)}(\rho),
\end{align*}
where the contour encloses all nonzero eigenvalues of $\rho$ (the appended zero eigenvalue is not a pole of $\ee_k[\bsR_{\rho\oplus 0}]$, since $\bsR_{\rho\oplus 0}$ has no pole at $z=0$; cf. the remark in the proof of Proposition~\ref{prop:eigfor} that a zero eigenvalue contributes the factor $\frac z{z-0}=1$ and no pole). The two contour integrals are exactly the definitions of $\op{Ent}_k^{(d)}$ and $\op{Ent}_{k-1}^{(d)}$ in ambient dimension $d$, with the convention $\op{Ent}_0^{(d)}\equiv 0$ (i.e. $\ee_0 \equiv 1$ and $\oint (\log z)\dif z = 0$, as shown in the proof of Proposition~\ref{prop:eigfor}). 

Finally, from $\op{Ent}_k$ to the normalized $S_k$ via $\op{Ent}^{(D)}_k=\binom{D-1}{k-1}S^{(D)}_k$:
\begin{align*}
S^{(d+1)}_k(\rho\oplus0)&=\binom{d}{k-1}^{-1}\binom{d-1}{k-1}S^{(d)}_k(\rho)+\binom{d}{k-1}^{-1}\binom{d-1}{k-2}S^{(d)}_{k-1}(\rho)\\
&= \frac{d-k+1}dS^{(d)}_k(\rho)+\frac{k-1}dS^{(d)}_{k-1}(\rho)
\end{align*}
gives exactly \eqref{eq:Sd1kpluszero}.
\end{proof}

\subsection{The rank-dependent augmentation formula}\label{app:rankdepen}

\begin{lem}[zero-padding of a single summand]\label{applem:0padding}
For $1\leqslant j\leqslant k$ and distinct nonzero $x_1,\dots,x_j$,
\begin{equation}\label{eq:zero-padding}
\Phi_k(x_1,\dots,x_j,\underbrace{0,\ldots,0}_{k-j})=\Phi_j(x_1,\dots,x_j).
\end{equation}
More generally, \eqref{eq:zero-padding} holds for arbitrary (possibly coincident, possibly zero) arguments by continuity, with the convention $x\ln x\to 0$ as $x\to 0^+$.
\end{lem}

\begin{proof}
The divided-difference product rule gives
$$
 [0,x_1,\ldots,x_j](xf(x))=[x_1,\ldots,x_j]f(x) \quad\text{if }f(0)=0.
$$
Apply this rule successively to $x^k\ln x$, removing one zero node and
one power of $x$ at each step. The functions at each intermediate step
have the continuous derivatives required by the corresponding confluent
divided differences. Hence
$$
 [x_1,\ldots,x_j,0,\ldots,0](x^k\log x) =[x_1,\ldots,x_j](x^j\log x).
$$
This proves \eqref{eq:zero-padding} for distinct positive $x_i$.
Continuity permits repeated positive nodes and additional zero nodes
among the $x_i$. The all-zero value is zero by continuity.
\end{proof}

The endpoints are consistent:$\S(\rho\oplus0)=\S(\rho)$ and $\Q(\rho\oplus0)=\Q(\rho)$. More generally, if $\rho$ has rank $r$ and $\rho_{\op{supp}}$ is its restriction to its support, then we have the following result.
\begin{prop}
Let $\rho \in \rD(\bbC^d)$ have rank $r$ (so its spectrum has exactly $r$ nonzero entries, listed in nonincreasing order), and let $\rho_{\mathrm{supp}}$ be the restriction of $\rho$ to its support (a state on a space of dimension $r$). Then for every $1 \leqslant k \leqslant d$,
$$
\op{Ent}_k^{(d)}(\rho)=\sum_{j=1}^{\min(k,r)}\binom{d-r}{k-j}\op{Ent}_j^{(r)}(\rho_{\mathrm{supp}}).
$$
Equivalently, in normalized form,
$$
\sfS_k^{(d)}(\rho)=\binom{d-1}{k-1}^{-1}\sum_{j=1}^{\min(k,r)}\binom{d-r}{k-j}\binom{r-1}{j-1} \sfS_j^{(r)}(\rho_{\mathrm{supp}}),
$$
with the conventions $\op{Ent}_0^{(r)}\equiv 0$ and $\op{Ent}_j^{(r)}=0$ for $j>r$.
\end{prop}

\begin{proof}
Let the spectrum of $\rho$ be
\begin{equation*}
\bdlambda(\rho)=(\lambda_1,\dots,\lambda_r,\underbrace{0,\ldots,0}_{d-r}),
\qquad \lambda_1\geqslant\cdots\geqslant\lambda_r>0,\qquad \sum_{i=1}^r\lambda_i=1 .
\end{equation*}
By Proposition~\ref{prop:eigfor} (the eigenvalue formula, Eq.~\eqref{eq:Entkrho}), the unnormalized $k$-entropy in ambient dimension $d$ is the sum of $\Phi_k$ over all $k$-element subsets of the $d$ eigenvalues:
\begin{equation*}
\op{Ent}_k^{(d)}(\rho)=\sum_{\cI\subseteq[d]:|\cI|=k}\Phi_k(\bdlambda_{\cI}),
\qquad
\Phi_k(x_1,\dots,x_k)=-\sum_{i=1}^k\frac{x_i^k}{\prod_{j\neq i}(x_i-x_j)}\ln x_i ,
\end{equation*}
where repeated or zero eigenvalues are handled by continuity. The support restriction is the state on $\bbC^r$ with spectrum
$$
\bdlambda(\rho_{\mathrm{supp}})=(\lambda_1,\dots,\lambda_r),
$$
so by the same formula in ambient dimension $r$,
$$
\op{Ent}_j^{(r)}(\rho_{\mathrm{supp}})=\sum_{\mathcal{J}\subseteq[r]:|\cJ|=j}\Phi_j(\bdlambda_{\cJ}).
$$
In divided-difference language,  \eqref{eq:zero-padding} is the statement
$$
[x_1,\ldots,x_j,0,\ldots,0]\Pa{x^k\ln x}=[x_1,\ldots,x_j]\Pa{x^j\log x},
$$
i.e. inserting $k-j$ zero nodes and lowering the power by $k-j$ leaves the divided difference invariant---the standard recursion identity for divided differences.

Partition the $k$-element subsets $\cI\subseteq[d]$ according to how many of their elements lie in the support $[r]=\set{1,\dots,r}$:
\begin{equation*}
\cI=\cJ\cup\cZ,\qquad \cJ\subseteq[r],\quad \cJ\cap\cZ=\varnothing,\quad \cZ\subseteq\set{r+1,\ldots,d},
\end{equation*}
with $|\cJ|=j$ and $|\cZ|=k-j$. Such a subset exists only if $0\leqslant j\leqslant r$ and $0\leqslant k-j\leqslant d-r$; the number of choices of $\cZ$ is $\binom{d-r}{k-j}$. On $\cI$ the eigenvalues are $(\bdlambda_{\cJ},0,\dots,0)$, so by Lemma~\ref{applem:0padding},
\begin{equation*}
\Phi_k(\bdlambda_{\mathcal{I}})=\Phi_k(\bdlambda_{\cJ},0,\dots,0)=\Phi_j(\bdlambda_{\cJ})
\qquad\text{whenever}\quad j=|\mathcal{J}|\geqslant 1,
\end{equation*}
while for $j=0$ (i.e. $\cI$ entirely inside the zero block) all eigenvalues are zero, and $\Phi_k(0,\ldots,0)=0$ by continuity (the convention $\op{Ent}_0\equiv 0$ absorbs this).

Summing over all subsets:
\begin{align*}
\op{Ent}_k^{(d)}(\rho)&=\sum_{\cI\subseteq[d]:|\cI|=k}\Phi_k(\bdlambda_{\cI})=\sum_{j=0}^{\min(k,r)}\ \sum_{\cJ\subseteq[r]:|\cJ|=j}\ \sum_{\substack{\cZ\subseteq\set{r+1,\dots,d}\\ |\cZ|=k-j}}\Phi_k(\bdlambda_{\cJ},0,\ldots,0)\\
&=\sum_{j=1}^{\min(k,r)}\binom{d-r}{k-j}\sum_{\cJ\subseteq[r]: |\cJ|=j}\Phi_j(\bdlambda_{\cJ})=\sum_{j=1}^{\min(k,r)}\binom{d-r}{k-j}\op{Ent}_j^{(r)}(\rho_{\mathrm{supp}}),
\end{align*}
where we used \eqref{eq:zero-padding}, counted the zero-block choices, and then applied Proposition~\ref{prop:eigfor} in ambient dimension $r$. 

Using $\op{Ent}_k^{(d)}=\binom{d-1}{k-1}\S_k^{(d)}$ (Definition~\ref{def:kentropy}) and $\op{Ent}_j^{(r)}=\binom{r-1}{j-1}\S_j^{(r)}$, the above equality becomes
\begin{equation*}
\binom{d-1}{k-1}\S_k^{(d)}(\rho)
=\sum_{j=1}^{\min(k,r)}\binom{d-r}{k-j}\binom{r-1}{j-1}\S_j^{(r)}(\rho_{\mathrm{supp}}),
\end{equation*}
which is the normalized formula stated above. This completes the proof.
\end{proof}

\subsection{Confluent evaluation at repeated zero nodes}\label{app:confluent}

The confluent identity in Lemma~\ref{applem:0padding} gives the
augmentation formula for arbitrary spectra, including repeated zero nodes.
Classifying subsets by the $j$ nodes chosen from the
original $l$ coordinates yields the normalized augmentation identity
\begin{equation}\label{eq:padding-normalized}
\S_k^{(D)}(\sigma)= \binom{D-1}{k-1}^{-1}\sum_{j=1}^{\min(k,l)}\binom{D-l}{k-j}\binom{l-1}{j-1}
\S_j^{(l)}(\sigma).
\end{equation}
This is the general augmentation law of Nichols and
Wootters~\cite[Eq. (37)]{Nichols2003}, expressed in the present notation;
the preceding argument includes coincident and zero nodes. Its
coefficients sum to one by Vandermonde's identity. If $k\leqslant l\leqslant D$,
the hierarchy gives
\begin{equation}\label{eq:padding-monotone}
\S_k^{(D)}(\sigma)\geqslant \S_k^{(l)}(\sigma).
\end{equation}
For $k=D$ only the $j=l$ term survives and gives padding-invariance of
$\Q$; for intermediate $k$ this reduction to one term does not occur in general.

\section{Haar integral over the complex Stiefel manifold}\label{app:B}

The Haar integral over the complex Stiefel manifold is a central tool in random matrix theory, quantum information, and multivariate statistics. It provides a natural, invariant way to average functions over the space of complex matrices with orthonormal columns.

This appendix collects the definitions, elementary properties, and the representation-theoretic integration formula for the complex Stiefel manifold that are used in Section~\ref{proof:compression-pinching}. The central result is the Schur-Weyl averaging identity \eqref{eq:Schurintegral}, which is the technical heart of the positive compression representation \eqref{eq:kentropy-intrep} and, consequently, of the pinching lemma and the subadditivity theorem.

\subsection{The complex Stiefel manifold and its Haar measure}

For integers $1 \leqslant \ell \leqslant d$, the complex Stiefel manifold is the set of complex $d \times \ell$ matrices with orthonormal columns:
$$
\op{St}(d,\ell) = \Set{\bsV \in M_{d,\ell}(\bbC) : \bsV^{\dagger}\bsV = \I_{\ell} }.
$$
It is a compact connected real manifold of dimension
\[
d^2-(d-\ell)^2=2d\ell-\ell^2,
\]
diffeomorphic to the homogeneous space $\rU(d)/\rU(d-\ell)$. The dimension formula follows from the quotient description, since $\dim_{\mathbb R}\rU(q)=q^2$. The unitary group $\rU(d)$ acts transitively on $\op{St}(d,\ell)$ by left multiplication, $\bsV \mapsto \bsU \bsV$. This transitive action carries a unique $\rU(d)$-invariant probability measure, denoted $\mu_{\ell}$, which we call the normalized Haar measure on $\op{St}(d,\ell)$. The invariance property is
$$
\int_{\op{St}(d,\ell)} f(\bsU\bsV) \dif\mu_{\ell}(\bsV) = \int_{\op{St}(d,\ell)} f(\bsV) \dif\mu_{\ell}(\bsV)
$$
for every integrable $f$ and every $\bsU \in \rU(d)$. Right multiplication by a fixed unitary $\bsW \in \rU(\ell)$ also preserves the measure, since the pushforward under $\bsV\mapsto\bsV\bsW$ is again a left $\rU(d)$-invariant probability measure, and such a measure is unique.

\paragraph{Relation to the unitary group.} A Haar-distributed $\bsV \in \op{St}(d,\ell)$ can be generated by taking a Haar-distributed $\bsU \in \rU(d)$ and extracting its first $\ell$ columns. The resulting probability law is exactly $\mu_{\ell}$. Consequently, the projection $\Pi = \bsV\bsV^{\dagger}$ is a rank-$\ell$ orthogonal projection in $\bbC^d$ distributed according to the image of the Haar measure under the map $\bsU \mapsto \bsU \Pi_0 \bsU^{\dagger}$, where $\Pi_0 = \diag(\I_{\ell}, 0_{d-\ell})$. This identification is used in the proof of the Schur-Weyl formula below.

\subsection{The Schur-Weyl averaging formula}

The key computational tool is the evaluation of Haar averages of Schur polynomials of compressed matrices. Let $\nu$ be a partition of a nonnegative integer with at most $d$ parts. The Schur polynomial $s_{\nu}(\bsX)$ evaluated on the eigenvalues of a Hermitian matrix $\bsX$ is a symmetric polynomial. The following identity holds for every $\bsX \geqslant \zero$ acting on $\bbC^d$:
\begin{eqnarray}\label{eq:Schurintegral}
\int_{\op{St}(d,\ell)} s_{\nu}(\bsV^{\dagger} \bsX \bsV) \dif \mu_{\ell}(\bsV) = \frac{s_{\nu}(1^{\ell})}{s_{\nu}(1^{d})} s_{\nu}(\bsX). 
\end{eqnarray}
When the compressed $\ell\times\ell$ matrix is viewed in the ambient
$d$-dimensional space, $s_{\nu}(\bsV^\dagger\bsX\bsV)$ is evaluated on
its $\ell$ eigenvalues after adjoining $d-\ell$ zero variables. This is the
same zero-variable convention used for the right-hand side. Here
$s_{\nu}(1^{m})$ denotes the dimension of the irreducible polynomial representation of $\rG\rL(m,\bbC)$ labelled by $\nu$, evaluated at the identity, i.e., the number of semistandard Young tableaux of shape $\nu$ with entries in $\set{1,\ldots,m}$. If $\nu$ has more than $\ell$ rows, both sides vanish: the left-hand side because $\bsV^{\dagger}\bsX\bsV$ has rank at most $\ell$, and the right-hand side because $s_{\nu}(1^{\ell}) = 0$.

\begin{proof}[Proof sketch]
The proof is representation-theoretic and follows the argument given in Section~\ref{subsubsect:4.1.1}. Let $\pi_{\nu}$ be the irreducible polynomial representation of $\rG\rL(d,\bbC)$ with character $s_{\nu}$, constructed via the Schur functor. Its restriction to $\rU(d)$ is irreducible. The average of $\pi_{\nu}(\Pi)$ over the unitary orbit of the projection $\Pi = \bsV\bsV^{\dagger}$ is a scalar operator by Schur's lemma, because the Haar measure is invariant under conjugation. Taking traces determines the scalar: the trace of $\pi_{\nu}(\Pi)$ is $s_{\nu}(\bsV^{\dagger}\bsV) = s_{\nu}(\I_{\ell}) = s_{\nu}(1^{\ell})$, while the trace of the identity on the representation space is $s_{\nu}(1^{d})$. Hence
$$
\int_{\rU(d)} \pi_{\nu}(\bsU \Pi_0 \bsU^{\dagger}) \dif \mu_{\haar}(\bsU) = \frac{s_{\nu}(1^{\ell})}{s_{\nu}(1^{d})} \I.
$$
The nonzero eigenvalues of $\bsV^{\dagger}\bsX\bsV$ coincide with those of $\bsX^{\frac12} \Pi\bsX^{\frac12}$, where $\Pi = \bsV\bsV^{\dagger}$. By the multiplicativity and cyclicity of the trace,
$$
s_{\nu}(\bsV^{\dagger}\bsX\bsV) = \Tr{\pi_{\nu}(\bsX^{\frac12}) \pi_{\nu}(\Pi) \pi_{\nu}(\bsX^{\frac12}) } = \Tr{\pi_{\nu}(\bsX) \pi_{\nu}(\Pi)}.
$$
Averaging over $\Pi$ and using the scalar identity proves \eqref{eq:Schurintegral}. The extension to singular $\bsX$ and to partitions with more than $\ell$ rows follows by polynomial continuation and the vanishing convention.
\end{proof}

\paragraph{Completeness of the Schur basis.} The Schur polynomials $s_{\nu}$, as $\nu$ ranges over partitions with at most $d$ rows, form a linear basis for the space of symmetric polynomials in $d$ variables. Therefore, an identity of the form \eqref{eq:Schurintegral} for all Schur polynomials extends by linearity to every symmetric polynomial in the eigenvalues of $\bsV^{\dagger}\bsX\bsV$. The continuous-function identity \eqref{eq:continuous-function-identity} used in the proof of the positive compression representation is obtained in two steps: first express each power sum $\Tr{(\bsV^{\dagger}\bsX\bsV)^n}$ in the Schur basis and apply \eqref{eq:Schurintegral} term by term; then use uniform polynomial approximation on the compact spectral interval $[0,\norm{\bsX}]$. Both sides are positive linear functionals with total mass $\binom{d}{k}$, so the uniform approximation error tends to zero on both sides.

\subsection{The Pure-state expectation and the functional $\G_{\ell}$}

The formula \eqref{eq:Schurintegral} yields the pure-state value of the compressed entropy functional $\G_{\ell}$. For a unit vector $\ket{\psi}\in \bbC^d$, the only possible nonzero eigenvalue of $\bsV^{\dagger}\proj{\psi}\bsV$ is $T = \norm{\bsV^{\dagger}\psi}^2$, which is the squared norm of the projection of $\ket{\psi}$ onto the column space of $\bsV$. For $\ell < d$, the random variable $T$ has the $\op{Beta}(\ell, d-\ell)$ distribution:
$\bbE[T^s] = \frac{\Gamma(\ell+s)\Gamma(d)}{\Gamma(\ell)\Gamma(d+s)}$
for $s>-\ell$. Differentiating at $s=1$ gives
$\bbE[T \log T] =  - \frac{\ell}{d}(H_d - H_{\ell})$, where $H_m = \sum_{j=1}^{m} 1/j$ is the harmonic number. Hence
$$
\G_{\ell}(\proj{\psi}) = \int_{\op{St}(d,\ell)} \Tr{\eta(\bsV^{\dagger}\proj{\psi}\bsV)} \dif\mu_{\ell}(\bsV) = \frac{\ell}{d}(H_d - H_{\ell}),
$$
which is the equation \eqref{eq:value@purestate} in the main text. For $\ell = d$, $T=1$ deterministically and both sides vanish. This calculation is used to define the shifted functional $\widetilde{\G}_{\ell}(\rho) = \G_{\ell}(\rho) - g_{\ell}$, which vanishes on pure states and is nonnegative and concave on density matrices.

\subsection{Bibliographical remarks}

The Schur-Weyl averaging formula \eqref{eq:Schurintegral} is a standard consequence of the representation theory of the unitary group; a detailed account of the Schur functor construction and the character theory used in the proof can be found in Fulton and Harris \cite{Fulton1991}. The broader program of computing integrals over Stiefel manifolds via invariant differential operators and Pizzetti-type formulas is developed by Coulembier and Kieburg \cite{Coulembier2015}, who also treat real and quaternionic analogues and applications to Itzykson-Zuber integrals. For polynomial integration over the unitary group itself, the Weingarten calculus of Collins and \'{S}niady provides the general moment formulas that underlie the Schur-Weyl identity when expressed in entry coordinates \cite[Sections 2.1--2.2]{Collins2006}. The specific use of these tools in the present paper---the compression identity \eqref{eq:kentropy-intrep} and its consequences for entropy inequalities---is self-contained and does not require the full machinery of those references beyond formula \eqref{eq:Schurintegral} and the elementary Beta integral for the pure-state calculation.

\section{Bregman divergence of the negative $k$-entropy}\label{app:C}

The Bregman divergence was introduced by Bregman \cite{Bregman1967} for convex functions $f:\bbR^d\to\bbR$ with gradient $\nabla f$, as the $f$-depending nonnegative measure of discrepancy
\begin{eqnarray}\label{appeq:BregDiv}
\B_f(\bsx||\bsy):=f(\bsx) - f(\bsy) - \Inner{\nabla f(\bsy)}{\bsx-\bsy},\qquad (\forall\bsx,\bsy\in\bbR^d).
\end{eqnarray}
This notion is extended to operators by Petz \cite{Petz2007}: 
\begin{definition}
If $\cC$ denotes a convex set in a Banach space and $\lin{\cH}$ denotes the bounded linear operators on the Hilbert space $\cH$, for an operator-valued smooth function $\Psi:\cC\to\lin{\cH}$, we define the \emph{Bregman divergence} of $\bsu,\bsv\in\cC$ as
\begin{eqnarray}\label{appeq:BregDivMatrix}
\sfB_{\Psi}(\bsu||\bsv):=\Psi(\bsu) - \Psi(\bsv) - \lim_{t\to 0} \frac{\Psi(\bsv + t(\bsu-\bsv)) - \Psi(\bsv)}t.
\end{eqnarray}
\end{definition}
The case is very important for us when $\cC=\rD(\bbC^d)$ and $\Psi(\rho)=-\eta(\rho)$, where $\Psi(t)=-\eta(t)=t\log t$ is a convex function defined on $\bbR_{\geqslant0}$ and $\eta(\rho)$ is defined by the functional calculus on $\rD(\bbC^d)$. Consider two invertible density matrices $\rho,\sigma\in\rD(\bbC^d)$, and the negative von Neumann entropy $\phi(\rho)=-\S(\rho) = - \Tr{\eta(\rho)}$, we can derive that 
\begin{eqnarray}
\B_{\phi}(\rho||\sigma) =\Tr{\rho(\log\rho - \log\sigma)},
\end{eqnarray}
the relative entropy. 

With this information, we can introduce the Bregman divergence for the negative $k$-entropy.
\begin{definition}
For $\bsX,\bsY\in \pd{\bbC^d}$ (the set of all positive definite matrices acting on $\bbC^d$), the Bregman divergence $\sfB^{(d)}_k$ associated with the negative $k$-entropy function $F_k(\rho)=-\sfS^{(d)}_k(\rho)$ is defined by
\begin{eqnarray}
\B^{(d)}_k(\bsX||\bsY) = F_k(\bsX) - F_k(\bsY) - \Inner{\sL_k(\bsY)}{\bsX-\bsY}.
\end{eqnarray}
Here $\sL_k(\bsY)=\nabla F_k(\bsY)$ is the gradient with respect to the real Hilbert-Schmidt inner product on Hermitian matrices.
\end{definition}

So the Bregman divergence of $F_k = -\S^{(d)}_k$ is the adapted divergence $\B_k^{(d)}$, built as a positive weighted Haar average of the standard Bregman divergence $\B_+$ over compressed matrices $\bsV^\dagger \bsX \bsV, \bsV^\dagger \bsY \bsV$.

From the above definition, we see that if two convex functions $f,g:\bbR^d\to\bbR$ with gradients $\nabla f,\nabla g$, then, for $\alpha,\beta\in\bbR_{\geqslant0}$
\begin{eqnarray*}
\B_{\alpha f+\beta g}(\bsx||\bsy) &=& (\alpha f+\beta g)(\bsx) -  (\alpha f+\beta g)(\bsy) - \Inner{\nabla(\alpha f+\beta g)(\bsy)}{\bsx-\bsy}\\
&=& \alpha [f(\bsx)-f(\bsy)-\Inner{\nabla f(\bsy)}{\bsx-\bsy}] + \beta [g(\bsx)-g(\bsy)-\Inner{\nabla g(\bsy)}{\bsx-\bsy}]\\
&=& \alpha \B_f(\bsx||\bsy) +  \beta \B_g(\bsx||\bsy).
\end{eqnarray*}
That is,
\begin{eqnarray}\label{appeq:additive-of-(f+g)}
\B_{\alpha f+\beta g}(\bsx||\bsy)=\alpha\B_f(\bsx||\bsy) +\beta \B_g(\bsx||\bsy)
\end{eqnarray}
is still Bregman divergence.

\subsection{Proof of Proposition~\ref{prop:GREproperty} }

\begin{enumerate}
\item \textbf{Finiteness and the support condition.} Let $\bsP=\op{supp}[\bsY]$, so that $\bsP$ is the orthogonal projection onto $(\ker [\bsY])^\perp$, and $\bsY_{\bsP}:=\bsY|_{\bsP}>0$. If $\op{supp}[\bsX]\subseteq \bsP$, then $\bsX=\bsP\bsX\bsP$, and
$$
\B_+(\bsX|| \bsY)=\Tr{\bsX(\log \bsX-\log \bsY_{\bsP})}-\Tr{\bsX-\bsY}.
$$
Here $\bsX\log \bsX$ is finite because $t\log t$ is continuous at $0$, and $\bsX\log \bsY_{\bsP}$ is finite because $\bsX$ is supported on $\bsP$ and $\bsY_{\bsP}>0$. Hence $\B_+(\bsX|| \bsY)<\infty$.

If $\op{supp}[\bsX]\not\subseteq \bsP$, then $\bsX$ has nonzero mass on $\ker [\bsY]$. On that subspace $\log \bsY=-\infty$, so the term $-\Tr{\bsX\log \bsY}$ is $+\infty$, while the remaining terms are finite. Hence
$\B_+(\bsX|| \bsY)=+\infty$. Thus $\B_+$ is finite exactly when $\op{supp}[\bsX]\subseteq\op{supp}[\bsY]$.
\item \textbf{Nonnegativity.} Assume first that $\op{supp}[\bsX]\subseteq\op{supp}[\bsY]$, otherwise the assertion is trivial because the value is $+\infty$.  If $\bsY=0$, then the support condition gives $\bsX=\zero$, and $\B_+(\zero||\zero)=0$. Hence assume $\bsY\neq0$. If $\bsX=\zero$, then
$$
\B_+(\zero||\bsY)=-\Tr{\zero-\bsY}=\Tr{\bsY}\geqslant0.
$$
So assume $\bsX\neq\zero$. Put $\Tr{\bsX}=x,\Tr{\bsY}=y$, and define the normalized states
$\rho=\frac{\bsX}{x}, \sigma=\frac{\bsY}{y}$. Then
$$
\log \bsX=\log x+\log\rho,\qquad \log \bsY=\log y+\log\sigma
$$
on the relevant supports, and therefore
\begin{eqnarray*}
\B_+(\bsX|| \bsY) &=& x\Tr{\rho(\log\rho-\log\sigma)} + x\log\frac{x}{y}-(x-y)\\
&=& x\B_+(\rho||\sigma)+y\Pa{\frac{x}{y}\log\frac{x}{y}-\frac{x}{y}+1}.
\end{eqnarray*}
The first term is nonnegative by Klein's inequality,
$$
\B_+(\rho||\sigma)=\Tr{\rho(\log\rho-\log\sigma)}\geqslant0.
$$
For the second term, the scalar function $\varphi(r)=r\log r-r+1$, where $r\geqslant0$, satisfies $\varphi(r)\geqslant0$, with equality only at $r=1$. Hence $\B_+(\bsX|| \bsY)\geqslant0$.
\item \textbf{Joint convexity.} We use the standard fact that the Umegaki relative entropy
$$
\K\L(\bsX|| \bsY)=\Tr{\bsX(\log \bsX-\log \bsY)}
$$
is jointly convex on pairs of positive definite matrices of arbitrary trace;
see Tropp~\cite[Definition 2 and Fact 4]{Tropp2012} for the equivalent
formulation with the affine trace correction. Since
$$
\B_+(\bsX|| \bsY)=\K\L(\bsX||\bsY)-\Tr{\bsX-\bsY},
$$
and the correction $-\Tr{\bsX-\bsY}$ is affine, joint convexity of $\K\L$ implies joint convexity of $\B_+$ on the positive definite cone.

For $t\in[0,1]$ and positive definite pairs $(\bsX_0,\bsY_0),(\bsX_1,\bsY_1)$, we therefore have
$$
\B_+((1-t)\bsX_0+t\bsX_1|| (1-t)\bsY_0+t\bsY_1)\leqslant
(1-t)\B_+(\bsX_0|| \bsY_0)+t\B_+(\bsX_1|| \bsY_1).
$$
To extend this to positive semidefinite matrices, fix $\bsX_0,\bsX_1,\bsY_0,\bsY_1\geqslant\zero$ and put
$$
\bsX_t=(1-t)\bsX_0+t\bsX_1,\qquad \bsY_t=(1-t)\bsY_0+t\bsY_1.
$$
At $t=0$ and $t=1$, the inequality reduces to an identity, with the usual convention $0\cdot(+\infty)=0$. Thus assume $0<t<1$. If one of $\B_+(\bsX_0|| \bsY_0)$ and $\B_+(\bsX_1|| \bsY_1)$ is infinite, the right-hand side is $+\infty$, so the inequality is immediate. Hence assume both values are finite. For $\varepsilon>0$, the matrices $\bsX_i+\varepsilon\I$ and $\bsY_i+\varepsilon\I$ are positive definite, so the positive definite inequality gives
$$
\B_+(\bsX_t+\varepsilon\I || \bsY_t+\varepsilon\I) \leqslant
(1-t)\B_+(\bsX_0+\varepsilon\I || \bsY_0+\varepsilon\I) + t\B_+(\bsX_1+\varepsilon\I|| \bsY_1+\varepsilon\I).
$$
Letting $\varepsilon\downarrow0$ and using the regularization identity proved in item 5 below yields
$$
\B_+(\bsX_t|| \bsY_t)\leqslant
(1-t)\B_+(\bsX_0|| \bsY_0)+t\B_+(\bsX_1|| \bsY_1).
$$
Thus $\B_+$ is jointly convex on the whole positive semidefinite cone.
\item \textbf{Lower semicontinuity.} We give a proof using joint convexity and the regularization identity in item 5 below. For positive definite $\bsX,\bsY$ and $a>0$, the identity $\log(a\bsX)=(\log a)\I+\log\bsX$ gives
$$
\B_+(a\bsX||a\bsY)=a\B_+(\bsX||\bsY).
$$
Combining this positive homogeneity with joint convexity at $t=\frac12$, we obtain, for positive definite $\bsX,\bsY,\bsZ,\bsW$,
\begin{eqnarray*}
\B_+(\bsX+\bsZ||\bsY+\bsW)
&=&2\B_+\Pa{\frac{\bsX+\bsZ}{2}||\frac{\bsY+\bsW}{2}}\\
&\leqslant&\B_+(\bsX||\bsY)+\B_+(\bsZ||\bsW).
\end{eqnarray*}
For $\varepsilon>0$ and positive semidefinite $\bsX,\bsY$, define
$$
f_\varepsilon(\bsX,\bsY):=\B_+(\bsX+\varepsilon\I||\bsY+\varepsilon\I).
$$
If $0<\delta<\varepsilon$, apply the preceding inequality with the first pair $(\bsX+\delta\I,\bsY+\delta\I)$ and the second pair $((\varepsilon-\delta)\I,(\varepsilon-\delta)\I)$. All four matrices are positive definite, and the divergence of the second pair is zero. Hence
$$
f_\varepsilon(\bsX,\bsY)\leqslant f_\delta(\bsX,\bsY).
$$
Thus $f_\varepsilon(\bsX,\bsY)$ increases as $\varepsilon\downarrow0$. The regularization identity proved in item 5 implies, with $+\infty$ allowed,
$$
\B_+(\bsX||\bsY)=\sup_{\varepsilon>0}f_\varepsilon(\bsX,\bsY).
$$
Indeed, for every fixed $\varepsilon>0$, letting $\delta\downarrow0$ in the preceding inequality gives $f_\varepsilon(\bsX,\bsY)\leqslant\B_+(\bsX||\bsY)$; conversely, the supremum is at least the limit along $\varepsilon\downarrow0$.

For each fixed $\varepsilon>0$, the function $f_\varepsilon$ is continuous on pairs of positive semidefinite matrices: both regularized matrices are positive definite, the matrix logarithm is continuous there, and matrix multiplication and trace are continuous. Consequently, if $\bsX_n\to\bsX$ and $\bsY_n\to\bsY$ in the positive semidefinite cone, then
$$
f_\varepsilon(\bsX,\bsY)
=\lim_{n\to\infty}f_\varepsilon(\bsX_n,\bsY_n)
\leqslant\liminf_{n\to\infty}\B_+(\bsX_n||\bsY_n).
$$
Taking the supremum over $\varepsilon>0$ yields
$$
\B_+(\bsX||\bsY)\leqslant\liminf_{n\to\infty}\B_+(\bsX_n||\bsY_n),
$$
which proves lower semicontinuity on the whole product of the positive semidefinite cones, without any equal-trace condition. In particular, if $\op{supp}[\bsX]\not\subseteq\op{supp}[\bsY]$, then $\B_+(\bsX||\bsY)=+\infty$, so the same inequality gives $\B_+(\bsX_n||\bsY_n)\to+\infty$.
\item \textbf{The regularization identity.} We prove $\B_+(\bsX|| \bsY) =\lim_{\varepsilon\downarrow0}\B_+(\bsX+\varepsilon\I|| \bsY+\varepsilon\I)$.
Let $\bsP=\op{supp}[\bsY]$ and $\bsQ=\I-\bsP$.

\paragraph{Case 1: $\op{supp}[\bsX]\subseteq \bsP$.} Then $\bsX=\bsP\bsX\bsP, \bsY=\bsP\bsY\bsP$, and $\bsY_{\bsP}=\bsY|_{\bsP}>0$. Decompose
$$
\bsX+\varepsilon\I=(\bsX+\varepsilon \bsP)\oplus \varepsilon \bsQ,
\qquad
\bsY+\varepsilon\I=(\bsY+\varepsilon \bsP)\oplus \varepsilon \bsQ.
$$
Since both matrices are block diagonal in this decomposition, and since $\B_+(\varepsilon \bsQ|| \varepsilon \bsQ)=0$, we get
$$
\B_+(\bsX+\varepsilon\I|| \bsY+\varepsilon\I) =
\B_+(\bsX+\varepsilon \bsP|| \bsY+\varepsilon \bsP).
$$
On $\bsP, \bsY+\varepsilon \bsP\to \bsY_{\bsP}>0$, and $\bsX+\varepsilon \bsP\to \bsX$. The function $t\mapsto t\log t$ is continuous at $0$, so
$(\bsX+\varepsilon \bsP)\log(\bsX+\varepsilon \bsP)\to \bsX\log \bsX$ in trace norm, and
$(\bsX+\varepsilon \bsP)\log(\bsY+\varepsilon \bsP)\to \bsX\log \bsY_{\bsP}$ because $\log$ is continuous on the positive definite cone. Hence
$\B_+(\bsX+\varepsilon \bsP|| \bsY+\varepsilon \bsP)\to \B_+(\bsX|| \bsY)$. Therefore
$\B_+(\bsX+\varepsilon\I|| \bsY+\varepsilon\I)\to \B_+(\bsX|| \bsY)$.

\paragraph{Case 2: $\op{supp}[\bsX]\not\subseteq \bsP$.} Then the right-hand side is $+\infty$. We show that the left-hand side diverges. Since $\bsX$ has nonzero mass on $\bsQ=\ker [\bsY]$, we have $c:=\Tr{\bsQ\bsX\bsQ}>0$. For $\varepsilon>0$,
$-\Tr{(\bsX+\varepsilon\I)\log(\bsY+\varepsilon\I)}$ contains the contribution from the block $\bsQ$, where $\bsY+\varepsilon\I=\varepsilon \bsI_{\bsQ}$. That contribution is
$$
-\Tr{(\bsQ\bsX\bsQ+\varepsilon \bsI_{\bsQ})\log(\varepsilon \bsI_{\bsQ})}
=
-\Tr{\bsQ\bsX\bsQ+\varepsilon \bsI_{\bsQ}}\log\varepsilon
\to+\infty
$$
as $\varepsilon\downarrow0$. The remaining terms are bounded or converge. Hence
$\B_+(\bsX+\varepsilon\I|| \bsY+\varepsilon\I)\to+\infty=\B_+(\bsX|| \bsY)$. Combining both cases proves
$$
\B_+(\bsX|| \bsY)=\lim_{\varepsilon\downarrow0}\B_+(\bsX+\varepsilon\I|| \bsY+\varepsilon\I)
$$
for all positive semidefinite $\bsX,\bsY$, with the support convention understood.
\end{enumerate}
This completes the proof of Proposition~\ref{prop:GREproperty}.

\subsection{The generalized relative entropy and the trace correction}

The Haar integral also appears in the Bregman divergence adapted to $\S_k^{(d)}$. The generalized relative entropy
$$
\B_{+}(\bsX||\bsY) = \Tr{\bsX(\log \bsX - \log \bsY)} - \Tr{\bsX - \bsY}
$$
is the correct divergence for compressed matrices because $\bsV^{\dagger}\bsX\bsV$ and $\bsV^{\dagger}\bsY\bsV$ generally have different traces. The trace correction $-\Tr{\bsX-\bsY}$ ensures nonnegativity and preserves joint convexity on the positive semidefinite cone, properties that are inherited by the adapted divergence
$$
\B_k^{(d)}(\bsX||\bsY) = \sum_{\ell=1}^{d-k+1} \mathfrak{C}_{d,k,\ell} \int_{\op{St}(d,\ell)} \B_{+}(\bsV^{\dagger}\bsX\bsV|| \bsV^{\dagger}\bsY\bsV)  \dif\mu_{\ell}(\bsV).
$$
The coefficients $\mathfrak{C}_{d,k,\ell}$ are those of \eqref{eq:constantC}. The proof that the pinching gain is convex in the input state relies on the joint convexity of $\B_{+}$ and the linearity of the compression maps $\bsX \mapsto \bsV^{\dagger}\bsX\bsV$. This is the mechanism behind the pinching lemma (Lemma~\ref{lem:pinching}) and the entropy-gain bounds for unital channels (Theorem~\ref{th:channel}).

\subsection{Proof of Proposition~\ref{prop:BD-for-kentropy}}

We recall the positive-compression representation from Proposition~\ref{prop:posrep}. For $2\leqslant k\leqslant d$ and $\bsX\geqslant \zero$,
$$
\S_k^{(d)}(\bsX) = \sum_{\ell=1}^{d-k+1}\mathfrak C_{d,k,\ell}\G_\ell(\bsX)
-(H_k-1)\Tr{\bsX},
$$
where
$$
\G_\ell(\bsX)=\int_{\op{St}(d,\ell)}\Tr{\eta(\bsV^\dagger \bsX\bsV)}\dif\mu_\ell(\bsV),
\qquad
\eta(x)=-x\log x,
$$
and
$$
\mathfrak C_{d,k,\ell}=\frac{k}{\binom{d-1}{k-1}}\binom{d-\ell-1}{k-2}>0,
\qquad 1\leqslant \ell\leqslant d-k+1.
$$
For $k=1$, $\S_1^{(d)}=\G_d$. Let $F_k:=-\S_k^{(d)}$. The Bregman divergence associated with $F_k$ is
$$
\B_k^{(d)}(\bsX|| \bsY) = F_k(\bsX)-F_k(\bsY) - \Inner{\sL_k(\bsY)}{\bsX-\bsY}, \qquad
\sL_k(\bsY)=\nabla F_k(\bsY),
$$
where the gradient is taken with respect to the real Hilbert--Schmidt inner product on Hermitian matrices.

We prove that for $\bsX,\bsY>0$ and $k\geqslant 2$,
$$
\B_k^{(d)}(\bsX|| \bsY) = \sum_{\ell=1}^{d-k+1}
\mathfrak C_{d,k,\ell}\int_{\op{St}(d,\ell)}\B_+(\bsV^\dagger \bsX\bsV|| \bsV^\dagger \bsY\bsV)\dif\mu_\ell(\bsV).
$$

\paragraph{Bregman divergence of the negative compressed functional $-\G_\ell$.} Fix $1\leqslant\ell\leqslant d$. For $\bsZ>0$, define $g(\bsZ) :=\Tr{\eta(\bsZ)} = -\Tr{\bsZ\log\bsZ}$. Its gradient is $\nabla g(\bsZ)=-\log\bsZ-\I$.

We first justify differentiation under the Haar integral. Fix $\bsY>\zero$ and set $a=\lambda_{\min}(\bsY)/2>0$ and $b=\norm{\bsY}+a$, where $\norm{\cdot}$ denotes the operator norm. For every Hermitian $\bsZ$ satisfying $\norm{\bsZ-\bsY}\leqslant a$ and every $\bsV\in\op{St}(d,\ell)$, we have
$$
a\I_\ell\leqslant\bsV^\dagger\bsZ\bsV\leqslant b\I_\ell.
$$
The compressed integrand and its Fr\'{e}chet derivative are jointly continuous in $(\bsZ,\bsV)$ on this neighborhood. Moreover, for every Hermitian $d\times d$ matrix $\bsH$, the derivative in direction $\bsH$ satisfies
\begin{eqnarray*}
\left|\Tr{\Pa{\log(\bsV^\dagger\bsZ\bsV)+\I_\ell}\bsV^\dagger\bsH\bsV}\right|
&\leqslant&\ell\Pa{1+\max\{\lvert\log a\rvert,\lvert\log b\rvert\}}\norm{\bsH},
\end{eqnarray*}
since $\norm{\bsV^\dagger\bsH\bsV}\leqslant\norm{\bsH}$. This bound is uniform in $\bsV$ and integrable with respect to the Haar probability measure. Thus Fr\'{e}chet differentiation may be passed under the integral. Using the cyclicity of the trace, the gradient of $\G_\ell$ at $\bsY>\zero$ is therefore
\begin{eqnarray}\label{appeq:gradient-of-Gell}
\nabla \G_\ell (\bsY) = \int_{\op{St}(d,\ell)} \bsV\Pa{-\log(\bsV^\dagger \bsY\bsV)-\I_\ell}\bsV^\dagger
\dif\mu_\ell(\bsV).
\end{eqnarray}
Therefore, for $\bsX,\bsY>0$, $-\G_\ell$ induces the following Bregman divergence:
\begin{eqnarray}\label{appeq:Bregman-Gell}
&&\B_{-\G_\ell}(\bsX||\bsY) := [-\G_\ell(\bsX)] - [-\G_\ell(\bsY)] - \Tr{\nabla (-\G_\ell)(\bsY)(\bsX-\bsY)}\\
&&= \G_\ell(\bsY)-\G_\ell(\bsX) + \Tr{\nabla \G_\ell(\bsY)(\bsX-\bsY)} \notag\\
&&= \int_{\op{St}(d,\ell)} \Br{g(\bsV^\dagger\bsY\bsV) - g(\bsV^\dagger\bsX\bsV) - \Tr{\bsV\Pa{\log(\bsV^\dagger \bsY\bsV)+\I_\ell}\bsV^\dagger (\bsX-\bsY)}}\dif\mu_\ell(\bsV)\notag\\
&&=  \int_{\op{St}(d,\ell)} \Br{g(\bsV^\dagger\bsY\bsV) - g(\bsV^\dagger\bsX\bsV) - \Tr{\Pa{\log(\bsV^\dagger \bsY\bsV)+\I_\ell}\Pa{\bsV^\dagger \bsX\bsV-\bsV^\dagger\bsY\bsV}}}\dif\mu_\ell(\bsV).\notag
\end{eqnarray}
For fixed $\bsV\in\op{St}(d,\ell)$, write $\bsX_{\bsV}:=\bsV^\dagger\bsX\bsV$ and  $\bsY_{\bsV}:=\bsV^\dagger\bsY\bsV$. Using $g(\bsZ)=-\Tr{\bsZ\log\bsZ}$, the integrand in the above equals
\begin{eqnarray*}
&&\Tr{\bsX_{\bsV}\log\bsX_{\bsV}} - \Tr{\bsY_{\bsV}\log\bsY_{\bsV}} - \Tr{(\log\bsY_{\bsV}+\I_\ell)(\bsX_{\bsV}-\bsY_{\bsV})}\\
&&=\Tr{\bsX_{\bsV}(\log\bsX_{\bsV}-\log\bsY_{\bsV})} - \Tr{\bsX_{\bsV}-\bsY_{\bsV}} = \B_+(\bsX_{\bsV}||\bsY_{\bsV}).
\end{eqnarray*}
Thus \eqref{appeq:Bregman-Gell} is reduced to
$$
\B_{-\G_\ell}(\bsX||\bsY) = \int_{\op{St}(d,\ell)}\B_+(\bsX_{\bsV}||\bsY_{\bsV})\dif\mu_\ell(\bsV)=\int_{\op{St}(d,\ell)}\B_+(\bsV^\dagger \bsX\bsV|| \bsV^\dagger \bsY\bsV)\dif\mu_\ell(\bsV).
$$
\paragraph{Bregman divergence of the negative $k$-entropy $-\S^{(d)}_k$.}

Since 
$$
F_k(\bsX) = -\S^{(d)}_k(\bsX) = \sum^{d-k+1}_{\ell=1}\mathfrak{C}_{d,k,\ell}[-\G_\ell(\bsX)] + (H_k-1)\Tr{\bsX},
$$
it follows from \eqref{appeq:additive-of-(f+g)} that
\begin{eqnarray*}
\B_{F_k} &=& \sum^{d-k+1}_{\ell=1}\mathfrak{C}_{d,k,\ell} \B_{-\G_\ell} + (H_k-1)\B_{\trace}=\sum^{d-k+1}_{\ell=1}\mathfrak{C}_{d,k,\ell} \B_{-\G_\ell},
\end{eqnarray*}
where we used the fact that $\B_{\trace}\equiv0$, as easily checked. Thus
\begin{eqnarray*}
\B^{(d)}_k(\bsX||\bsY)=\B_{F_k}(\bsX||\bsY) = \sum^{d-k+1}_{\ell=1}\mathfrak{C}_{d,k,\ell}\int_{\op{St}(d,\ell)}\B_+(\bsV^\dagger \bsX\bsV|| \bsV^\dagger \bsY\bsV)\dif\mu_\ell(\bsV).
\end{eqnarray*}
For $k=1$, one has $S^{(d)}_1(\bsX)=\Tr{\eta(\bsX)}$, so $F_1(\bsX)=\Tr{\bsX\log\bsX}$ and
$\nabla F_1(\bsY)=\log\bsY+\I_d$. Hence 
$$
\B_1^{(d)}(\bsX||\bsY) = \Tr{\bsX(\log\bsX-\log\bsY)}-\Tr{\bsX-\bsY}=\B_+(\bsX||\bsY).
$$
\paragraph{The Bregman identity \eqref{eq:adaptedBD}.} By definition,
$$
\sL_k(\bsY)=\nabla F_k(\bsY)=\nabla\Pa{-\S_k^{(d)}}(\bsY).
$$
From the previous discussion,
$$
\sL_k(\bsY)= \sum_{\ell=1}^{d-k+1}\mathfrak C_{d,k,\ell}\int_{\op{St}(d,\ell)} \bsV(\log(\bsV^\dagger \bsY\bsV)+\I_\ell)\bsV^\dagger
\dif\mu_\ell(\bsV) +(H_k-1)\I_d.
$$
For $k=1$, $\sL_1(\bsY)=\log \bsY+\I_d$. Therefore, for $\bsX,\bsY>0$,
$$
\B_k^{(d)}(\bsX|| \bsY)=F_k(\bsX)-F_k(\bsY)-\Tr{\sL_k(\bsY)(\bsX-\bsY)}
$$
which is exactly \eqref{eq:adaptedBD}. 

\paragraph{Nonnegativity and joint convexity.}

For each fixed $\bsV$, the map $(\bsX,\bsY)\mapsto \B_+(\bsV^\dagger \bsX\bsV|| \bsV^\dagger \bsY\bsV)$
is jointly convex in $(\bsX,\bsY)$, because compression is linear and $\B_+$ is jointly convex by Proposition~\ref{prop:GREproperty}. It is also nonnegative by Proposition~\ref{prop:GREproperty}. Since the coefficients $\mathfrak C_{d,k,\ell}$ are positive and Haar averaging preserves joint convexity and nonnegativity, the right-hand side of \eqref{eq:BregDivforkentropy} is nonnegative and jointly convex in $(\bsX,\bsY)$. For positive semidefinite $\bsX,\bsY$, we define $\B_k^{(d)}$ by the same Haar integral, allowing the value $+\infty$. The support convention for $\B_+$ is applied separately to each compressed pair $\bsV^\dagger\bsX\bsV,\bsV^\dagger\bsY\bsV$. Integrating its pointwise joint-convexity inequality proves joint convexity directly on the positive semidefinite cone, for arbitrary traces.

For completeness, this integral definition agrees with simultaneous regularization. Joint convexity and positive homogeneity of $\B_+$ imply subadditivity under simultaneous addition:
$$
\B_+(\bsA+\bsC||\bsB+\bsD)
\leqslant\B_+(\bsA||\bsB)+\B_+(\bsC||\bsD).
$$
Taking $\bsC=\bsD=\varepsilon\I_\ell$ and using Proposition~\ref{prop:GREproperty} gives the pointwise regularization limit. Since $\bsV^\dagger\I_d\bsV=\I_\ell$, subadditivity and $\B_+(\varepsilon\I_\ell||\varepsilon\I_\ell)=0$ give, for every $\varepsilon>0$,
\[
0\leqslant\B_+(\bsV^\dagger(\bsX+\varepsilon\I_d)\bsV||\bsV^\dagger(\bsY+\varepsilon\I_d)\bsV)
\leqslant\B_+(\bsV^\dagger\bsX\bsV||\bsV^\dagger\bsY\bsV).
\]
If the limiting Haar integral is finite, dominated convergence applies to these
nonnegative integrands. If it is infinite, Fatou's lemma applied to any
sequence $\varepsilon_n\downarrow0$ shows that the regularized integrals have
liminf $+\infty$. Hence, in the extended-real sense,
\begin{eqnarray*}
\B_k^{(d)}(\bsX||\bsY)=\lim_{\varepsilon\downarrow0}
\B_k^{(d)}(\bsX+\varepsilon\I_d||\bsY+\varepsilon\I_d).
\end{eqnarray*}

\paragraph{Lower semicontinuity.}

For fixed $\bsV$, the integrand $(\bsX,\bsY)\mapsto \B_+(\bsV^\dagger \bsX\bsV||\bsV^\dagger \bsY\bsV)$ is lower semicontinuous and nonnegative. Therefore, by Fatou's lemma,
$$
\B_k^{(d)}(\bsX|| \bsY) \leqslant \liminf_{n\to\infty} \B_k^{(d)}(\bsX_n|| \bsY_n)
$$
whenever $\bsX_n\to \bsX, \bsY_n\to \bsY$. Hence $\B_k^{(d)}$ is lower semicontinuous on pairs of positive semidefinite matrices. Its value is determined by the compressed integrands; failure of $\op{supp}[\bsX]\subseteq\op{supp}[\bsY]$ for the original matrices does not by itself force this Haar average to be infinite.

\paragraph{Unitary invariance.} Let $\bsU$ be unitary. In the Haar integral for $\B_k^{(d)}(\bsU\bsX\bsU^\dagger|| \bsU\bsY\bsU^\dagger)$, make the change of variables $\bsV=\bsU\bsW,\qquad \bsW\in \op{St}(d,\ell)$. Since $\bsW^\dagger \bsW=\bsV^\dagger \bsV=\I_\ell$ and the Haar measure is invariant, $\dif\mu_\ell(\bsV)=\dif\mu_\ell(\bsW)$. Moreover,
$$
\bsV^\dagger (\bsU\bsX\bsU^\dagger)\bsV=\bsW^\dagger \bsX\bsW,\qquad \bsV^\dagger (\bsU\bsY\bsU^\dagger)\bsV=\bsW^\dagger \bsY\bsW.
$$
Therefore $\B_k^{(d)}(\bsU\bsX\bsU^\dagger|| \bsU\bsY\bsU^\dagger)=\B_k^{(d)}(\bsX|| \bsY)$.

\paragraph{The endpoint $k=d$.}

For $k=d$, the sum in \eqref{eq:BregDivforkentropy} has only the term $\ell=1$, and from the definition of the constant $\mathfrak{C}_{d,k,\ell}$,
$$
\mathfrak C_{d,d,1}=\frac{d}{\binom{d-1}{d-1}}\binom{d-2}{d-2}=d.
$$
Hence
$$
\B_d^{(d)}(\bsX|| \bsY)=d\int_{\op{St}(d,1)} \B_+\Pa{\Innerm{\bsv}{\bsX}{\bsv}|| \Innerm{\bsv}{\bsY}{\bsv}}
\dif\mu_1(\bsv),
$$
which is the scalar-compression formula stated in Proposition~\ref{prop:BD-for-kentropy}, completing the proof.

\section{Exact polynomial certificates}\label{app:D}

The two pointwise polynomial comparisons used in Sections~\ref{subsubsect:4.5.4}
and~\ref{subsubsect:4.5.5} are exact integer calculations. The calculation
below reconstructs all four coefficient maps directly from the displayed
polynomial identities. The source package also contains the machine-readable
file \texttt{polynomial-certificates.json} and the same calculation as
\texttt{polynomial-certificate-check.py}.

Each nonzero monomial is specified by six integers
\begin{eqnarray*}
 (e_t,e_v,e_h,e_q,e_r,c),\qquad
 c\,t^{e_t}v^{e_v}h^{e_h}q^{e_q}r^{e_r},
\end{eqnarray*}
with variables in the order $(t,v,h,q,r)$ and $c\in\bbZ$.
In the JSON file a term is stored as an object with fields
\texttt{exponents}, whose value is the five-entry array
$[e_t,e_v,e_h,e_q,e_r]$, and \texttt{coefficient}, whose value is $c$.
The file contains the subentropy polynomial $\cP_\kappa$ and the $k=2$
polynomial $\cP^{(2)}_\kappa$ on each of the two charts. Its SHA-256 hash is
\begin{eqnarray*}
\texttt{b3640659c7b55e8a81531e5c781da10832f973e7875038c4322f11860db52930}.
\end{eqnarray*}

The calculation substitutes the corresponding row of the following table
into the polynomial identities \eqref{eq:certificate-identity} and
\eqref{eq:k2-certificate}, expands over $\bbZ$, and collects the
coefficients in the variables $(t,v,h,q,r)$:
\begin{center}
\begin{tabular}{c|cccc}
region & $x$ & $y$ & $z$ & $w$\\ \hline
$x\geqslant y$ & $v+h$ & $v$ & $v+q$ & $v+h+q+r$\\
$x\leqslant y$ & $v$ & $v+h$ & $v+h+q$ & $v+q+r$.
\end{tabular}
\end{center}
The parameters $v,h,q,r$ are nonnegative, and the two rows cover the whole
ordered region $0\leqslant x+y\leqslant x+z\leqslant\kappa/2$; the overlap
$h=0$ is harmless. The four maps have respectively
\begin{eqnarray*}
 (2657,2028)\quad\text{and}\quad(2664,2034)
\end{eqnarray*}
positive coefficients for $(\cP_\kappa,\cP^{(2)}_\kappa)$, with no negative
coefficient. Their total degrees and $t$-degrees are $(14,12)$ and
$(13,11)$, respectively. Thus every term is nonnegative for
$t,v,h,q,r\geqslant0$, which proves the sign assertion used in the two
sharp comparisons. The leading coefficients in
\eqref{eq:leading-certificate} and below \eqref{eq:k2-certificate}
give the strictness argument.

\newpage
The following Python~3 code uses SymPy only for exact expansion over the
integer polynomial ring. It contains every defining expression, checks the
full coefficient signs and degrees, and separately checks the two leading
coefficient identities. It does not read or assume the coefficient file.
In the code, \texttt{kap}, \texttt{a}, \texttt{b}, \texttt{A},
\texttt{B}, \texttt{C}, \texttt{N}, and \texttt{N2} denote
$\kappa,a',b',A_\kappa,B_\kappa,C_\kappa,N_\kappa$, and
$N^{(2)}_\kappa$, respectively.

\begin{samepage}
{\footnotesize
\begin{verbatim}
from sympy import Poly, symbols

t, v, h, q, r = symbols('t v h q r')
charts = [(v+h, v, v+q, v+h+q+r),
          (v, v+h, v+h+q, v+q+r)]
for j, (x, y, z, w) in enumerate(charts):
    kap = x+y+z+w
    a, b = x+y, x+z
    A = t*t+kap*t+a*(kap-a)
    B = t*t+kap*t+b*(kap-b)
    C = (t+x)*(t+y)*(t+z)*(t+w)
    U, V = (2*t+kap)**2, (4*t+kap)**4
    E = (128*t**4+64*kap*t**3+28*kap**2*t*t
         +12*kap**3*t+kap**4)
    N = (A*B*C-t*(t+kap)*(A+B)*C
         +t**3*(t+kap)*A*B)
    e2 = x*y+x*z+x*w+y*z+y*w+z*w
    N2 = (t*t*(6*t*t+3*kap*t+e2)*(t+kap)*A*B
          -t*(5*t+2*kap)*(t+kap)*(A+B)*C
          +(4*t+kap)*A*B*C)
    V2, E2 = (4*t+kap)**2, kap**3+12*kap*t*t+40*t**3
    exprs = [N*U*V-4*a*b*E*A*B*C,
             N2*U*V2-4*a*b*E2*A*B*C]
    lead = h*h+h*q+2*v*(q+r)+(2*h*r if j == 0 else 0)
    for i, expr in enumerate(exprs):
        p = Poly(expr, t, v, h, q, r, domain='ZZ')
        count = ((2657, 2028), (2664, 2034))[j][i]
        assert len(p.terms()) == count
        assert all(c > 0 for c in p.coeffs())
        assert all(sum(m) == (14, 13)[i] for m in p.monoms())
        assert p.degree(t) == (12, 11)[i]
        lc = Poly(p.as_expr(), t).LC()
        assert Poly(lc-(512, 160)[i]*lead, v, h, q, r).is_zero
        print(j+1, ('Q', 'k=2')[i], count,
              p.total_degree(), p.degree(t))
\end{verbatim}
}
\end{samepage}

\newpage
\begin{samepage}
For a direct comparison with the JSON file, construct the dictionary
\texttt{dict(p.terms())} in each iteration. Its keys are the exponent
five-tuples and its values are the exact integer coefficients. Compare it
with the dictionary formed from the \texttt{exponents} and
\texttt{coefficient} fields of the corresponding row of the file; equality
checks every term, including the absence of additional terms. The two
rows of \texttt{charts} correspond to \texttt{x\_ge\_y} and
\texttt{x\_le\_y}, and the two expressions correspond to
\texttt{subentropy} and \texttt{k2}. The four printed lines are
{\footnotesize
\begin{verbatim}
1 Q   2657 14 12
1 k=2 2028 13 11
2 Q   2664 14 12
2 k=2 2034 13 11
\end{verbatim}
}
\end{samepage}


\end{document}